\documentclass[12pt,a4paper]{article}
\pdfoutput=1

\usepackage[nosort]{cite}
\usepackage[height=8.85in,width=6.45in]{geometry}
\usepackage[whole]{bxcjkjatype} 

\usepackage{ascmac}

\usepackage{makecell}
\usepackage{multirow}
\usepackage{bbm}

\usepackage{pifont}

\newcommand{\zz}{$\mathbb{Z}_2$ }
\newcommand{\zzxzz}{$\mathbb{Z}_2 \times \mathbb{Z}_2$ }
\newcommand{ \lskip}{\vspace{\baselineskip}}

\usepackage{times}

\usepackage{comment}
\usepackage{graphicx}

\usepackage[utf8]{inputenc}
\usepackage{amsmath}
\usepackage{amssymb}
\usepackage{mathtools}
\numberwithin{equation}{section}
\usepackage{slashed}
\usepackage{braket}
\usepackage[svgnames,dvipsnames]{xcolor}
\usepackage[colorlinks,citecolor=DarkGreen,linkcolor=FireBrick,linktocpage,pagebackref]{hyperref}

\usepackage{cite}
\usepackage{graphicx}
\usepackage{tikz}
\usepackage{tikz-cd}
\usepackage{tikzit}

\tikzstyle{vertex}=[fill=black, draw=black, shape=circle]
\tikzstyle{dualvertex}=[fill=white, draw=black, shape=circle]
\tikzstyle{boundaryvertex}=[fill=red, draw=red, shape=circle]
\tikzstyle{boundarydualvertex}=[fill=white, draw=red, shape=circle]

\tikzstyle{boundary}=[-, draw=red, thick, dashed]
\tikzstyle{lineoperator}=[-, draw=blue, ultra thick]
\tikzstyle{dotline}=[-, dashed]

\usepackage{orcidlink}

\usepackage{courier}
\usepackage{bm}

\usepackage{dashbox}
\usepackage{subcaption}
\usepackage{enumitem}

\usepackage{mdframed}

\usepackage{blkarray}
\usepackage{arydshln}
\usepackage{dsfont}
\usetikzlibrary{patterns}
\usetikzlibrary{arrows.meta}

\usepackage{calc}
\usepackage{physics}

\newcommand{\vertex}{\mathbf{V}}
\newcommand{\dualvertex}{\mathbf{\Hat{V}}}

\newcommand*{\cub}[1]{\mathbf{Cub}(#1)}
\newcommand*{\vhat}{\hat{v}}
\newcommand*{\Xhat}{\hat{X}}
\newcommand*{\Zhat}{\hat{Z}}
\newcommand*{\Yhat}{\hat{Y}}
\newcommand*{\Uhat}{\hat{U}}

\newcommand*{\phat}{\hat{p}}
\newcommand{\Dee}{\mathrm{D}}
\newcommand*{\Deight}{\Dee_8}
\newcommand*{\Xtilde}{\tilde{X}}
\newcommand*{\Ztilde}{\tilde{Z}}
\newcommand*{\Ytilde}{\tilde{Y}}
\newcommand*{\Xeff}{\Xhat^{\mathrm{eff}}}
\newcommand*{\Yeff}{\Yhat^{\mathrm{eff}}}
\newcommand*{\Zeff}{\Zhat^{\mathrm{eff}}}
\newcommand*{\Xeffhatnasi}{X^{\mathrm{eff}}}
\newcommand*{\Yeffhatnasi}{Y^{\mathrm{eff}}}
\newcommand*{\Zeffhatnasi}{Z^{\mathrm{eff}}}

\newcommand*{\adj}[1]{\mathbf{Adj}(#1)}
\newcommand{\repde}{\mathrm{Rep} (\Dee_8 )}
\newcommand{\repdelike}{$\mathrm{Rep} (\Dee_8 )$-like }
\newcommand{\Deightlike}{$\Dee_8$-like }
\newcommand{\ztwothree}{\mathbb{Z}_2^3 }
\newcommand{\ztwoV}{\mathbb{Z}_2^{\vertex} }
\newcommand{\ztwodualV}{\mathbb{Z}_2^{\dualvertex} }
\newcommand{\ztwoCZ}{\mathbb{Z}_2^{\mathrm{CZ}} }
\newcommand{\VCZ}{V_{\mathrm{CZ}} }
\newcommand{\swap}{\mathbf{Swap} }
\newcommand{\Vhatinterface}{\dualvertex_{\mathrm{interface}}}
\newcommand{\Vinterface}{\vertex_{\mathrm{interface}}}
\newcommand{\Vhateff}{\dualvertex_{\mathrm{eff}}}
\newcommand{\Veff}{\vertex_{\mathrm{eff}}}
\newcommand{\Ztrans}{\mathbb{Z}_{\mathrm{trans}}}
\newcommand{\even}{\mathrm{e}}
\newcommand{\odd}{\mathrm{o}}
\newcommand{\vece}{\mathbf{e}}
\newcommand{\setto}{\mathrm{S}}
\newcommand{\Ve}{\vertex_{\mathrm{e}}}
\newcommand{\Vo}{\vertex_{\mathrm{o}}}

\newif\ifincludeFile
\includeFilefalse

\newcommand{\preprint}[1]{\begin{table}[t]
    \begin{flushright}
    {#1}
    \end{flushright}
    \end{table}}
\renewcommand{\title}[1]{\vbox{\center\LARGE{#1}}\vspace{5mm}}
\renewcommand{\author}[1]{\vbox{\center#1}\vspace{5mm}}

\begin{document}

 \begin{titlepage}

\title{\mbox{} \\\mbox{} \\\mbox{} \\\mbox{} \\\mbox{} \\\mbox{} \\Lattice models with subsystem/weak non-invertible symmetry-protected topological order\mbox{} \\\mbox{} \\\mbox{} }

 ~~\\

\preprint{UT-Komaba/25-3}

\author{Yuki Furukawa \\ \lskip Graduate School of Arts and Sciences, University of Tokyo\\ Komaba, Meguro-ku, Tokyo 153-8902, Japan}
		
	~~	\\

\abstract{\noindent We construct a family of lattice models which possess subsystem non-invertible symmetry-protected topological (SPT) order and analyze their interface modes protected by the symmetry, whose codimension turns out to be more than one. We also propose 2+1d lattice models which belong to two different weak SPT phases distinguished by a combination of translational symmetry and non-invertible symmetry. We show that the interface between them exhibits an exotic Lieb-Schultz-Mattis (LSM) anomaly associated with a modulated symmetry, which cannot be factorized into a direct product of internal and translational symmetries.}

 \end{titlepage}
\tableofcontents

\bigskip

\newpage

\section{Introduction}
Symmetry plays a fundamental role in understanding quantum many-body systems, from field theories to lattice models.
One crucial aspect of symmetry is its ability to classify different phases of matter.
Among these phases, Symmetry-Protected Topological (SPT) phases \cite{Gu:2009dr,Pollmann:2009mhk,Chen:2010zpc,Chen:2011pg} form a class of quantum phases that go beyond Landau's theory of spontaneous symmetry breaking.
While a system with nontrivial SPT order appears gapped and featureless in the bulk, its topological nature manifests through boundary modes that are protected by the symmetry.

At the same time, the concept of symmetry itself has been generalized in multiple directions by reinterpreting it in terms of topological operators.
Higher-form ($p$-form) symmetry extends conventional symmetry by considering codimension-$(p+1)$ topological operators \cite{Gaiotto:2014kfa}.
Non-invertible symmetry is described by fusion category and its generalizations, where topological operators do not necessarily have an inverse \cite{Bhardwaj:2017xup,Chang:2018iay}.
Even conventional $0$-form symmetries can be naturally described in this framework, incorporating 't Hooft anomalies in a unified manner.
Subsystem symmetry arises when the topological nature of operators is restricted to deformations along specific submanifolds in space-time.

These generalized symmetries naturally lead to generalized SPT phases.
In particular, SPT phases protected by non-invertible symmetries have been actively studied in both field theories and lattice models\cite{Thorngren:2019iar,Inamura:2021wuo,Bhardwaj:2023idu,Fechisin:2023dkj,Bhardwaj:2024qrf,Seifnashri:2024dsd,Jia:2024bng,Choi:2024rjm,Li:2024fhy,Inamura:2024jke,Li:2024gwx,Jia:2024zdp,Meng:2024nxx,Cao:2025qhg,Bhardwaj:2025piv,Maeda:2025rxc,Lu:2025rwd}.
Notably, the 2+1d lattice models studied in \cite{Choi:2024rjm} can be viewed as higher-dimensional analogues of the 1+1d constructions introduced in \cite{Seifnashri:2024dsd}.

In this paper, we further extend the construction of non-invertible SPT models \cite{Seifnashri:2024dsd,Choi:2024rjm} by considering more exotic generalizations:
\begin{description}
    \item[Subsystem non-invertible SPT:] Subsystem non-invertible symmetry \cite{Cao:2023doz,Ebisu:2024lie} is characterized by non-group-like fusion rules of symmetry operators, together with restricted mobility.
    We propose new lattice models with subsystem non-invertible SPT order in general dimensions.
    We find that their interface modes are localized on lower-dimensional regions such as corners, hinges and so on.
    \item[Weak non-invertible SPT:] Weak SPT phases are SPT phases which involve both internal symmetry and lattice translational symmetry for their classification\cite{Fu:2006djh,Chen:2011pg}.
    In 2+1d, we propose models that belong to the same non-invertible SPT phase but realizes distinct \emph{weak} non-invertible SPT phases.
    That is, these models cannot be distinguished by non-invertible symmetry alone; rather, both the non-invertible symmetry and the lattice translation symmetry are required to differentiate them.
    The interface mode between them is protected by an exotic Lieb-Schultz-Mattis (LSM) anomaly involving a modulated symmetry, which cannot be factorized into a direct product of internal and translational symmetries.
\end{description}

Let us make some comments:
\begin{itemize}
    \item General non-invertible symmetries lack a well-defined stacking operation which combines two systems with a symmetry into another symmetric system. As a result, unlike conventional SPT phases, we cannot define stacking non-invertible SPT phases and, consequently, the notion of a trivial phase in the usual sense\cite{Thorngren:2019iar,Seifnashri:2024dsd}.
    \item At the interface between subsystem non-invertible SPTs, we find a corner mode with a codimension greater than one. A similar phenomenon can also occur at the boundary of certain (invertible) subsystem SPTs\cite{You:2019bvu,Yamaguchi:2021xeq,You:2024syf}.
    \item Unlike conventional weak SPTs, weak \emph{non-invertible} SPT phases cannot be constructed by stacking lower-dimensional SPTs, due to the absence of a stacking operation.
    \item While 1+1d examples for weak non-invertible SPTs were recently discussed in \cite{Pace:2024acq}, this work presents the first realization in 2+1d.
\end{itemize}

The rest of this paper is organized as follows.
In Section \ref{section_of_bipartite_graphs}, we review qubit systems on bipartite graphs and their various properties, including non-invertible symmetry and dualities.
In Section \ref{Section_of_Non-invertible_SPTs}, we construct a family of lattice models exhibiting subsystem non-invertible SPT phases. We also analyze interfaces between different SPT phases, showing that the symmetry protects interface modes.
In Section \ref{Section_of_weak_SPT}, we explore the interplay between generalized symmetries and lattice translational symmetry. In particular, we propose two lattice models that belong to distinct SPT phases involving a combination of non-invertible symmetry and lattice translational symmetry.
Between these models, we find a symmetry-protected interface mode. We show that the interface theory exhibits the LSM anomaly associated with a modulated symmetry. We also discuss examples of SPT phases distinguished by a combination of higher-form symmetry and translational symmetry.
Section \ref{Section_of_conclusion} summarizes our results and outlines several directions for future study.
In Appendix \ref{appendix_interface}, we present technical details of the analysis of interfaces, including the derivation of non-invertible symmetry operator in the low-energy limit and how the symmetry action on the interface can be extracted from it.

\section{Formulation on bipartite graphs}
\label{section_of_bipartite_graphs}
In this section, we warm up with a review of qubit models and their properties in terms of bipartite graphs.
\subsection{Setup and conventions}

We formulate qubit systems based on biparitite graphs, following the framework of \cite{Gorantla:2024ocs}.
A graph is called bipartite if and only if the set of vertices can be partitioned into two disjoint subsets, $\vertex$ and $\dualvertex$, so that no edge connects two vertices within the same subset.
For simplicity, we assume that the graph does not have multiple edges, i.e., no pair of vertices is connected by more than one edge.
Unless otherwise specified, all the statements in this section can be applied to general bipartite graphs.
We denote the set of vertices adjacent to $v \in \vertex \cup \dualvertex$ by $\adj{v}$.
In a bipartite graph, we have $\adj{v}\subset\dualvertex$ for $v\in\vertex$, and $\adj{\vhat}\subset\vertex$ for $\vhat\in\dualvertex$.

Throughout this paper, we study models on the graph denoted by $\cub{d,p,q}$, where $0\leq p < q \leq d$ are integers.
$d$ denotes the spatial dimension.
The graph $\cub{d,p,q}$ is defined on a $d$-dimensional hypercubic lattice with \begin{itemize}
    \item $\vertex$ = \{$p$-cell\}
    \item $\dualvertex$ = \{$q$-cell = $(d-q)$-cell of the dual lattice\}
\end{itemize}
The set of edges is defined so that $v \in \vertex$ and $\vhat \in \dualvertex$ are adjacent to each other iff $v$ is on the $(q-1)$-dimensional boundary of $\vhat$.
Several examples are shown in Figure \ref{picture_of_cubs}.
\begin{figure}[htbp]
    \begin{center}
    \input{pictures/101and201and202.tikz}
    \caption{Several examples of $\cub{d,p,q}$. Black and white dots indicate the vertices in $\vertex$ and $\dualvertex$, respectively.}
    \label{picture_of_cubs}
    \end{center}
\end{figure}

In Section \ref{Section_of_Ztwo}, we will discuss lattice models with a qubit for each $v\in\vertex$.
Pauli matrices acting on the qubit are $X_v$ and $Z_v$.
They satisfy
\begin{align}
    X_{v}^2 = Z_{v}^2 = 1,\; Z_v X_v = -X_v Z_v.
\end{align}
We also define $Y_{v} = iX_v Z_v$ for convenience.
Similarly, we consider models with a qubit for each $\vhat\in\dualvertex$ whose Pauli matrices are $\Xhat_{\vhat}$ and $\Zhat_{\vhat}$.

In most of this paper — specifically in Sections \ref{Section_of_Ztwotwo}, \ref{Section_of_Ztwothree}, \ref{section_of_repde}, \ref{Section_of_Non-invertible_SPTs}, and \ref{Section_of_weak_SPT} — we will define quantum systems by assigning a qubit for each vertex in $\vertex \cup \dualvertex$.
For $v \in \vertex$, $X_{v}$ and $Z_{v}$ are Pauli operators which act on the qubit defined on $v$.
Similarly, we also consider $\Xhat_{\vhat}, \Yhat_{\vhat}, \Zhat_{\vhat}$ for each (dual) vertex $\vhat \in \dualvertex$.

In Sections \ref{subsection_of_Deight} and \ref{section_of_weak_duality_argument}, we will consider models with \textbf{two} qubits for each vertex in $\vertex$ and \textbf{no} qubits for each vertex in $\dualvertex$.
In this case, we denote by $\Xtilde_{v}, \Ytilde_{v}, \Ztilde_{v}$ the operators acting on the second qubit at $v \in \vertex$, while $X_{v}, Y_{v}, Z_{v}$ denote the Pauli operators acting on the first qubit.

When we work on the square or cubic lattice, we use $s, \ell, p,$ and $c$ to denote a site, a link, a plaquette, and a cube in the square or cubic lattice, respectively.
$\sum_{s}$/$\sum_{\ell}$/$\sum_{p}$/$\sum_{c}$ indicates a summation over all sites/links/plaquettes/cubes.
The symbol $\partial$ denotes the set of one-dimensional lower-dimensional cells located on the boundary.
For example, $\partial p$ is the set of links surrounding the plaquette $p$.

\subsection{$\mathbb{Z}_2$ symmetry and gauging}
\label{Section_of_Ztwo}

In this section, we introduce $\ztwoV$ symmetry and its gauging based on a given bipartite graph.
We consider a Hamiltonian which is an element of the bond algebra\cite{Cobanera:2011wn} defined below,
i.e., it can be written as a linear combination of products of its generators,
\begin{align}
    \label{ztwobondalg}
    \mathcal{B}_{\vertex} = \braket{X_v, \prod_{v'\in\adj{\vhat}} Z_{v'}}{v\in \vertex,\vhat\in \dualvertex}.
\end{align}
The Hamiltonian commutes with
\begin{align}
    U_p &= \prod_{v\in V} X_v^{p_v},
\end{align}
for $p_v \in \{0,1\},\; v \in V$ which satisfy
\begin{align}
    \label{pcond}
    \sum_{v\in \adj{\vhat}}p_v = 0 \mod{2}.
\end{align}
This implies that the system has a $\mathbb{Z}_2$ symmetry.
We denote this symmetry by $\ztwoV$.
However, although $U_{p}$ commutes with the Hamiltonian, it is not necessarily topological in the sense that a $U_{p}$ associated with a contractible object (such as a loop or a closed surface) acts nontrivially on the Hilbert space.
In other words, we cannot freely deform the symmetry operator along spatial directions.
In continuum field theories, when the support of the $\mathbb{Z}_2$ symmetry operator can be shrunk without intersecting with other operators, we can simply shrink it and get a c-number.
This comes from the requirement for the topological nature of the symmetry operator to be regarded as the manifestation of a generalized symmetry.
Coming back to our lattice model, to be precise, the constraint for the topologicalness must be further imposed by restricting the Hilbert space so that the symmetry operator for a contractible cycle acts trivially on it.
The restricted Hilbert space is regarded as the physical one.
This constraint is called the magnetic Gauss law.
Practically, we impose the condition energetically by adding the following flux term\footnote{This term is dual to a c-number via gauging we introduce later.} for each independent contractible object $M$:
\begin{align}
    \label{fluxterms}
    -g A_{M} = - g \prod_{v \in M} X_{v}
\end{align}
to the Hamiltonian. Here, $g>0$ is a coupling constant, and we assume $g$ is sufficiently large.
In the limit $g\to \infty$, this term enforces $A_{M} = 1$ in the low-energy sector, particularly in the ground states.
For some models, these constraints are automatically satisfied for the ground states without adding the flux terms.
We omit the flux terms in this case, since we are mainly interested in the ground states of the model.

Similarly, when we have a qubit for each $\vhat\in\dualvertex$ (without qubits on $\vertex$), we can consider a Hamiltonian which is an element of the bond algebra
\begin{align}
    \mathcal{B}_{\dualvertex} = \braket{\Xhat_{\vhat}, \prod_{\vhat'\in\adj{v}} \Zhat_{\vhat'}}{v\in \vertex,\vhat\in \dualvertex}.
\end{align}
Again, we have the $\ztwodualV$ symmetry generated by
\begin{align}
    \Uhat_{\phat} &= \prod_{\vhat\in \dualvertex} \Xhat_{\vhat}^{\phat_{\vhat}}
\end{align}
for $\phat_{\vhat} \in \{0,1\},\; \vhat \in \dualvertex$ which satisfy
\begin{align}
    \label{phatcond}
    \sum_{\vhat\in \adj{v}}\phat_{\vhat} = 0 \mod{2}.
\end{align}

\paragraph{Gauging}
We review gauging procedure on biparite graphs as a map of operators and models, mimicking the conventional one (See \cite{Shirley:2018vtc,Moradi:2023dan} for example).
In particular, for \emph{physical graphs} on which locality is meaningful, this map does not change local dynamics at all, since it can be regarded as an isomorphism of the bond algebra\cite{Cobanera:2011wn,Moradi:2023dan}.
However, it changes the global properties of the system. For example, it maps one gapped phase to another nontrivially like the prototypical example of Kramers-Wannier duality.

Let us gauge $\ztwoV$ symmetry of the system with qubits on $\vertex$.
Specifically, we introduce gauge fields to promote the global $\ztwoV$ symmetry to a local one.
To implement this, we introduce a qubit for each $\vhat \in \dualvertex$ with Pauli matrices $\sigma_{\vhat}^{x,y,z}$ as a gauge field, and couple them as follows:
\begin{align}
    \begin{split}
        X_{v} &\mapsto X_{v},\\
        \prod_{v\in\adj{\vhat}} Z_{v} &\mapsto \sigma_{\vhat}^{x}\prod_{v\in\adj{\vhat}} Z_{v}.
    \end{split}
\end{align}
This procedure is possible since the Hamiltonian is written in terms of the bond algebra (\ref{ztwobondalg}).
As a result, the system is invariant under the \emph{local gauge transformation} generated by
\begin{align}
    \mathcal{G}_{v} = X_{v} \prod_{\vhat\in \adj{v}} \sigma_{\vhat}^{z}.
\end{align}
We restrict the Hilbert space by imposing the Gauss's law constraints $\mathcal{G}_{v}=1$ for all $v\in\vertex$.
Furthermore, the constraints can be resolved by applying a unitary transformation:
\begin{align}
    V_{\mathrm{CZ}}^{\sigma} = \prod_{v\in V} \prod_{\vhat \in \adj{v}} \mathrm{CZ}^{\sigma}_{v,\vhat},
\end{align}
where
\begin{align}
    \mathrm{CZ}^{\sigma}_{v,\vhat} = \frac{1+Z_v+\sigma_{\vhat}^{z}-Z_v \sigma_{\vhat}^{z}}{2}.
\end{align}
$V_{\mathrm{CZ}}^{\sigma}$ maps operators as follows:
\begin{align}
    \begin{split}
        V_{\mathrm{CZ}}^{\sigma} X_v \left(V_{\mathrm{CZ}}^{\sigma}\right)^\dagger &= X_{v} \prod_{\vhat\in \adj{v}} \sigma_{\vhat}^{z} =\mathcal{G}_v,\\
        V_{\mathrm{CZ}}^{\sigma} Z_v \left(V_{\mathrm{CZ}}^{\sigma}\right)^\dagger &= Z_v,\\
        V_{\mathrm{CZ}}^{\sigma} \sigma_{\vhat}^{x} \left(V_{\mathrm{CZ}}^{\sigma}\right)^\dagger &= \sigma_{\vhat}^{x} \prod_{v\in\adj{\vhat}} Z_{v},\\
        V_{\mathrm{CZ}}^{\sigma} \sigma_{\vhat}^{z} \left(V_{\mathrm{CZ}}^{\sigma}\right)^\dagger &= \sigma_{\vhat}^{z}.
    \end{split}
\end{align}
After changing the frame of the Hilbert space with the unitary operator $V_{\mathrm{CZ}}^{\sigma}$, the system is described by the bond algebra
\begin{align}
    \tilde{\mathcal{B}}_{\vertex} = \braket{X_v \prod_{\vhat\in \adj{v}}\sigma^z_{\vhat},\; \sigma_{\vhat}^{x}}{v\in \vertex,\vhat\in \dualvertex},
\end{align}
with Gauss's law constraint $X_v = 1$. Now, we can simply remove the qubits on $v\in\vertex$ by this constraint. Finally, we obtain the gauged system generated by
\begin{align}
    \mathcal{B}^{\mathrm{gauged}}_{\vertex} = \braket{\prod_{\vhat\in \adj{v}}\sigma^z_{\vhat},\; \sigma_{\vhat}^{x}}{v\in \vertex,\vhat\in \dualvertex}.
\end{align}

Through the gauging procedure, $\ztwoV$ symmetric local operators are mapped as follows:
\begin{align}
    \begin{split}
        X_{v} &\mapsto \prod_{\vhat\in \adj{v}}\sigma^z_{\vhat}, \\
        \prod_{v\in\adj{\vhat}} Z_{v} &\mapsto \sigma_{\vhat}^{x}.
    \end{split}
\end{align}
According to \cite{Gorantla:2024ocs}, by utilizing ZX-calculus\cite{Coecke:2008lcg,Duncan:2009ocf}, we can construct an operator $\Dee^{\dualvertex \leftarrow \vertex}$ implementing this map as
\begin{align}
    \begin{split}
        \Dee^{\dualvertex \leftarrow \vertex} X_{v} &= \left(\prod_{\vhat\in \adj{v}}\sigma^z_{\vhat}\right) \Dee^{\dualvertex \leftarrow \vertex}, \\
        \Dee^{\dualvertex \leftarrow \vertex} \left(\prod_{v\in\adj{\vhat}} Z_{v}\right) &= \sigma_{\vhat}^{x} \Dee^{\dualvertex \leftarrow \vertex}.
    \end{split}
\end{align}
The operator satisfy the following algebra \cite{Gorantla:2024ocs}:
\begin{align}
    \begin{split}
        \Dee^{\dualvertex \leftarrow \vertex} U_p = \Uhat_{\phat}^{\sigma} \Dee^{\dualvertex \leftarrow \vertex} = \Dee^{\dualvertex \leftarrow \vertex}.
    \end{split}
\end{align}

Notably, the resulting system has a \emph{dual} $\mathbb{Z}_2$ symmetry generated by
\begin{align}
    \Uhat_{\phat}^{\sigma} &= \prod_{\vhat\in \dualvertex} (\sigma_{\vhat}^{x})^{\phat_{\vhat}}
\end{align}
for $\phat_{\vhat} \in \{0,1\},\; \vhat \in \dualvertex$ which satisfies
\begin{align}
    \sum_{\vhat\in \adj{v}}\phat_{\vhat} = 0 \mod{2}.
\end{align}
It can be identified as $\ztwodualV$ symmetry.
However, at this point, the dual symmetry is not topological since its operator cannot be deformed freely in the spatial directions.
As reviewed above, we impose the magnetic Gauss law by adding the flux terms which are dual to c-numbers.

Similarly, we define gauging $\ztwodualV$ by exchanging the role of $\vertex$ and $\dualvertex$.
Starting from a $\ztwoV$ symmetric system, we can sequentially gauge $\ztwoV$ and then its dual symmetry $\ztwodualV$, which brings us back to the original system (with the magnetic Gauss law imposed).

\paragraph{Examples}
We review several lattice models from the viewpoint of our framework.
Given a bipartite graph, we define the Ising-type model as
\begin{align}
    H_{\text{Ising-type}} = - \sum_{\vhat\in\dualvertex} \prod_{v\in\adj{\vhat}} Z_{v} + (\text{flux terms}).
\end{align}
For example, the Ising-type model for $\cub{d,0,1}$
\begin{align}
    H_{\mathrm{Ising}} = - \sum_{\ell} \prod_{s\in\partial\ell} Z_{s}
\end{align}
is the usual $(d+1)$-dimensional Ising model with $\ztwoV$ 0-form symmetry in the ferromagnetic phase.

In the case of $\cub{2,1,2}$, the Ising-type model is
\begin{align}
    H_{\mathrm{TC}} = - \sum_{p} \prod_{\ell \in\partial p} Z_{\ell} - \sum_{s} \prod_{\partial\ell \ni s} X_{\ell},
\end{align}
which is known as the toric code model\cite{Kitaev:1997wr}.
The model has \emph{two} $\mathbb{Z}_2$ 1-form symmetries, one of which is the $\ztwoV$ symmetry. The other is generated by a product of $Z_{\ell}$s over the loop running on the square lattice.

Furthermore, the Ising-type model for $\cub{3,1,3}$ is given by
\begin{align}
    H_{\text{X-cube}} = - \sum_{c} \prod_{\ell\in c} Z_{\ell} - \sum_{s} \sum_{\mu = x,y,z} \prod_{\partial\ell\ni s, \ell \perp \hat{\mu}} X_{\ell}
\end{align}
which is known as the X-cube model\cite{Vijay:2016phm}.\footnote{In our convention, this model might more appropriately be called the "Z-cube model" rather than the X-cube model, but the two are essentially equivalent.}
$\hat{\mu}$ is a unit vector directing in the $\mu$ direction.

Another prototypical Hamiltonian is the trivial one given by
\begin{align}
    H_{\mathrm{trivial}} = -\sum_{v\in\vertex} X_v.
\end{align}
Gauging $\ztwodualV$ symmetry of the trivial Hamiltonian for $\dualvertex$ qubits
\begin{align}
    \hat{H}_{\mathrm{trivial}} = -\sum_{\vhat\in\dualvertex} \Xhat_{\vhat}.
\end{align}
yields the Ising-type model\cite{Shirley:2018vtc}.

\subsection{Two $\mathbb{Z}_2$ symmetries and Kennedy-Tasaki transformation}
\label{Section_of_Ztwotwo}
In this section, we introduce a $\mathbb{Z}_2 \times \mathbb{Z}_2$ symmetry and Kennedy-Tasaki transformation we use throughout this paper.
The $\mathbb{Z}_2 \times \mathbb{Z}_2$ symmetry is determined by the bipartite graph we choose.
The gauging procedure is also naturally discussed based on the bipartite graph.

From now on, we discuss models which are elements of
\begin{align}
    \label{zzxzz_bondalg}
    \mathcal{B}_{\vertex\cup\dualvertex} = \braket{X_v,\; \prod_{v\in\adj{\vhat}} Z_{v},\; \Xhat_{\vhat},\; \prod_{\vhat\in\adj{v}} \Zhat_{\vhat}}{v\in \vertex,\vhat\in \dualvertex}.
\end{align}
The models have both of $\ztwoV$ symmetry and $\ztwodualV$ symmetry defined in Section \ref{Section_of_Ztwo}.
We denote them $\ztwoV\times\ztwodualV$ as a whole.
In the following sections, we introduce several other symmetries, all of which contain the $\ztwoV \times \ztwodualV$ symmetry.

Let us see some examples. $\cub{d,p,p+1}$ has $\ztwoV$ $p$-form symmetry and $\ztwodualV$ $(d-p-1)$-form symmetry generated by
\begin{align}
    \begin{split}
        U(\mathcal{M}_{d-p}) &= \prod_{v\in \mathcal{M}_{d-p}} X_v, \\
        \Uhat(\hat{\mathcal{M}}_{p+1}) &= \prod_{\vhat\in \hat{\mathcal{M}}_{p+1}} \Xhat_{\vhat},
    \end{split}
\end{align}
where $\mathcal{M}_{d-p}$ is a closed $(d-p)$-dimensional subregion on the lattice and $\hat{\mathcal{M}}_{p+1}$ is a closed $(p+1)$-dimensional subregion on the dual lattice.
For the case where $q > p + 1$, $\cub{d,p,q}$ leads to $\ztwoV\times\ztwodualV$ subsystem symmetry.
In this case, $\ztwoV$ symmetry operator $U_p$ has a support on $(d-q+1)$-dimensional subregion, and $\ztwodualV$ symmetry operator $\Uhat_{\phat}$ has a support on $(p+1)$-dimensional subregion.
However, they have restricted mobilities.
Note that $\cub{d,p,q}$ and $\cub{d,d-q,d-p}$ are essentially equivalent since they are related by a diagonal translation which swaps $\vertex$ and $\dualvertex$.

\paragraph{Gauging}

Now, let us consider gauging $\ztwoV \times \ztwodualV$.
After the gauging procedure for both of $\ztwoV$ and $\ztwodualV$, which is reviewed in the last section, and the identification of $\sigma^x$s and $\sigma^z$s as $\Xhat$s and $\Zhat$s and so on, we have the following map:
\begin{align}
    \begin{split}
        \label{map_of_zzxzz_gauging}
        X_{v} &\overset{\Dee}{\longleftrightarrow} \prod_{\vhat\in \adj{v}} \Zhat_{\vhat},\\
        \Xhat_{\vhat} &\overset{\Dee}{\longleftrightarrow} \prod_{v\in\adj{\vhat}} Z_v.
    \end{split}
\end{align}
We denote the operator implementing this map as $\Dee$.
It satisfies
\begin{align}
    \begin{split}
        \Dee X_{v} = \left(\prod_{\vhat\in\adj{v}} \Zhat_{\vhat}\right) \Dee,\; \Dee \left(\prod_{\vhat\in\adj{v}} \Zhat_{\vhat}\right) = X_v \Dee,\\
        \Dee \Xhat_{\vhat} = \left(\prod_{v\in\adj{\vhat}} Z_{v}\right) \Dee,\; \Dee \left(\prod_{v\in\adj{\vhat}} Z_{v}\right) = \Xhat_{\vhat} \Dee.
    \end{split}
\end{align}
More concretely, the matrix element of $\Dee$ is given by
\begin{align}
    \Braket{\{s,\hat{s}\}|\Dee|\{s',\hat{s}'\}} = K (-1)^{\sum_{v\in\vertex}s_v\sum_{\vhat\in\adj{v}}\hat{s}'_{\vhat}+\sum_{\vhat\in\dualvertex}\hat{s}_{\vhat}\sum_{v\in\adj{\vhat}}s'_{v}},
\end{align}
where $K$ is a positive real number and $\ket{\{s',\hat{s}'\}}$ is a basis vector which satisfies $Z_{v}\ket{\{s',\hat{s}'\}}=(-1)^{s_v}\ket{\{s',\hat{s}'\}},\,\Zhat_{\vhat}\ket{\{s',\hat{s}'\}}=(-1)^{s'_{\vhat}}\ket{\{s',\hat{s}'\}}$.
Importantly, $\Dee$ absorbs the generators of the $\ztwoV\times\ztwodualV$ symmetry,
\begin{align}
    \Dee U_p = U_p \Dee = \Dee \Uhat_{\phat} = \Uhat_{\phat} \Dee = \Dee.
\end{align}
This implies that $\Dee$ removes the states with a nontrivial $\ztwoV\times\ztwodualV$ charge.
As shown in \cite{Gorantla:2024ocs}, $\Dee^2$ amounts to a condensation operator\cite{Roumpedakis:2022aik}, which is the projection operator onto the space with a trivial $\ztwoV\times\ztwodualV$ charge ($U_p=\Uhat_{\phat}=1$).
\paragraph{Stacking SPT}
Next, we consider a unitary transformation
\begin{align}
    \label{definition_of_VCZ}
    \VCZ = \prod_{v\in\vertex}\prod_{\vhat\in\adj{v}} \mathrm{CZ}_{v,\vhat},
\end{align}
where
\begin{align}
    \mathrm{CZ}_{v,\vhat} = \frac{1+Z_v+\Zhat_{\vhat}-Z_v\Zhat_{\vhat}}{2}.
\end{align}
$\VCZ$ acts on $\ztwoV\times\ztwodualV$ symmetric operators as follows:
\begin{align}
    \begin{split}
        \label{action_of_VCZ}
        X_v \overset{\VCZ}{\longleftrightarrow} X_v \prod_{\vhat\in\adj{v}} \Zhat_{\vhat},\\
        \Xhat_{\vhat} \overset{\VCZ}{\longleftrightarrow} \Xhat_{\vhat} \prod_{v\in\adj{\vhat}} Z_{v}.
    \end{split}
\end{align}
Let us see its physical meaning.
Precisely, (\ref{action_of_VCZ}) is an automorphism of the $\ztwoV\times\ztwodualV$ symmetric bond algebra (\ref{zzxzz_bondalg}) and does not change the local dynamics but rather maps a gapped phase of the system to another, similar to gauging.
Consider the 1+1d chain $\cub{1,0,1}$ as an example\cite{Scaffidi:2017ppg,Li:2023ani}. 
The sites are labeled as $j = 0, 1, 2, \dots, L-1$, with a periodic boundary condition that identifies $j$ with $j+L$. Here, $L$ is assumed to be even.
Under this setup, $\vertex$ is the set of sites with even indices, and $\dualvertex$ is the set of sites with odd indices.
We omit hats of Pauli operators in $\dualvertex$ here.
We begin with the trivial Hamiltonian
\begin{align}
    H_{\mathrm{triv}} = -\sum_{j=0}^{L-1} X_{j}.
\end{align}
Its ground state is a tensor product of local states satisfying $X_j=+1$ at each site $j$, and therefore is in the trivial phase.

Applying $\VCZ$ to the trivial Hamiltonian, we obtain
\begin{align}
    H_{\mathrm{cluster}} = \VCZ H_{\mathrm{triv}} \VCZ^\dagger = - \sum_{j=0}^{L-1} Z_{j-1}X_j Z_{j+1}.
\end{align}
This is nothing but the cluster model, which is known as a prototypical model in the non-trivial 1+1d \zzxzz SPT phase\cite{Chen:2014zvm}.
This implies that this operator implements stacking a nontrivial SPT\cite{Chen:2011pg}.
Since $\VCZ^2=1$, applying $\VCZ$ to $H_{\mathrm{cluster}}$ returns it to $H_{\mathrm{triv}}$.
This is consistent with the fact that \zzxzz bosonic SPT phases in 1+1d are classified by \zz. In other words, stacking the SPT twice is equivalent to doing nothing.
On the other hand, the Hamiltonian
\begin{align}
    H_{\mathrm{SSB}} = -\sum_{j=0}^{L-1} Z_{j-1}Z_{j+1}
\end{align}
belongs to the phase where \zzxzz is fully broken. After applying $\VCZ$ to this Hamiltonian, the model is still in the SSB phase, which is again consistent with interpreting $\VCZ$ as stacking an SPT.

For general bipartite graphs, there is no guarantee that this picture is appropriate.
However, as we will review in Section \ref{section_of_repde}, the cluster model as a nontrivial SPT phase admits some generalizations.

\paragraph{Kennedy-Tasaki transformation}
Kennedy-Tasaki (KT) transformation was originally proposed as a unitary transformation which interchanges the \zzxzz SSB phase and the \zzxzz SPT phase in $S=1$ spin chains\cite{Kennedy:1992tke,Kennedy:1992ifl}, and was later extended to arbitrary integer spin\cite{Oshikawa1992}.
Recently, \cite{Li:2023ani} pointed out that the KT transformation, viewed as a mapping between gapped phases, can be expressed as a combination of the topological manipulations reviewed above:
\begin{align}
    \mathrm{KT} = \mathrm{STS} = \mathrm{TST},
\end{align}
where $\mathrm{S}$ denotes gauging \zzxzz and $\mathrm{T}$ represents stacking the non-trivial SPT protected by \zzxzz.
Note that $\mathrm{S}^2$, $\mathrm{T}^2$ and $\mathrm{KT}^2$ are equivalent to doing nothing.
In $S=1/2$ spin chains\footnote{In this setup, the realization of $\mathbb{Z}_2 \times \mathbb{Z}_2$ symmetry differs from that in \cite{Kennedy:1992tke,Kennedy:1992ifl,Oshikawa1992}.}, such as in the cluster model, this mapping has also been explicitly constructed at the operator level in \cite{Li:2023ani}.
A generalization to subsystem symmetry has been discussed in \cite{You:2018oai,ParayilMana:2024txy}, including the cases of $\cub{2,0,2}$ and $\cub{3,0,3}$.\footnote{\cite{ParayilMana:2024txy} also discusses the case with a single \zz symmetry which is beyond the scope of our framework.}
Additionally, \cite{Choi:2024rjm} considers the version of $\cub{2,0,1}$ with $\mathbb{Z}_2$ 0-form symmetry and $\mathbb{Z}_2$ 1-form symmetry.
At the operator level in our setup, the KT transformation is implemented by the operator $\VCZ \Dee \VCZ$, which is equivalent to $\Dee \VCZ \Dee$ up to a constant factor.
For general bipartite graphs, the KT transformation maps $\ztwoV\times\ztwodualV$ operators as follows:
\begin{align}
    \begin{split}
        X_{v} &\overset{\mathrm{KT}}{\mapsto} X_v,\; \prod_{v\in \adj{\vhat}} Z_{v} \overset{\mathrm{KT}}{\mapsto} \Xhat_{\vhat} \prod_{v\in \adj{\vhat}} Z_v, \\
        \Xhat_{\vhat} &\overset{\mathrm{KT}}{\mapsto} \Xhat_{\vhat},\; \prod_{\vhat\in \adj{v}} \Zhat_{\vhat} \overset{\mathrm{KT}}{\mapsto} X_{v} \prod_{\vhat\in \adj{v}} \Zhat_{\vhat}.
    \end{split}
\end{align}

\paragraph{Example}
We take an example from the case of $\cub{1,0,1}$, where $\vertex$ is the set of sites labeled by an even integer, and $\dualvertex$ is the set of sites labeled by an odd integer.
In this case, the $\ztwoV\times\ztwodualV$ symmetry is generated by
\begin{align}
    U = \prod_{n} X_{2n},\; \Uhat = \prod_{n} \Xhat_{2n+1}.
\end{align}
Consider the $\ztwoV\times\ztwodualV$ symmetric Hamiltonian
\begin{align}
    \label{example_Hamiltonian}
    \begin{split}
        H(h_{\even},J_{\even},K_{\even},h_{\odd},J_{\odd},K_{\odd}) =& - \sum_{n} \left(h_{\even} X_{2n} + J_{\even} \Zhat_{2n-1}\Zhat_{2n+1} + K_{\even}\Zhat_{2n-1}X_{2n}\Zhat_{2n+1}\right)\\
        & - \sum_{n} \left(h_{\odd} \Xhat_{2n+1} + J_{\odd} Z_{2n}Z_{2n+2} + K_{\odd}Z_{2n}\Xhat_{2n+1}Z_{2n+2}\right).
    \end{split}
\end{align}
Under the topological manipulations we reviewed above, the parameters are transformed as
\begin{align}
    \begin{split}
    \mathrm{S} \;:&\; (h_{\even},J_{\even},K_{\even},h_{\odd},J_{\odd},K_{\odd}) \mapsto (J_{\even},h_{\even},K_{\even},J_{\odd},h_{\odd},K_{\odd}),\\
    \mathrm{T} \;:&\; (h_{\even},J_{\even},K_{\even},h_{\odd},J_{\odd},K_{\odd}) \mapsto (K_{\even},J_{\even},h_{\even},K_{\odd},J_{\odd},h_{\odd}),\\
    \mathrm{KT} \;:&\; (h_{\even},J_{\even},K_{\even},h_{\odd},J_{\odd},K_{\odd}) \mapsto (h_{\even},K_{\even},J_{\even},h_{\odd},K_{\odd},J_{\odd}).
    \end{split}
\end{align}

\subsection{$\mathbb{Z}_2^3$ symmetry and its type III anomaly in $\cub{d,p,q}$}
\label{Section_of_Ztwothree}
We introduce an additional $\mathbb{Z}_2$ symmetry, denoted by $\ztwoCZ$, which is generated by $\VCZ$ (\ref{definition_of_VCZ}).
This $\ztwoCZ$ symmetry implies the invariance under (\ref{action_of_VCZ}).
From the conditions (\ref{pcond}) and (\ref{phatcond}), it follows that $U_p$ and $\Uhat_{\phat}$ commute with $\VCZ$.

\paragraph{Type III anomaly}
Intuitively, since $\VCZ$ is not onsite (unlike $U_p$ and $\Uhat_{\phat}$) and its local components $\mathrm{CZ}_{v,\vhat}$ break $\ztwoV\times\ztwodualV$ symmetry,
the full $\ztwoV\times\ztwodualV\times\ztwoCZ$ symmetry is typically anomalous in the sense that the whole symmetry cannot be gauged consistently.
This anomaly prohibits a unique gapped ground state.

In the case of the graph $\cub{1,0,1}$, this anomaly is known as type III anomaly and has been discussed in lattice models in \cite{Li:2022nwa,Seifnashri:2024dsd}.
Variants of the type III anomaly appear in the case of $\cub{2,0,1}$ as discussed in \cite{Li:2022nwa,Choi:2024rjm},
and in the $\cub{2,0,2}$ as considered in \cite{Li:2022nwa}.\footnote{\cite{Li:2022nwa} also analyzes the case related to \cite{Yoshida:2015cia}.}
In particular, the $\cub{2,0,2}$ systems exhibit a mixed anomaly between $\ztwoCZ$ 0-form symmetry and $\ztwoV \times \ztwodualV$ linear subsystem symmetry.
We refer to this anomaly in all such cases as type III anomaly.
Here, we generalize the analysis to arbitrary $\cub{d,p,q}$.

We work on a torus, where each $p/q$-dimensional hypercube is labeled by the coordinates of its center. The sets of vertices are defined as
\begin{align}
    \begin{split}
        \vertex &= \bigcup_{1 \leq k_1 < k_2 < \dots k_p \leq d} \left\{\left(n_1,n_2,\cdots,n_d \right) + \frac{1}{2}\sum_{j=1}^{p} \vece_{k_j} \middle|n_i \in \{0,1,\cdots,L_i - 1\}\right\}, \\
        \dualvertex &= \bigcup_{1 \leq k_1 < k_2 < \dots k_q \leq d} \left\{\left(n_1,n_2,\cdots,n_d \right)+ \frac{1}{2}\sum_{j=1}^{q} \vece_{k_j}\middle|n_i \in \{0,1,\cdots,L_i - 1\}\right\},
    \end{split}
\end{align}
where $\vece_{k} = (\underbrace{0,\cdots,0}_{(k-1)\text{-times}},1,\underbrace{0,\cdots,0}_{(d-k)\text{-times}})$ is a unit (row) vector in the $k$-th direction,
and $L_i$ is the system size in the $i$-th direction.
Periodic boundary conditions are imposed in all directions.
The bipartite graph $\cub{d,p,q}$ is constructed by connecting $v \in \vertex$ and $\vhat \in \dualvertex$ with an edge if and only if $v$ lies on the boundary of $\vhat$.
Concretely, for $\vhat = \mathbf{n} + \frac{1}{2}\sum_{j=1}^{q} \vece_{k_j}\in\dualvertex, \mathbf{n} = (n_1,\cdots n_d)$, we have
\begin{align}
    \begin{split}
        &\adj{\vhat} \\ =& \left\{\mathbf{n} + \frac{1}{2}\sum_{j=1}^{q} \vece_{k_j} + \frac{1}{2}\sum_{j=1}^{q-p} \sigma_{j}\vece_{\ell_j}\middle| \ell_1,\cdots,\ell_{q-p} \in \{k_1,\cdots,k_q \},\ell_1<\cdots<\ell_{q-p}, \sigma_1,\cdots,\sigma_{q-p}\in\{1,-1\}\right\}.
    \end{split}
\end{align}
We also have $\adj{v}$ for $v\in\vertex$ by $\adj{v} = \{\vhat\in\dualvertex | v \in \adj{\vhat}\}$.

Let us consider a model constructed from the algebra (\ref{zzxzz_bondalg}).
The model has $\ztwoV\times\ztwodualV$ symmetry.
For example, the $\ztwoV$ symmetry contains the operator
\begin{align}
    \label{cubU}
    U = \prod_{n_1,n_{p+2}\cdots,n_{d+p-q+1}} X_{(n_1,N_{2}+\frac{1}{2},\cdots,N_{p+1}+\frac{1}{2},n_{p+2},\cdots,n_{d+p-q+1},N_{d+p-q+2},\cdots,N_{d})},
\end{align}
and $\ztwodualV$ symmetry includes the operator
\begin{align}
    \label{cubUhat}
    \Uhat = \prod_{n_1,\cdots,n_{p+1}} \Xhat_{(n_{1}+\frac{1}{2},\cdots,n_{p+1}+\frac{1}{2},N_{p+2},\cdots,N_{d+p-q+1},N_{d+p-q+2}+\frac{1}{2},\cdots,N_{d}+\frac{1}{2})}.
\end{align}
Here, $N_2,\cdots,N_d$ are fixed integers.

We now consider a system on $\cub{d,p,q}$ with a Hamiltonian which preserves the $\ztwoV\times\ztwodualV\times\ztwoCZ$ symmetry, by imposing the invariance under (\ref{action_of_VCZ}) in addition to $\ztwoV\times\ztwodualV$.
Our goal is to show that the system is anomalous and cannot have a unique gapped ground state.

The outline of our argument is as follows\cite{Cheng:2022sgb}. See also \cite{Yao:2020xcm}.
If the bulk is trivially gapped, the system should remain trivially gapped under a boundary condition twisted by a (unitary) symmetry operator.
However, we will observe that the symmetry operators form a nontrivial projective representation under the symmetry-twisted boundary condition, indicating that all states, including the ground states, are doubly degenerate.
This contradiction implies that the original system cannot be trivially gapped.
The symmetry-twisted boundary condition can be viewed as an insertion of a symmetry defect along the time direction.
On the other hand, the symmetry operator itself can be interpreted as a symmetry defect along the spatial directions.
From this viewpoint, the resulting projective representation signals an obstruction to gauge these symmetries simultaneously, i.e. an 't Hooft anomaly.

In our case, we consider a boundary condition twisted by $\ztwoCZ$.
It is implemented by applying $\VCZ$ to the system in a halfway manner.
Specifically, we take $L_1 \to\infty$ and forget the periodicity $x_1\sim x_1 + L_1$. Then, we focus on the region around $x_1 = 0$.
We apply $\VCZ$ to the region $x_1 \leq 0$. Precisely, we transform our Hamiltonian by the operator
\begin{align}
    \label{halfVCZ}
    \VCZ^{\mathrm{half}} = \prod_{\vhat\in\{x_1 \leq 0\}\cap \dualvertex} \prod_{v\in\adj{\vhat}} \mathrm{CZ}_{v,\vhat}.
\end{align}
We can modify the definition of $\VCZ^{\mathrm{half}}$ up to a finite-depth unitary transformation around the location of the defect, which does not affect our argument.

Under the twisted boundary condition defined by (\ref{halfVCZ}), the original operator (\ref{cubUhat}) no longer commutes with the Hamiltonian.
Conjugating (\ref{cubU}) and (\ref{cubUhat}) by (\ref{halfVCZ}) and restoring finite $L_1$, we obtain the modified symmetry operators with a proper redefinition of the junction between the $\ztwoCZ$ defect given by
\begin{align}
    \begin{split}
    U^{\mathrm{defect}} &= U,\\
    \Uhat^{\mathrm{defect}} &= \Uhat \prod_{n_2,\cdots,n_{p+1}}\prod_{\tau_{d+p-q+2},\cdots,\tau_{d}\in\{0,1\}} Z_{(0,n_2+\frac{1}{2},\cdots,n_{p+1}+\frac{1}{2},N_{p+2},\cdots,N_{d+p-q+1},N_{d+p-q+2}+\tau_{d+p-q+2},\cdots,N_{d}+\tau_{d})},
    \end{split}
\end{align}
which commute with the defect Hamiltonian with the twisted boundary condition.
Finally, we have the projective representation
\begin{align}
    \label{type3anomprojective}
    U^{\mathrm{defect}}\Uhat^{\mathrm{defect}} = -\Uhat^{\mathrm{defect}}U^{\mathrm{defect}}.
\end{align}
The sign factor arises from the anticommutation of $X_v$ and $Z_v$ at the vertex $v = (0,N_{2}+\frac{1}{2},\cdots,N_{p+1}+\frac{1}{2},N_{p+2},\cdots,N_d)$.

Therefore, any energy levels of the defect Hamiltonian, including the ground states, must be doubly degenerate.
If the untwisted system is trivially gapped, the twisted system should be as well.
We conclude that the symmetry prevents the system from having a unique gapped ground state.
The nontrivial sign factor in (\ref{type3anomprojective}) also implies an obstruction to gauging $\ztwoV\times\ztwodualV\times\ztwoCZ$.
Note that our argument does not depend on the choice of $\VCZ^{\mathrm{half}}$, and hence, it is independent of the specific definition of the $\ztwoCZ$ defect.
These are manifestations of the anomaly of the $\ztwoV\times\ztwodualV\times\ztwoCZ$ symmetry.

\paragraph{Example}
Let us return to the Hamiltonian (\ref{example_Hamiltonian}).
By setting $h_{\even}=K_{\even}$ and $h_{\odd}=K_{\odd}$, we have
\begin{align}
    \label{example_ztwothree}
    \begin{split}
    H^{\ztwothree}(J_{\even},K_{\even},J_{\odd},K_{\odd}) =& -\sum_{n} \left\{J_{\even}\Zhat_{2n-1}\Zhat_{2n+1} + K_{\even} (X_{2n}+\Zhat_{2n-1}X_{2n}\Zhat_{2n+1})\right\} \\
    &-\sum_{n} \left\{J_{\odd} Z_{2n} Z_{2n+2} + K_{\odd} (\Xhat_{2n+1}+Z_{2n}\Xhat_{2n+1}Z_{2n+2})\right\},
    \end{split}
\end{align}
and the symmetry of the Hamiltonian enhances to $\ztwoV\times\ztwodualV\times\ztwoCZ$.

\subsection{Non-invertible \repdelike symmetry}
\label{section_of_repde}
Starting from a $\ztwothree$ symmetric model, we apply the KT transformation to it.
The resulting model is invariant under (\ref{map_of_zzxzz_gauging}), coming from the fact that the original Hamiltonian is invariant under (\ref{action_of_VCZ}).
The model has a symmetry generated by $U_p$ and $\Uhat_{\phat}$, where $p$ and $\phat$ satisfy conditions (\ref{pcond}) and (\ref{phatcond}), respectively, along with the non-invertible operator $\Dee$.
The operator algebra, as discussed in \cite{Gorantla:2024ocs}, is given by
\begin{align}
    \label{D_absorbs_U}
    U_p \Dee = \Dee U_p = \Uhat_{\phat} \Dee = \Dee \Uhat_{\phat} = \Dee,\;\Dee^2 = \mathrm{C}.
\end{align}
where $\mathrm{C}$ is a condensation operator which projects onto the subspace with the constraint $U_p = \Uhat_{\phat} = 1$\footnote{The overall factor can be determined via a careful computation using ZX-calculus\cite{Gorantla:2024ocs}.  We are sloppy about that here and only use the fact that it is nonzero.}.
In the case of $\cub{1,0,1}$, the symmetry is described by the fusion category $\repde$\footnote{In this paper, we use the convention that $\Deight$ denotes the dihedral group of order 8.}\cite{Seifnashri:2024dsd}.
For $\cub{2,0,1}$, the models possess the fusion 2-category symmetry $2\text{-Rep}((\mathbb{Z}_2^{(1)}\times \mathbb{Z}_2^{(1)})\rtimes \mathbb{Z}_{2}^{(0)})$ \cite{Choi:2024rjm}.
The cases of $\cub{2,0,2}$ and $\cub{3,0,2}$ are constructed in \cite{Ebisu:2024lie}, where the symmetry becomes subsystem non-invertible symmetry.
In general, we refer to such non-invertible symmetries as \repdelike symmetries.
In turn, if we apply the KT transformation to a system with \repdelike symmetry, then we obtain a model with $\ztwoV\times\ztwodualV\times\ztwoCZ$ symmetry reviewed in Section \ref{Section_of_Ztwothree}.
\paragraph{Cluster models}
Given a (bipartite) graph, the associated cluster model given by
\begin{align}
    H_{\mathrm{cluster}} = -\sum_{v\in\vertex} X_v \prod_{\vhat\in\adj{v}} \Zhat_{\vhat} - \sum_{\vhat\in\dualvertex} \Xhat_{\vhat} \prod_{v\in\adj{\vhat}} Z_{v}.
\end{align}
It has \repdelike non-invertible symmetry\cite{Seifnashri:2024dsd} when the graph is bipartite.
It is also known to be a nontrivial $\ztwoV \times \ztwodualV$ SPT in the case of $\cub{1,0,1}$.
The cluster model as an SPT can be generalized to higher-form symmetries\footnote{$(d+1)$-dimensional lattice models for SPTs protected by higher-form symmetries are first constructed in \cite{Yoshida:2015cia} on $(d+1)$-colorable graphs.}.
The cluster model for $\cub{3,1,2}$ is known as the Raussendorf-Bravyi-Harrington (RBH) model\cite{Raussendorf:2005dmx}, which is in a nontrivial SPT phase protected by $\mathbb{Z}_2 \times \mathbb{Z}_2$ 1-form symmetry.
\cite{Sukeno:2022pmx} proposes the cluster states for $\cub{d,p,p+1}$ as a nontrivial SPT protected by $\mathbb{Z}_2$ $p$-form symmetry and $\mathbb{Z}_2$ $(d-p-1)$-form symmetry.
A generalization to subsystem symmetry for $\cub{2,0,2}$ and $\cub{3,0,3}$ is discussed in \cite{You:2018oai}.

\paragraph{Example}
Let us apply the KT transformation to the Hamiltonian (\ref{example_ztwothree}).
We obtain
\begin{align}
    \label{example_repde}
    \begin{split}
    H^{\repde}(J_{\even},K_{\even},J_{\odd},K_{\odd}) =& -\sum_{n} \left\{J_{\even}\Zhat_{2n-1}X_{2n}\Zhat_{2n+1} + K_{\even} (X_{2n}+\Zhat_{2n-1}\Zhat_{2n+1})\right\} \\
    &-\sum_{n} \left\{J_{\odd} Z_{2n}\Xhat_{2n+1} Z_{2n+2} + K_{\odd} (\Xhat_{2n+1}+Z_{2n}Z_{2n+2})\right\},
    \end{split}
\end{align}
which is $\repde$ invariant as expected.

\subsection{\Deightlike symmetry}
\label{subsection_of_Deight}
Before moving on to the construction of SPT phases, we briefly discuss \Deightlike symmetry obtained by a duality map from \repdelike symmetric systems.
This duality on $\cub{2,0,1}$ is discussed in \cite{Choi:2024rjm}.
Starting from a system with \repdelike symmetry, we gauge the $\ztwodualV$ symmetry.
After gauging, the qubits originally located on $\dualvertex$ are removed, and instead, each vertex $v\in\vertex$ acquires a new qubit, whose Pauli operators are $\Xtilde_{v}$ and $\Ztilde_{v}$.
Under the duality transformation, the operators are mapped as
\begin{align}
    \begin{split}
        \Xhat_{\vhat} &\mapsto \prod_{v\in\adj{\vhat}} \Ztilde_{v},\\
        \prod_{\vhat\in\adj{v}} \Zhat_{\vhat} &\mapsto \Xtilde_{v}.
    \end{split}
\end{align}
The dual symmetry $\tilde{\mathbb{Z}}^{\vertex}_2$ is generated by
\begin{align}
    \tilde{U}_{p} = \prod_{v\in\vertex} \Xtilde_{v}^{p_v}
\end{align}
for $p$ satisfying (\ref{pcond}).
The $\ztwoV$ symmetry remains after gauging.

Moreover, since the original system with \repdelike symmetry is invariant under (\ref{map_of_zzxzz_gauging}), the gauged system is also invariant under
\begin{align}
    X_v \leftrightarrow \Xtilde_{v},\;\; \prod_{v\in\adj{\vhat}}Z_v \leftrightarrow \prod_{v\in\adj{\vhat}}\Ztilde_{v},
\end{align}
which is realized by swapping two qubits on the same vertex. We denote the unitary operator which implements this map as $\swap$.
The symmetry operators satisfy
\begin{align}
    U_p^2 = \tilde{U}_p^2 = \swap^2 = 1, \;\; U_p \tilde{U}_{p'} = \tilde{U}_{p'} U_p, \;\; \swap\cdot U_p = \tilde{U}_p \cdot \swap.
\end{align}
In the case of $\cub{d,0,1}$, the total symmetry is $\Deight$ 0-form symmetry.
For general bipartite graphs, we refer to this as \Deightlike symmetry.

\paragraph{Example}
Let us gauge $\ztwodualV$ of the Hamiltonian (\ref{example_repde}).
Then, we have
\begin{align}
    \label{example_deight}
    \begin{split}
    H^{\repde}(J_{\even},K_{\even},J_{\odd},K_{\odd}) =& -\sum_{n} \left\{J_{\even}X_{2n}\Xtilde_{2n} + K_{\even} (X_{2n}+\Xtilde_{2n})\right\} \\
    &-\sum_{n} \left\{J_{\odd} Z_{2n}Z_{2n+2}\Ztilde_{2n}\Ztilde_{2n+2} + K_{\odd} (\Ztilde_{2n}\Ztilde_{2n+2}+Z_{2n}Z_{2n+2})\right\},
    \end{split}
\end{align}
which is $\Deight$ invariant as expected.

\section{Subsystem non-invertible SPTs and interface modes}
\label{Section_of_Non-invertible_SPTs}

In this section, we study a family of lattice models exhibiting subsystem non-invertible SPT order, which generalize the $\cub{1,0,1}$ models\cite{Seifnashri:2024dsd} and the $\cub{2,0,1}$ models\cite{Choi:2024rjm}.
In Section \ref{section_of_models}, we explicitly define the lattice models studied throughout this section and in Section \ref{Section_of_weak_SPT}.
Once the models are defined, the remaining task is to demonstrate that certain pairs of models belong to distinct phases.
To achieve this, we employ two main strategies:
\begin{description}
    \item[Interface mode :] We analyze the interface between two models and show that it hosts a symmetry-protected interface mode.
    \item[Duality argument :] We apply a duality transformation, compare the resulting systems, and demonstrate that they are in distinct gapped phases.
\end{description}
As explained in Section \ref{section_of_interface}, the former approach can be systematically carried out by utilizing the structure of the bipartite graphs on which the systems are defined.
With this preparation, we are ready to discuss, for each graph, the SPT phases to which these models belong.
As a warm-up, Section \ref{Section_of_d01} generalizes the constructions of non-invertible SPTs in \cite{Seifnashri:2024dsd,Choi:2024rjm} to higher dimensions.
In Section \ref{Section_of_d0d}, \ref{Section_of_302} and \ref{Section_of_d0q}, we study lattice models realizing subsystem non-invertible SPT phases.

\subsection{The models}
\label{section_of_models}
Here, we consider two \repdelike symmetric lattice models, which we call the $\alpha$-model and the $\beta$-model, defined on a given bipartite graph.
They are written as commuting projector Hamiltonians.
Although we can define them for general $\cub{d,p,q}$, it turns out that the $\alpha$-model and the $\beta$-model are in distinct phases protected by the \repdelike symmetry only when $p=0$.

\paragraph{$\alpha$-model (cluster model)}
The $\alpha$-model is a cluster model associated with the given bipartite graph. The Hamiltonian is given by
\begin{align}
    \label{firstdefofalpha}
    H_\alpha^{\repde} = - \sum_{v\in \vertex} X_v \prod_{\vhat\in \adj{v}} \Zhat_{\vhat} - \sum_{\vhat\in \dualvertex} \Xhat_{\vhat} \prod_{v\in \adj{\vhat}} Z_v.
\end{align}
The unique gapped ground state is specified by
\begin{align}
    \begin{split}
        X_v \prod_{\vhat\in \adj{v}} \Zhat_{\vhat} = 1,\\
        \Xhat_{\vhat} \prod_{v\in \adj{\vhat}} Z_v = 1.
    \end{split}
\end{align}

\paragraph{$\beta$-model}
The Hamiltonian for the $\beta$-model is given by\footnote{From now on, we use the fact that $\abs{\adj{v}} \in 2\mathbb{Z}$ for any $v \in \vertex \cup \dualvertex$ in $\cub{d,0,q}$, which ensures that the $\beta$-model is frustration-free.}
\begin{align}
    \label{firstdefofbeta}
    \begin{split}
        H_{\beta}^{\repde} =& + \sum_{v\in \vertex} X_v \prod_{\vhat\in\adj{v}} \Zhat_{\vhat}
        - \sum_{\vhat\in\dualvertex} \Xhat_{\vhat} \left(\prod_{v\in \adj{\vhat}} Y_v\right) \left\{1 + \prod_{v\in\adj{\vhat}}\left(X_v \prod_{\vhat' \in \adj{v}} \Zhat_{\vhat'}\right)\right\}\\
        =& + \sum_{v\in \vertex} X_v \prod_{\vhat\in\adj{v}} \Zhat_{\vhat} \\&- \sum_{\vhat\in\dualvertex} (-1)^{\frac{\abs{\adj{\vhat}}}{2}} \Xhat_{\vhat} \left(\prod_{v\in \adj{\vhat}} Z_v\right) \left\{\prod_{v\in\adj{\vhat}} X_v + \prod_{v\in\adj{\vhat}}\left(\prod_{\vhat' \in \adj{v}} \Zhat_{\vhat'}\right)\right\},
    \end{split}
\end{align}
which is a natural generalization of lattice models introduced in \cite{Seifnashri:2024dsd,Choi:2024rjm}.
The unique gapped ground state is specified by
\begin{align}
    \begin{split}
        X_v \prod_{\vhat\in\adj{v}} \Zhat_{\vhat} = -1,\\
        \Xhat_{\vhat} \prod_{v\in \adj{\vhat}} Y_v = +1.
    \end{split}
\end{align}

We will discuss concrete examples of these models in later sections.

\subsection{$\alpha\beta$-interface}
\label{section_of_interface}

Here, we define a Hamiltonian for the interface between the $\alpha$-model and the $\beta$-model.
To do so, we first separate $\vertex$ and $\dualvertex$ as follows:
\begin{itemize}
    \item $\vertex = \vertex_{\alpha} \sqcup \vertex_{\beta}$
    \item $\dualvertex = \dualvertex_{\alpha} \sqcup \Vhatinterface\sqcup\dualvertex_{\beta}$
    \item $\vertex_{\alpha}$ and $\dualvertex_{\beta}$ are not directly connected by any edge.
    \item $\vertex_{\beta}$ and $\dualvertex_{\alpha}$ are not directly connected by any edge.
\end{itemize}
$\vertex_{\alpha}$ and $\dualvertex_{\alpha}$ belong to the bulk of the $\alpha$-model.
Similarly, $\vertex_{\beta}$ and $\dualvertex_{\beta}$ are in the bulk of the $\beta$-model.
$\Vhatinterface$ are the set of the qubits located at the interface.

Then, we consider the bulk part of the Hamiltonian
\begin{align}
    \begin{split}
        \label{alphabetainterface}
        H_{\alpha|\beta,0}^{\repde} =& -\sum_{v\in \vertex_\alpha} X_v \prod_{\vhat\in\adj{v}} \Zhat_{\vhat} - \sum_{\vhat\in\dualvertex_{\alpha}} \Xhat_{\vhat} \prod_{v\in \adj{\vhat}} Z_v \\
        &+ \sum_{v\in \vertex_{\beta}} X_v \prod_{\vhat \in \adj{v}} \Zhat_{\vhat} - \sum_{\vhat\in \dualvertex_{\beta}} \Xhat_{\vhat} \left(\prod_{v\in \adj{\vhat}} Y_v\right) \left\{1 + \prod_{v\in\adj{\vhat}} \left(X_{v}\prod_{\vhat'\in \adj{v}}\Zhat_{\vhat'}\right)\right\}.
    \end{split}
\end{align}
The ground states of $H_{\alpha|\beta,0}^{\repde}$ are specified by
\begin{align}
    \label{bulk_condition_alpha}
    \begin{split}
        X_v \prod_{\vhat'\in \adj{v}} \Zhat_{\vhat'} = +1,\\
        \Xhat_{\vhat} \prod_{v'\in \adj{\vhat}} Z_{v'} = +1,
    \end{split}
\end{align}
for each $v \in \vertex_{\alpha}$ and $\vhat \in \dualvertex_{\alpha}$, and
\begin{align}
    \label{bulk_condition_beta}
    \begin{split}
        X_v \prod_{\vhat'\in\adj{v}} \Zhat_{\vhat'} = -1,\\
        \Xhat_{\vhat} \prod_{v'\in \adj{\vhat}} Y_{v'} = +1,
    \end{split}
\end{align}
for each $v \in \vertex_{\beta}$ and $\vhat \in \dualvertex_{\beta}$.

We study the action of the symmetry operators on the low-energy effective Hilbert space which satisfies the conditions (\ref{bulk_condition_alpha}) and (\ref{bulk_condition_beta}).
The remaining degrees of freedom at this point are effective qubits on the interface. Their Pauli operators are given by
\begin{align}
    \Xeff_{\vhat} &= \Xhat_{\vhat} \left(\prod_{v_{\alpha}\in \vertex_{\alpha}\cap\adj{\vhat}} Z_{v_{\alpha}}\right) \left(\prod_{v_{\beta}\in \vertex_{\beta}\cap\adj{\vhat}} Y_{v_{\beta}}\right),\\
    \Yeff_{\vhat} &= \Yhat_{\vhat} \left(\prod_{v_{\alpha}\in \vertex_{\alpha}\cap\adj{\vhat}} Z_{v_{\alpha}}\right) \left(\prod_{v_{\beta}\in \vertex_{\beta}\cap\adj{\vhat}} Y_{v_{\beta}}\right),\\
    \Zeff_{\vhat} &= \Zhat_{\vhat},
\end{align}
for each $\vhat \in \Vhatinterface$.
The $\ztwoV$ symmetry acts as
\begin{align}
    \label{upaction_on_alphabeta_interface}
    U_{p} = \prod_{v\in\vertex} \left(X_v \prod_{\vhat \in \adj{v}}\Zhat_{\vhat}\right)^{p_v} \sim \prod_{v \in \vertex_{\beta}}(-1)^{p_v}.
\end{align}
Here, $\sim$ implies an equality in the low-energy effective Hilbert space defined by (\ref{bulk_condition_alpha}) and (\ref{bulk_condition_beta}).
On the other hand, the $\ztwodualV$ symmetry acts as
\begin{align}
    \label{uhatphataction_on_alphabeta_interface}
    \Uhat_{\hat{p}} = \prod_{\vhat\in\dualvertex_{\alpha}} \left(\Xhat_{\vhat} \prod_{v \in \adj{\vhat}} Z_{v}\right)^{\hat{p}_{\vhat}} \prod_{\vhat\in\dualvertex_{\beta}} \left(\Xhat_{\vhat} \prod_{v \in \adj{\vhat}} Y_{v}\right)^{\hat{p}_{\vhat}} \prod_{\vhat\in\Vhatinterface} (\Xeff_{\vhat})^{\hat{p}_{\vhat}}\sim \prod_{\vhat\in\Vhatinterface} (\Xeff_{\vhat})^{\hat{p}_{\vhat}}.
\end{align}
The non-invertible symmetry acts effectively on the interface qubits as
\begin{align}
    \begin{split}
    \Dee \Xeff_{\vhat} &= \Dee \Xhat_{\vhat} \left(\prod_{v\in\adj{\vhat}} Z_{v} \right) \left(\prod_{v_{\beta}\in\vertex_{\beta}\cap\adj{\vhat}} \left(-iX_{v_{\beta}}\right)\right)\\
    &= \Xhat_{\vhat} \left(\prod_{v\in\adj{\vhat}} Z_{v} \right) \left(\prod_{v_{\beta}\in\vertex_{\beta}\cap\adj{\vhat}} \left(-i\prod_{\vhat'\in\adj{v_{\beta}}}\Zhat_{\vhat'}\right)\right) \Dee\\
    &= \Xeff_{\vhat} \prod_{v_{\beta}\in \vertex_{\beta}\cap\adj{\vhat}} \left(X_{v_{\beta}}\prod_{\vhat'\in\adj{v_{\beta}}}\Zhat_{\vhat'}\right) \Dee\\
    &\sim (-1)^{|\vertex_{\beta}\cap\adj{\vhat}|}\Xeff_{\vhat} \Dee.
    \end{split}
\end{align}

It implies that some of the qubits in $\Vhatinterface$ can be gapped out by adding the term
\begin{align}
    H_{\mathrm{interface}} = -\sum_{\vhat \in \Vhatinterface} \frac{1+(-1)^{|\vertex_{\beta}\cap\adj{\vhat}|}}{2} \Xeff_{\vhat}
\end{align}
to the Hamiltonian $H_{\alpha|\beta,0}^{\repde}$.
The remaining qubits are living on the vertices
\begin{align}
    \label{definition_of_vhateff}
    \Vhateff = \left\{ \vhat\in \Vhatinterface \middle| (-1)^{|\vertex_{\beta}\cap\adj{\vhat}|} = -1\right\}.
\end{align}
Finally, since $\Dee$ acts effectively on the ground space of the total Hamiltonian $H_{\alpha|\beta,0}^{\repde} + H_{\mathrm{interface}}$ as
\begin{align}
    \label{eq_of_D_flips_X}
    \Xeff_{\vhat} \Dee \sim - \Dee \Xeff_{\vhat},
\end{align}
for $\vhat\in\dualvertex$, it can be written as
\begin{align}
    \label{eq_of_noninv_general}
    \Dee \sim \prod_{\vhat\in\Vhateff} \Zeff_{\vhat} \cdot (\text{a polynomial of $\Xeff$s}).
\end{align}
We will determine the second factor case-by-case from (\ref{D_absorbs_U}) and the way the operators composed of $\Zeff$s are mapped, as explained in Appendix \ref{appendix_of_derivation_of_noninv}.
Focusing on the region around the interface and neglecting degrees of freedom far from it, we locally obtain the algebra
\begin{align}
    \label{equation_of_projective_algebra}
    \Dee \Uhat_{\hat{p}} \sim (-1)^{\sum_{\vhat\in\Vhateff}\hat{p}_{\vhat}}\Uhat_{\hat{p}} \Dee.
\end{align}
If there is an appropriate $\hat{p}$ which satisfies (\ref{phatcond}) and yields a nontrivial sign factor in (\ref{equation_of_projective_algebra})\footnote{If we do not neglect the qubits located far from the interface, their contributions to the phase factor cancel those from the interface region, resulting in an overall trivial phase. However, the sign factor localized around the interface is crucial to protect the degenerate mode.}, the interface typically has a degenerate mode protected by the symmetry.
However, caution is required: $\Dee = 0$ also satisfies the above equation, but in that case, it does not guarantee the presence of a symmetry-protected degenerate edge mode.
To ensure that $\Dee$ is non-zero and actually protects the edge mode, we have to identify the operator realization on the effective Hilbert space.
In other words, since (\ref{D_absorbs_U}) implies that any state which violates $U_p = \Uhat_{\phat} = 1$ vanishes under $\Dee$, we choose the system size carefully so that (\ref{upaction_on_alphabeta_interface}) is $1$ for any $p$ satisfying (\ref{pcond}).
Once we get the expression of $\Dee$ on the effective Hilbert space, we can determine the symmetry action on a group of qubits which are spatially well separated from the others by using a method discussed in Appendix \ref{appendix_of_decomposing_symmetry}.

\subsection{Non-invertible SPTs : $\cub{d,0,1}$ case}
\label{Section_of_d01}
Here, we consider $\cub{d,0,1}$, a $d$-dimensional hypercubic lattice with qubits living on sites and links, on a $d$-dimensional torus obtained by imposing periodic boundary conditions $x_i \sim x_i + L_i$.
Specifically, we define $\vertex$ and $\dualvertex$ as
\begin{align}
    \begin{split}
        \vertex &= \left\{(n_1,n_2,\cdots,n_d)\middle|n_i \in \{0,1,\cdots,L_i - 1\}\right\}, \\
        \dualvertex &= \bigcup_{k=1}^d \left\{\left(n_1,n_2,\cdots,n_{k-1},n_k+\frac{1}{2},n_{k+1},\cdots,n_d\right)\middle|n_i \in \{0,1,\cdots,L_i - 1\}\right\}.
    \end{split}
\end{align}
$\vertex$ and $\dualvertex$ are respectively the set of sites and links on the hypercubic lattice.
The bipartite graph is constructed by connecting a site $s \in \vertex$ and a link $\ell \in \dualvertex$ if and only if the site $s$ is an endpoint of the link $\ell$.
Equivalently, $s \in \vertex$ and $\ell \in \dualvertex$ are connected if and only if the distance between $s$ and (the center of) $\ell$ is $\frac{1}{2}$.

The system constructed from the bond algebra (\ref{zzxzz_bondalg}) has $\ztwoV$ 0-form symmetry generated by
\begin{align}
    U = \prod_{s\in \vertex} X_s,
\end{align}
and $\ztwodualV$ $(d-1)$-form symmetry generated by
\begin{align}
    \Uhat(C) = \prod_{\ell\in C} \Xhat_\ell,
\end{align}
where $C \subset \dualvertex$ is a 1-dimensional closed loop running on the lattice.
In this section, we show that the $\alpha$- and the $\beta$-models for $\cub{d,0,1}$ are in different SPT phases distinguished by the \repdelike symmetry introduced in Section \ref{section_of_repde}.

\subsubsection{Interface modes}
Here, we show that the interface between the $\alpha$-model and the $\beta$-model has a non-trivial degenerate mode protected by the symmetry.
\paragraph{$\alpha\beta$-interface}
\begin{figure}[htbp]
    \begin{center}
    \begin{tikzpicture}[scale=2]
	\begin{pgfonlayer}{nodelayer}
		\node [style=none] (0) at (0, 0) {};
		\node [style=none] (2) at (0, 1) {};
		\node [style=none] (3) at (0, 2) {};
		\node [style=none] (4) at (1, 2) {};
		\node [style=none] (6) at (2, 1) {};
		\node [style=none] (7) at (2, 2) {};
		\node [style=none] (8) at (-2, 2) {};
		\node [style=none] (9) at (-1, 2) {};
		\node [style=none] (10) at (-2, 1) {};
		\node [style=none] (11) at (-2, 0) {};
		\node [style=none] (12) at (-1, 0) {};
		\node [style=none] (13) at (-2, -1) {};
		\node [style=none] (14) at (-2, -2) {};
		\node [style=none] (15) at (-1, -2) {};
		\node [style=none] (16) at (0, -2) {};
		\node [style=none] (17) at (0, -1) {};
		\node [style=none] (18) at (1, -2) {};
		\node [style=none] (19) at (2, -2) {};
		\node [style=none] (20) at (2, -1) {};
		\node [style=none] (21) at (1, 0) {};
		\node [style=none] (22) at (-3, 0) {};
		\node [style=none] (23) at (-3, 2) {};
		\node [style=none] (24) at (-2, 3) {};
		\node [style=none] (25) at (0, 3) {};
		\node [style=none] (26) at (2, 3) {};
		\node [style=none] (27) at (3, 2) {};
		\node [style=none] (28) at (3, -2) {};
		\node [style=none] (29) at (2, -3) {};
		\node [style=none] (30) at (-2, -3) {};
		\node [style=none] (31) at (-3, -2) {};
		\node [style=none] (32) at (3, 0) {};
		\node [style=none] (34) at (0, -3) {};
		\node [style=none] (35) at (2, 0) {};
		\node [style=none] (36) at (-2, -4) {};
		\node [style=none] (37) at (0, -4) {};
		\node [style=none] (38) at (2, -4) {};
		\node [style=none] (39) at (4, -4) {};
		\node [style=none] (40) at (4, -2) {};
		\node [style=none] (41) at (5, -2) {};
		\node [style=none] (42) at (4, 0) {};
		\node [style=none] (43) at (5, 0) {};
		\node [style=none] (44) at (4, 2) {};
		\node [style=none] (45) at (5, 2) {};
		\node [style=none] (46) at (-2, -5) {};
		\node [style=none] (47) at (0, -5) {};
		\node [style=none] (48) at (2, -5) {};
		\node [style=none] (49) at (4, -3) {};
		\node [style=none] (50) at (3, -4) {};
		\node [style=none] (51) at (5, -4) {};
		\node [style=none] (52) at (4, 3) {};
		\node [style=none] (53) at (4, -5) {};
		\node [style=none] (54) at (-1, -4) {};
		\node [style=none] (55) at (1, -4) {};
		\node [style=none] (56) at (-3, -4) {};
		\node [style=none] (57) at (4, 1) {};
		\node [style=none] (58) at (4, -1) {};
		\node [style=none] (63) at (1, -5) {};
		\node [style=none] (64) at (1, 3) {};
		\node [style=dualvertex] (65) at (-1, 2) {};
		\node [style=dualvertex] (66) at (-2, 1) {};
		\node [style=dualvertex] (67) at (0, 1) {};
		\node [style=dualvertex] (68) at (-1, 0) {};
		\node [style=dualvertex] (69) at (-2, -1) {};
		\node [style=dualvertex] (70) at (0, -1) {};
		\node [style=dualvertex] (71) at (-1, -2) {};
		\node [style=dualvertex] (72) at (-2, -3) {};
		\node [style=dualvertex] (73) at (0, -3) {};
		\node [style=dualvertex] (74) at (-1, -4) {};
		\node [style=dualvertex] (76) at (2, -3) {};
		\node [style=dualvertex] (79) at (2, -1) {};
		\node [style=dualvertex] (81) at (2, 1) {};
		\node [style=dualvertex] (82) at (3, 2) {};
		\node [style=dualvertex] (83) at (4, 1) {};
		\node [style=dualvertex] (84) at (3, 0) {};
		\node [style=dualvertex] (85) at (4, -1) {};
		\node [style=dualvertex] (86) at (3, -2) {};
		\node [style=dualvertex] (87) at (4, -3) {};
		\node [style=dualvertex] (88) at (3, -4) {};
		\node [style=vertex] (89) at (-2, 2) {};
		\node [style=vertex] (90) at (0, 2) {};
		\node [style=vertex] (91) at (0, 0) {};
		\node [style=vertex] (92) at (-2, 0) {};
		\node [style=vertex] (93) at (-2, -2) {};
		\node [style=vertex] (94) at (0, -2) {};
		\node [style=vertex] (95) at (-2, -4) {};
		\node [style=vertex] (96) at (0, -4) {};
		\node [style=vertex] (97) at (2, -4) {};
		\node [style=vertex] (98) at (4, -4) {};
		\node [style=vertex] (99) at (4, -2) {};
		\node [style=vertex] (100) at (4, 0) {};
		\node [style=vertex] (101) at (2, -2) {};
		\node [style=vertex] (102) at (2, 0) {};
		\node [style=vertex] (103) at (4, 2) {};
		\node [style=vertex] (104) at (2, 2) {};
		\node [style=boundarydualvertex] (105) at (1, 2) {};
		\node [style=boundarydualvertex] (106) at (1, 0) {};
		\node [style=boundarydualvertex] (107) at (1, -2) {};
		\node [style=boundarydualvertex] (108) at (1, -4) {};
		\node [style=none] (109) at (1.5, -5.5) {\color{red}$\Vhatinterface$\color{black}};
		\node [style=none] (110) at (-3, -1) {\Huge $\alpha$};
		\node [style=none] (111) at (5, -1) {\Huge $\beta$};
		\node [style=none] (112) at (5.5, -2) {};
		\node [style=none] (113) at (5.5, -2) {\color{blue} $\hat{U}_{\hat{p}}$\color{black}};
	\end{pgfonlayer}
	\begin{pgfonlayer}{edgelayer}
		\draw (21.center) to (0.center);
		\draw (0.center) to (12.center);
		\draw (12.center) to (11.center);
		\draw (11.center) to (22.center);
		\draw (6.center) to (7.center);
		\draw (7.center) to (26.center);
		\draw (7.center) to (27.center);
		\draw (7.center) to (4.center);
		\draw (4.center) to (3.center);
		\draw (3.center) to (25.center);
		\draw (3.center) to (2.center);
		\draw (2.center) to (0.center);
		\draw (3.center) to (9.center);
		\draw (9.center) to (8.center);
		\draw (8.center) to (24.center);
		\draw (8.center) to (10.center);
		\draw (8.center) to (23.center);
		\draw (10.center) to (11.center);
		\draw (11.center) to (13.center);
		\draw (13.center) to (14.center);
		\draw (14.center) to (31.center);
		\draw (14.center) to (30.center);
		\draw (0.center) to (17.center);
		\draw (17.center) to (16.center);
		\draw (14.center) to (15.center);
		\draw (15.center) to (16.center);
		\draw (16.center) to (18.center);
		\draw (18.center) to (19.center);
		\draw (19.center) to (20.center);
		\draw (19.center) to (28.center);
		\draw (19.center) to (29.center);
		\draw (16.center) to (34.center);
		\draw (35.center) to (21.center);
		\draw (35.center) to (32.center);
		\draw (35.center) to (6.center);
		\draw (35.center) to (20.center);
		\draw (30.center) to (36.center);
		\draw (56.center) to (36.center);
		\draw (36.center) to (46.center);
		\draw (36.center) to (54.center);
		\draw (54.center) to (37.center);
		\draw (34.center) to (37.center);
		\draw (37.center) to (47.center);
		\draw (29.center) to (38.center);
		\draw (37.center) to (55.center);
		\draw (55.center) to (38.center);
		\draw (38.center) to (48.center);
		\draw (38.center) to (50.center);
		\draw (50.center) to (39.center);
		\draw (39.center) to (53.center);
		\draw (39.center) to (51.center);
		\draw (27.center) to (44.center);
		\draw (44.center) to (45.center);
		\draw (32.center) to (42.center);
		\draw (42.center) to (43.center);
		\draw (28.center) to (40.center);
		\draw (40.center) to (41.center);
		\draw (52.center) to (44.center);
		\draw (44.center) to (57.center);
		\draw (57.center) to (42.center);
		\draw (42.center) to (58.center);
		\draw (58.center) to (40.center);
		\draw (40.center) to (49.center);
		\draw (49.center) to (39.center);
		\draw [style=boundary] (64.center) to (4.center);
		\draw [style=boundary] (55.center) to (63.center);
		\draw [style=boundary] (105) to (106);
		\draw [style=boundary] (106) to (107);
		\draw [style=boundary] (107) to (108);
		\draw [style=lineoperator] (22.center) to (102);
		\draw [style=lineoperator] (102) to (101);
		\draw [style=lineoperator] (101) to (41.center);
	\end{pgfonlayer}
\end{tikzpicture}
    \caption{The $\alpha\beta$-interface in the case of $d=2$.}
    \label{201alphabetainterface}
    \end{center}
\end{figure}
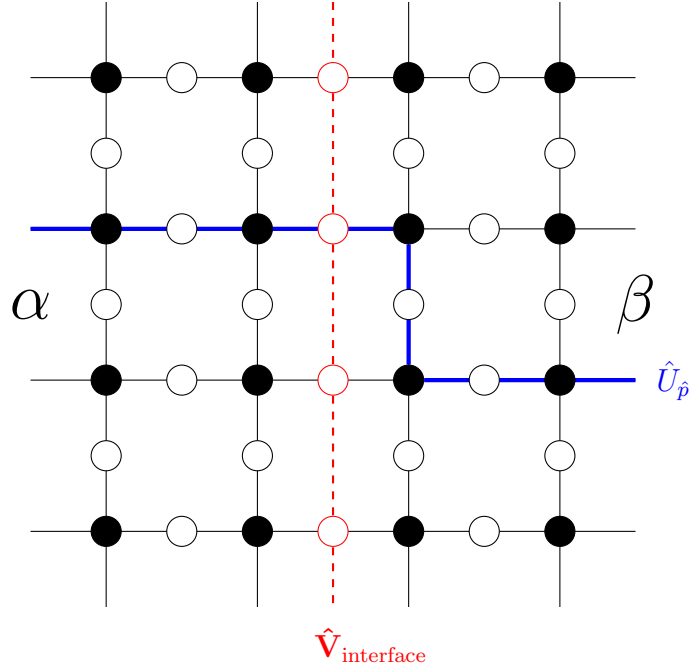

We discuss an $\alpha\beta$-interface and the interface mode. We assume that $L_1,\cdots,L_d$ are even.
To characterize the interface, it is convenient to define the bulk of the $\beta$-model $\mathcal{D}$ as
\begin{align}
    \mathcal{D} = \left\{(x_1,x_2,\cdots ,x_d)\middle|\frac{1}{2} < x_1 < \ell_1 + \frac{1}{2}\right\},
\end{align}
where $\ell_1$ is an even integer with $2\leq\ell_1<L_1$. We denote the surface of $\mathcal{D}$ as $\partial\mathcal{D}$.
We then consider the $\alpha\beta$-interface, as defined in Section \ref{section_of_interface}, which is specified by
\begin{align}
    \begin{split}
        \Vhatinterface = \dualvertex \cap \partial\mathcal{D},
        \dualvertex_{\beta} = \dualvertex \cap \mathcal{D},
        \vertex_{\beta} = \vertex \cap \mathcal{D},\\
        \vertex_{\alpha} = \vertex \setminus \vertex_{\beta},
        \dualvertex_{\alpha} = \dualvertex \setminus (\dualvertex_{\beta}\cup \Vhatinterface).
    \end{split}    
\end{align}
The effective qubits live on $\Vhateff = \Vhatinterface$.
$\ztwodualV$ symmetry operators piercing the interface (as illustrated in Fig. \ref{201alphabetainterface} for $d=2$) act nontrivially on $\Vhateff$.
From (\ref{D_absorbs_U}) and
\begin{align}
    \begin{split}
    \Dee \prod_{\vhat\in \Vhateff} \Zeff_{\vhat} = \Dee \prod_{v\in\vertex_{\mathrm{\alpha}}} \left(\prod_{\vhat\in\adj{v}}\Zhat_{\vhat}\right)
    = \prod_{v\in\vertex_{\mathrm{\alpha}}} \left(\prod_{\vhat\in\adj{v}}\Zhat_{\vhat}\right) \cdot \prod_{v\in\vertex_{\mathrm{\alpha}}} \left(X_{v} \prod_{\vhat\in\adj{v}}\Zhat_{\vhat}\right) \Dee
    \sim \prod_{\vhat\in \Vhateff} \Zeff_{\vhat} \Dee,
    \end{split}
\end{align}
the non-invertible symmetry on the interface is identified as
\begin{align}
    \label{eq_of_noninv_d01}
    \Dee \sim \left(\prod_{\vhat\in \Vhateff} \Zeff_{\vhat}\right)\left(\prod_{\vhat \in \Vhateff} \frac{1+\Xeff_{\vhat}}{2} + \prod_{\vhat \in \Vhateff} \frac{1-\Xeff_{\vhat}}{2}\right),
\end{align}
up to a nonzero constant factor. See Appendix \ref{appendix_of_derivation_of_noninv} for the derivation of (\ref{eq_of_noninv_d01}).
We then focus on the interface around $x_1 = \frac{1}{2}$. After the procedure shown in Appendix \ref{appendix_of_decomposing_symmetry}, we obtain two symmetries realized on the interface. One is generated by $\Uhat_{\vhat}^{\mathrm{eff}}=\Xeff_{\vhat}$ for each $\vhat \in \Vhateff$ around $x_1 = \frac{1}{2}$, coming from bulk $\ztwodualV$ $(d-1)$-form symmetry.
Note that $\Uhat_{\vhat}^{\mathrm{eff}}$ is physically independent of $\vhat\in\Vhateff$ due to the magnetic Gauss law.
The other symmetry is
\begin{align}
    \Uhat_{\Dee}^{\mathrm{eff}} = \prod_{\vhat \in \Vhateff \;\text{such that}\; x_1 = \frac{1}{2}} \Zeff_{\vhat},
\end{align}
coming from the non-invertible symmetry.
Note that the non-invertible symmetry reduces to a unitary (and hence invertible) operator on the interface.
Then we obtain the projective algebra
\begin{align}
    \Uhat_{\Dee}^{\mathrm{eff}} \Uhat_{\vhat}^{\mathrm{eff}} = - \Uhat_{\vhat}^{\mathrm{eff}} \Uhat_{\Dee}^{\mathrm{eff}},
\end{align}
which protects the interface mode.

\subsubsection{Duality argument}
We argue that the $\alpha$- and $\beta$-models are in different phases by comparing these two after applying the KT transformation.
The duality argument has been given for $\cub{1,0,1}$ \cite{Seifnashri:2024dsd} and $\cub{2,0,1}$ \cite{Choi:2024rjm}.
As reviewed in Section \ref{Section_of_Ztwothree}, the resulting models have $\ztwoV\times\ztwodualV\times\ztwoCZ$ symmetry.
\paragraph{$\alpha$-model}
Applying the KT transformation to the $\alpha$-model, we obtain
\begin{align}
    H_{\alpha}^{\ztwothree} = - \sum_{s\in\vertex} \prod_{\partial\ell\ni s} \Zhat_{\ell} - \sum_{\ell\in\dualvertex} \prod_{s\in\partial\ell} Z_s - g\sum_{p}\prod_{\ell \in \partial p} \Xhat_{\ell}.
\end{align}
$\sum_{p}$ is a summation over all the plaquettes.
The last term energetically imposes the magnetic Gauss law, which lifts degeneracies without breaking the symmetry.
If we take $g\to\infty$, in the low-energy limit, $\ztwodualV$ $(d-1)$-form symmetry becomes topological.
The ground states satisfy
\begin{align}
    \begin{split}
        \prod_{\partial\ell\ni s} \Zhat_{\ell} &= +1,\; \prod_{\ell \in \partial p} \Xhat_{\ell} = +1,\\
        \prod_{s\in\partial\ell} Z_s &= +1.
    \end{split}
\end{align}
The first line results in a toric code state for the $\dualvertex$ sector, which spontaneously breaks $\ztwodualV$ $(d-1)$-form symmetry. The second line implies the $\vertex$ sector is in the ferromagnetic phase spontaneously breaking $\ztwoV$ 0-form symmetry.
\paragraph{$\beta$-model}
For the $\beta$-model, we obtain
\begin{align}
    \begin{split}
        H_{\beta}^{\ztwothree} = + \sum_{s\in\vertex} \prod_{\partial\ell\ni s} \Zhat_{\ell} - \sum_{\ell\in\dualvertex} \left(\prod_{s\in\partial\ell} Y_s\right) \left\{1 + \prod_{s\in\adj{\ell}}\left(\prod_{\partial \ell'\ni s}\Zhat_{\ell'}\right)\right\} - g\sum_{p}\prod_{\ell \in \partial p} \Xhat_{\ell}.
    \end{split}
\end{align}
The ground states satisfy
\begin{align}
    \begin{split}
        \prod_{\partial\ell\ni s} \Zhat_{\ell} &= -1,\; \prod_{\ell \in \partial p} \Xhat_{\ell} = +1,\\
        \prod_{s\in\partial\ell} Y_s &= +1.
    \end{split}
\end{align}
Again, the first line results in a toric code state for the $\dualvertex$ sector, which spontaneously breaks $\ztwodualV$ $(d-1)$-form symmetry. The second line implies the $\vertex$ sector is in the ferromagnetic phase breaking $\ztwoV$ 0-form symmetry.
However, for the $\vertex$ sector, $Y_s$ is charged under $\ztwoCZ$:
\begin{align}
    \label{eq_charged_under_VCZ}
    \VCZ Y_s \VCZ^\dagger = Y_s \prod_{\partial\ell\ni s} \Zhat_{\ell} \sim - Y_s.
\end{align}
Therefore, the system preserves $\mathbb{Z}_2$ symmetry generated by $U \VCZ$.
\paragraph{}
To summarize the above, after the KT transformation, we find that
\begin{description}
    \item[$\alpha$-model] The preserved internal symmetry is $\ztwoCZ$ 0-form symmetry generated by $\VCZ$. $\ztwoV$ 0-form symmetry and $\ztwodualV$ $(d-1)$-form symmetry are spontaneously broken.
    \item[$\beta$-model] The preserved internal symmetry is $\mathrm{diag}(\ztwoV\times\ztwoCZ)$ 0-form symmetry generated by $U \VCZ$. $\ztwoV$ 0-form symmetry and $\ztwodualV$ $(d-1)$-form symmetry are spontaneously broken.
\end{description}
Therefore, since the unbroken symmetries differ after the KT transformation, the original $\alpha$-model and $\beta$-model are in distinct \repdelike SPT phases.

\subsection{Subsystem non-invertible SPTs for $\cub{d,0,d}$}
\label{Section_of_d0d}

We consider systems on $\cub{d,0,d}$, where for each $n = (n_1,n_2,\cdots,n_d)\in \mathbb{Z}^d$, a qubit in $\vertex$ lives on $(n_1,n_2,\cdots,n_d)$
and a qubit in $\dualvertex$ lives on $(n_1+\frac{1}{2},n_2+\frac{1}{2},\cdots,n_d+\frac{1}{2})$.
We work on a $d$-dimensional torus with periodic boundary conditions imposed as $(x_1,x_2,\cdots,x_d) \sim (x_1,x_2,\cdots,x_j + L_j, \cdots,x_d)$ for each $j=1,2,\cdots,d$, where $L_i\in\mathbb{Z}$ denotes the length of the system in the $i$-th direction.
To summarize the above,
\begin{align}
    \begin{split}
        \vertex &= \left\{(n_1,n_2,\cdots,n_d)\middle|n_i \in \{0,1,\cdots,L_i - 1\}\right\}, \\
        \dualvertex &= \left\{\left(n_1+\frac{1}{2},n_2+\frac{1}{2},\cdots,n_d+\frac{1}{2}\right)\middle|n_i \in \{0,1,\cdots,L_i - 1\}\right\}.
    \end{split}
\end{align}

The system has two linear $\mathbb{Z}_2$ subsystem symmetries,
\begin{align}
    \begin{split}
    U_{i}(n_1,\cdots,n_{i-1}, n_{i+1},\cdots, n_d) &= \prod_{n_i=0}^{L_i-1} X_{(n_1,\cdots,n_d)},\\
    \Uhat_{i}(n_1,\cdots,n_{i-1}, n_{i+1}, \cdots, n_d) &= \prod_{n_i=0}^{L_i-1} \Xhat_{(n_1+\frac{1}{2},\cdots,n_d+\frac{1}{2})},
    \end{split}
\end{align}
for each $i = 1,2,\cdots,d$.
The \repdelike symmetry is discussed in \cite{Ebisu:2024lie}.

In this section, we show that the $\alpha$-, $\beta$-, and $\gamma$-models, where the $\gamma$-model will be introduced shortly, realize three distinct SPT phases protected by \repdelike subsystem non-invertible symmetry.

\paragraph{$\gamma$-model}
The $\gamma$-model with \repdelike symmetry is obtained by exchanging the role of $\vertex$ and $\dualvertex$ in $H_{\beta}^{\repde}$, such as
\begin{align}
    \label{firstdefofgamma}
    \begin{split}
        H_{\gamma}^{\repde} =& - \sum_{v\in\vertex} X_{v} \left(\prod_{\vhat\in \adj{v}} \Yhat_{\vhat}\right) \left\{1 + \prod_{\vhat\in\adj{v}}\Xhat_{\vhat}\left(\prod_{v' \in \adj{\vhat}} Z_{v'} \right)\right\} + \sum_{\vhat\in \dualvertex} \Xhat_{\vhat} \prod_{v\in\adj{\vhat}} Z_{v}.
    \end{split}
\end{align}
This can also be viewed as a natural generalization of the model in \cite{Seifnashri:2024dsd}.
The unique gapped ground state is specified by
\begin{align}
    \begin{split}
        X_{v} \prod_{\vhat\in \adj{v}} \Yhat_{\vhat} = +1,\\
        \Xhat_{\vhat} \prod_{v\in\adj{\vhat}} Z_{v} = -1.
    \end{split}
\end{align}

\subsubsection{Interface modes}
We analyze interfaces between two different models, such as the $\alpha\beta$-, $\alpha\gamma$- and $\beta\gamma$-interfaces.

\begin{figure}[htbp]
    \begin{center}
    \begin{tikzpicture}[scale=2]
	\begin{pgfonlayer}{nodelayer}
		\node [style=none] (0) at (0, 0) {};
		\node [style=none] (1) at (-2, 0) {};
		\node [style=none] (2) at (-2, 2) {};
		\node [style=none] (3) at (0, 2) {};
		\node [style=none] (4) at (2, 2) {};
		\node [style=none] (5) at (2, 0) {};
		\node [style=none] (6) at (2, -2) {};
		\node [style=none] (7) at (0, -2) {};
		\node [style=none] (8) at (-2, -2) {};
		\node [style=none] (9) at (-1, -1) {};
		\node [style=none] (10) at (1, -1) {};
		\node [style=none] (11) at (1, 1) {};
		\node [style=none] (12) at (-1, 1) {};
		\node [style=none] (13) at (-3, 3) {};
		\node [style=none] (14) at (-1, 3) {};
		\node [style=none] (15) at (1, 3) {};
		\node [style=none] (16) at (3, 3) {};
		\node [style=none] (17) at (3, 1) {};
		\node [style=none] (18) at (3, -1) {};
		\node [style=none] (19) at (3, -3) {};
		\node [style=none] (20) at (1, -3) {};
		\node [style=none] (21) at (-1, -3) {};
		\node [style=none] (22) at (-3, -3) {};
		\node [style=none] (23) at (-3, -1) {};
		\node [style=none] (24) at (-3, 1) {};
		\node [style=vertex] (25) at (-2, 2) {};
		\node [style=vertex] (26) at (0, 2) {};
		\node [style=vertex] (27) at (2, 2) {};
		\node [style=vertex] (28) at (2, 0) {};
		\node [style=vertex] (29) at (0, 0) {};
		\node [style=vertex] (30) at (-2, 0) {};
		\node [style=vertex] (31) at (-2, -2) {};
		\node [style=vertex] (32) at (0, -2) {};
		\node [style=vertex] (33) at (2, -2) {};
		\node [style=dualvertex] (34) at (1, -1) {};
		\node [style=dualvertex] (38) at (-3, -3) {};
		\node [style=dualvertex] (40) at (1, -3) {};
		\node [style=dualvertex] (41) at (3, -3) {};
		\node [style=dualvertex] (42) at (3, -1) {};
		\node [style=dualvertex] (44) at (3, 3) {};
		\node [style=dualvertex] (45) at (1, 3) {};
		\node [style=dualvertex] (46) at (-1, 3) {};
		\node [style=dualvertex] (47) at (-3, 3) {};
		\node [style=dualvertex] (48) at (-3, 1) {};
		\node [style=dualvertex] (49) at (-3, -1) {};
		\node [style=boundarydualvertex] (50) at (-1, 1) {};
		\node [style=boundarydualvertex] (51) at (1, 1) {};
		\node [style=boundarydualvertex] (52) at (3, 1) {};
		\node [style=boundarydualvertex] (53) at (-1, -1) {};
		\node [style=boundarydualvertex] (54) at (-1, -3) {};
		\node [style=none] (55) at (-4, 1.05) {};
		\node [style=none] (56) at (4, 1.05) {};
		\node [style=none] (57) at (-4, 4) {};
		\node [style=none] (58) at (-4, 2) {};
		\node [style=none] (59) at (-4, 0) {};
		\node [style=none] (60) at (-4, -2) {};
		\node [style=none] (61) at (-4, -4) {};
		\node [style=none] (62) at (-2, -4) {};
		\node [style=none] (63) at (0, -4) {};
		\node [style=none] (64) at (2, -4) {};
		\node [style=none] (65) at (4, -4) {};
		\node [style=none] (66) at (4, -2) {};
		\node [style=none] (67) at (4, 0) {};
		\node [style=none] (68) at (4, 2) {};
		\node [style=none] (69) at (4, 4) {};
		\node [style=none] (70) at (2, 4) {};
		\node [style=none] (71) at (0, 4) {};
		\node [style=none] (72) at (-2, 4) {};
		\node [style=none] (73) at (-1, -4) {};
		\node [style=none] (74) at (4, 1) {};
		\node [style=none] (75) at (5, 1.25) {\color{blue}$\hat{U}_{\hat{p}}$};
		\node [style=none] (76) at (5, 0.75) {\color{red} $\Vhatinterface$};
		\node [style=none] (77) at (-4, 3) {\Huge $\alpha$};
		\node [style=none] (78) at (4, -3) {\Huge $\beta$};
	\end{pgfonlayer}
	\begin{pgfonlayer}{edgelayer}
		\draw (23.center) to (21.center);
		\draw (24.center) to (20.center);
		\draw (13.center) to (19.center);
		\draw (14.center) to (18.center);
		\draw (15.center) to (17.center);
		\draw (24.center) to (14.center);
		\draw (23.center) to (15.center);
		\draw (22.center) to (16.center);
		\draw (21.center) to (17.center);
		\draw (20.center) to (18.center);
		\draw [style=boundary] (50) to (54);
		\draw [style=boundary] (50) to (52);
		\draw [style=lineoperator] (55.center) to (56.center);
		\draw (57.center) to (47);
		\draw (58.center) to (72.center);
		\draw (59.center) to (48);
		\draw (46) to (71.center);
		\draw (60.center) to (49);
		\draw (45) to (70.center);
		\draw (72.center) to (46);
		\draw (71.center) to (45);
		\draw (70.center) to (44);
		\draw (44) to (69.center);
		\draw (44) to (68.center);
		\draw (52) to (67.center);
		\draw (42) to (67.center);
		\draw (42) to (66.center);
		\draw (41) to (65.center);
		\draw (41) to (66.center);
		\draw (41) to (64.center);
		\draw (40) to (63.center);
		\draw (54) to (63.center);
		\draw (62.center) to (54);
		\draw (61.center) to (38);
		\draw (38) to (62.center);
		\draw (60.center) to (38);
		\draw (40) to (64.center);
		\draw (58.center) to (48);
		\draw (68.center) to (52);
		\draw (59.center) to (49);
		\draw [style=boundary] (52) to (74.center);
		\draw [style=boundary] (54) to (73.center);
	\end{pgfonlayer}
\end{tikzpicture}
    \caption{The $\alpha\beta$-interface in the case of $d=2$.}
    \label{202interface}
    \end{center}
\end{figure}
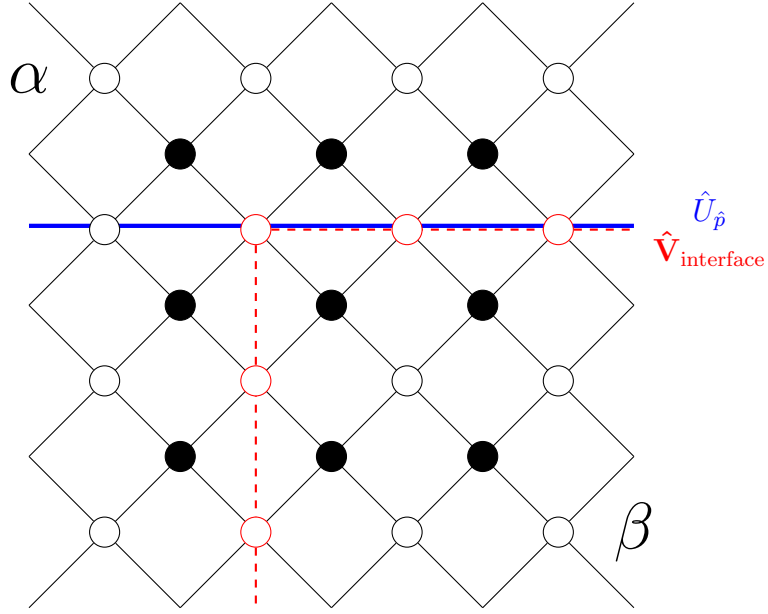

\paragraph{$\alpha\beta$-interface}
We analyze the interface between the $\alpha$-model and the $\beta$-model following Section \ref{section_of_interface}.
Let $\ell_i$ be even integers satisfying $2 \leq \ell_i < L_i$ for $i=1,\cdots,d$\footnote{We set $\ell_i$ to be even so that the non-invertible operator $\Dee$ acts as a non-zero operator on the low-energy subspace.}.
As in the previous case, we define the bulk of the $\beta$-model $\mathcal{D}$ as
\begin{align}
    \mathcal{D} = \left\{(x_1,x_2,\cdots ,x_d)\middle|\frac{1}{2} < x_k < \ell_k + \frac{1}{2},\,k=1,\cdots,d \right\}.
\end{align}
We define the interface by
\begin{align}
    \begin{split}
        \Vhatinterface = \dualvertex \cap \partial\mathcal{D},
        \dualvertex_{\beta} = \dualvertex \cap \mathcal{D},
        \vertex_{\beta} = \vertex \cap \mathcal{D},\\
        \vertex_{\alpha} = \vertex \setminus \vertex_{\beta},
        \dualvertex_{\alpha} = \dualvertex \setminus (\dualvertex_{\beta}\cup \Vhatinterface).
    \end{split}
\end{align}
In this case, some of the qubits in $\Vhatinterface$ can be gapped out. We finally obtain the set of remaining qubits
\begin{align}
    \Vhateff = \left\{\left(m_1 \ell_1 + \frac{1}{2}, m_2 \ell_2 + \frac{1}{2}, \cdots, m_d \ell_d + \frac{1}{2}\right)\middle| m_1,m_2,\cdots, m_d \in \{0,1\}\right\},
\end{align}
which consists of $2^d$ corners.
Each corner hosts an effective qubit.
$\ztwodualV$ symmetry operators hitting a corner (as illustrated in Fig. \ref{202interface} for $d=2$) act nontrivially on $\Vhateff$.
We can choose $\phat$ satisfying (\ref{phatcond}) so that $\Uhat_{\phat} \sim \Xeff_{\vhat_{1}} \Xeff_{\vhat_{2}}$ for any $\vhat_{1}, \vhat_{2} \in \Vhateff$.
We also have
\begin{align}
    \Dee \prod_{\vhat\in\Vhateff} \Zeff_{\vhat} \sim \prod_{\vhat\in\Vhateff} \Zeff_{\vhat} \Dee.
\end{align}
Therefore, the non-invertible symmetry acts on the interface as
\begin{align}
    \label{eq_of_noninv_d0d}
    \Dee \sim \left(\prod_{\vhat\in \Vhateff} \Zeff_{\vhat}\right)\left(\prod_{\vhat \in \Vhateff} \frac{1+\Xeff_{\vhat}}{2} + \prod_{\vhat \in \Vhateff} \frac{1-\Xeff_{\vhat}}{2}\right),
\end{align}
up to a nonzero constant factor. See Appendix \ref{appendix_of_derivation_of_noninv} for the derivation of (\ref{eq_of_noninv_d0d}).
Let us focus on the qubit at the corner $\vhat_{\mathrm{corner}} = (\frac{1}{2},\cdots,\frac{1}{2})\in\Vhateff$.
After the procedure shown in Appendix \ref{appendix_of_decomposing_symmetry}, we identify the symmetry action on the corner $\vhat_{\mathrm{corner}}$.
$\ztwodualV$ symmetry acts on $\vhat_{\mathrm{corner}}$ as $\Uhat_{\vhat_{\mathrm{corner}}}^{\mathrm{eff}}=\Xeff_{\vhat_{\mathrm{corner}}}$.
On the other hand, the subsystem non-invertible symmetry $\Dee$ is reduced to a $\mathbb{Z}_2$ symmetry generated by
\begin{align}
    \Uhat_{\Dee}^{\mathrm{eff}} = \Zeff_{\vhat_{\mathrm{corner}}}.
\end{align}
Finally, we have
\begin{align}
    \Uhat_{\Dee}^{\mathrm{eff}} \Uhat_{\vhat_{\mathrm{corner}}}^{\mathrm{eff}} = - \Uhat_{\vhat_{\mathrm{corner}}}^{\mathrm{eff}} \Uhat_{\Dee}^{\mathrm{eff}},
\end{align}
which protects the corner mode. The other corners also host symmetry-protected effective qubits in the same way.

\paragraph{$\alpha\gamma$-interface}
Since the $\beta$- and $\gamma$-models are related by a translation by $(\frac{1}{2}, \cdots, \frac{1}{2})$, and the $\alpha$-model is invariant under this translation, the $\alpha\gamma$-interface also hosts a nontrivial corner mode protected by the symmetry.

\paragraph{$\beta\gamma$-interface}
We consider the interface between the $\beta$-model and the $\gamma$-model.
We show that it has a nontrivial corner mode and therefore the models are in different phases.
The essence of the analysis is the same as the case of the $\alpha\beta$-interface in Section \ref{section_of_interface}.

Let $L_i$ for each $i=1,\cdots, d$ be even integers. 
We also set $2 \leq \ell_i \leq L_i-2$.
For convenience, we define the bulk of the $\gamma$-model as
\begin{align}
    \begin{split}
        \mathcal{D} &= \left\{(x_1,x_2,\cdots ,x_d)\middle|1 < x_k < \ell_k + 1,\;k=1,\cdots,d\right\},\\
        \tilde{\mathcal{D}} &= \left\{(x_1,x_2,\cdots ,x_d)\middle|\frac{1}{2} < x_k < \ell_k + \frac{3}{2},\;k=1,\cdots,d\right\}.
    \end{split}
\end{align}
We define the interface by
\begin{align}
    \begin{split}
        \Vinterface &= \vertex \cap \partial\mathcal{D},
        \Vhatinterface = \dualvertex \cap \partial\tilde{\mathcal{D}},
        \vertex_{\gamma} = \vertex \cap \mathcal{D},\\
        \dualvertex_{\gamma} &= \dualvertex \cap \tilde{\mathcal{D}},
        \vertex_{\beta} = \vertex \setminus (\vertex_{\gamma}\cup \Vinterface),
        \dualvertex_{\beta} = \dualvertex \setminus (\dualvertex_{\gamma}\cup \Vhatinterface).
    \end{split}
\end{align}
This division has the following properties:
\begin{itemize}
    \item $\vertex = \vertex_{\beta} \sqcup \Vinterface\sqcup\vertex_{\gamma}$
    \item $\dualvertex = \dualvertex_{\beta} \sqcup \Vhatinterface\sqcup\dualvertex_{\gamma}$
    \item $\vertex_{\beta}$ and $\dualvertex_{\gamma}$ are not directly connected by any edge.
    \item $\vertex_{\gamma}$ and $\dualvertex_{\beta}$ are not directly connected by any edge.
    \item $\Vinterface$ and $\dualvertex_{\beta}$ are not directly connected by any edge.
    \item $\Vhatinterface$ and $\vertex_{\gamma}$ are not directly connected by any edge.
\end{itemize}

We then define the bulk part of the Hamiltonian as
\begin{align}
    \begin{split}
        H_{\beta|\gamma,0}^{\repde} =& + \sum_{v\in \vertex_{\beta}} X_v \prod_{\vhat\in\adj{v}} \Zhat_{\vhat}
        - \sum_{\vhat\in\dualvertex_{\beta}} \Xhat_{\vhat} \left(\prod_{v\in \adj{\vhat}} Y_v\right) \left\{1 + \prod_{v\in\adj{\vhat}}\left(X_v \prod_{\vhat' \in \adj{v}} \Zhat_{\vhat'}\right)\right\}\\
        & - \sum_{v\in\vertex_{\gamma}} X_{v} \left(\prod_{\vhat\in \adj{v}} \Yhat_{\vhat}\right) \left\{1  + \prod_{\vhat\in\adj{v}}\left(\Xhat_{\vhat}\prod_{v' \in \adj{\vhat}} Z_{v'} \right)\right\} + \sum_{\vhat\in \dualvertex_{\gamma}} \Xhat_{\vhat} \prod_{v\in\adj{\vhat}} Z_{v}.
    \end{split}
\end{align}

The bulk state is specified by
\begin{align}
    \label{d0d_betagamma_bulk_condition_beta}
    \begin{split}
        X_{v} \prod_{\vhat\in \adj{v}} \Zhat_{\vhat} = -1,\\
        \Xhat_{\vhat} \prod_{v\in\adj{\vhat}} Y_{v} = +1,
    \end{split}
\end{align}
for each $v \in \vertex_{\beta}$ and $\vhat \in \dualvertex_{\beta}$, and
\begin{align}
    \label{d0d_betagamma_bulk_condition_gamma}
    \begin{split}
        X_{v} \prod_{\vhat\in \adj{v}} \Yhat_{\vhat} = +1,\\
        \Xhat_{\vhat} \prod_{v\in\adj{\vhat}} Z_{v} = -1,
    \end{split}
\end{align}
for each $v \in \vertex_{\gamma}$ and $\vhat \in \dualvertex_{\gamma}$.

Our next task is to compute the action of the \repdelike symmetry on the effective Hilbert space defined by (\ref{d0d_betagamma_bulk_condition_beta}) and (\ref{d0d_betagamma_bulk_condition_gamma}).
The $\ztwoV\times\ztwodualV$ symmetry acts as
\begin{align}
    \begin{split}
    U_p = \prod_{v\in\vertex} X_{v}^{p_v} \sim (-1)^{\sum_{v\in \vertex_{\beta}}{p_v}} \prod_{v\in\Vinterface} (\Xeffhatnasi_{v})^{p_v},\\
    \Uhat_{\phat} = \prod_{\vhat\in\dualvertex} \Xhat_{\vhat}^{\phat_{\vhat}} \sim (-1)^{\sum_{\vhat\in \dualvertex_{\gamma}}{\phat_{\vhat}}} \prod_{\vhat\in\Vhatinterface} (\Xeff_{\vhat})^{\phat_{\vhat}},
    \end{split}
\end{align}
where
\begin{align}
    \begin{split}
    \Xeffhatnasi_v &= X_v \left(\prod_{\vhat'\in \adj{v}\cap\Vhatinterface} \Zhat_{\vhat'}\right) \left(\prod_{\vhat'\in \adj{v}\cap\dualvertex_{\gamma}} \Yhat_{\vhat'} \right),\;\Zeffhatnasi_v = Z_v,\\
    \Xeff_{\vhat} &= \Xhat_{\vhat} \left(\prod_{v'\in \adj{\vhat}\cap\Vinterface} Z_{v'}\right) \left(\prod_{v'\in \adj{\vhat}\cap\vertex_{\beta}} Y_{v'} \right),\;\Zeff_{\vhat} = \Zhat_{\vhat},
    \end{split}
\end{align}
for each $v \in \Vinterface$ and $\vhat \in \Vhatinterface$.
The non-invertible operator $\Dee$ acts as
\begin{align}
    \Dee \Xeffhatnasi_{v} \sim (-1)^{|\dualvertex_{\gamma}\cap\adj{v}|}\Xeffhatnasi_{v} \Dee,\;
    \Dee \Xeff_{\vhat} \sim (-1)^{|\vertex_{\beta}\cap\adj{\vhat}|}\Xeff_{\vhat} \Dee,
\end{align}
for $v\in\Vinterface$ and $\vhat\in\Vhatinterface$.

Therefore we can gap out the qubits on the interface except for the corners of the bulk of the $\gamma$-model by introducing the term,
\begin{align}
    H_{\mathrm{interface}} = -\sum_{v \in \Vinterface} \frac{1+(-1)^{|\dualvertex_{\gamma}\cap\adj{v}|}}{2} \Xeffhatnasi_{v}-\sum_{\vhat \in \Vhatinterface} \frac{1+(-1)^{|\vertex_{\beta}\cap\adj{\vhat}|}}{2} \Xeff_{\vhat}.
\end{align}
We finally obtain the effective qubits on the corner as
\begin{align}
    \begin{split}
    \Veff &= \left\{\left(m_1 \ell_1 + 1, m_2 \ell_2 + 1, \cdots, m_d \ell_d + 1\right)\middle| m_1,m_2,\cdots, m_d \in \{0,1\}\right\},\\
    \Vhateff &= \left\{\left(m_1 (\ell_1+1) + \frac{1}{2}, m_2 (\ell_2+1) + \frac{1}{2}, \cdots, m_d (\ell_d+1) + \frac{1}{2}\right)\middle| m_1,m_2,\cdots, m_d \in \{0,1\}\right\}.
    \end{split}
\end{align}
The remaining degrees of freedom at this point are \emph{two} effective qubits for each corner.
We show that an interaction between a pair of the qubits for each corner reduces the degeneracy from four to two and we cannot lift the degeneracy further unless we break the symmetry.

We examine the action of the non-invertible symmetry $\Dee$ on the operators as
\begin{align}
    \begin{split}
    \Dee \prod_{v\in\Veff} \Zeffhatnasi_v &= \Dee \prod_{\vhat\in\dualvertex_{\gamma}} \prod_{v \in \adj{\vhat} }Z_v = \prod_{\vhat\in\dualvertex_{\gamma}} \Xhat_{\vhat} \Dee \\
    &= \prod_{v\in\Veff} \Zeffhatnasi_v \prod_{\vhat\in\dualvertex_{\gamma}}\left(\Xhat_{\vhat}\prod_{v'\in\adj{\vhat}}Z_{v'}\right)\Dee \\
    &\sim (-1)^{\prod_{j=1}^{d} \ell_j}\prod_{v\in\Veff} \Zeffhatnasi_v \Dee,
    \end{split}
\end{align}
and similarly,
\begin{align}
    \Dee \prod_{\vhat\in\Vhateff} \Zeff_{\vhat} \sim (-1)^{\prod_{j=1}^{d} L_j - \prod_{j=1}^{d} (\ell_j+1)} \prod_{\vhat\in\Vhateff} \Zeff_{\vhat} \Dee = (-1)^{\prod_{j=1}^{d} (\ell_j+1)} \prod_{\vhat\in\Vhateff} \Zeff_{\vhat} \Dee.
\end{align}
We note that $L_i$ are assumed to be even.
From the discussion similar to the one in Appendix \ref{appendix_of_derivation_of_noninv},
the non-invertible symmetry operator $\Dee$ on the ground space is given by
\begin{align}
    \label{eq_of_noninv_d0d2}
    \Dee \sim& \left(\prod_{v\in\Veff} \Zeffhatnasi_{v}\right) \left(\prod_{\vhat\in \Vhateff} \Zeff_{\vhat}\right) (\Xeffhatnasi_{(1,\cdots,1)})^{\prod_{j=1}^{d} \ell_j}(\Xeff_{(\frac{1}{2},\cdots,\frac{1}{2})})^{\prod_{j=1}^{d} (\ell_j+1)} \nonumber \\
    & \cdot \left(\prod_{v \in \Veff} \frac{1+\Xeffhatnasi_{v}}{2}+\prod_{v \in \Veff} \frac{1-\Xeffhatnasi_{v}}{2}\right)\left(\prod_{\vhat \in \Vhateff} \frac{1+\Xeff_{\vhat}}{2}+\prod_{\vhat \in \Vhateff} \frac{1-\Xeff_{\vhat}}{2}\right),
\end{align}
up to a nonzero constant factor.

Let us focus on the corner with the qubits at $v_{\mathrm{corner}} = (1,\cdots,1)\in\Veff$ and $\vhat_{\mathrm{corner}} = (\frac{1}{2},\cdots,\frac{1}{2})\in\Vhateff$.
After the procedure shown in Appendix \ref{appendix_of_decomposing_symmetry}, we identify the symmetry action on the corner.
The $\ztwoV$ symmetry acts on $v_{\mathrm{corner}}$ as $U_{v_{\mathrm{corner}}}^{\mathrm{eff}}=\Xeffhatnasi_{v_{\mathrm{corner}}}$.
The $\ztwodualV$ symmetry acts on $\vhat_{\mathrm{corner}}$ as $\Uhat_{\vhat_{\mathrm{corner}}}^{\mathrm{eff}}=\Xeff_{\vhat_{\mathrm{corner}}}$.
On the other hand, the subsystem non-invertible symmetry $\Dee$ is reduced to a $\mathbb{Z}_2$ symmetry generated by
\begin{align}
    V_{\Dee}^{\mathrm{eff}} = \Zeffhatnasi_{v_{\mathrm{corner}}}\Zeff_{\vhat_{\mathrm{corner}}} (\Xeffhatnasi_{(1,\cdots,1)})^{\prod_{j=1}^{d} \ell_j}(\Xeff_{(\frac{1}{2},\cdots,\frac{1}{2})})^{\prod_{j=1}^{d} (\ell_j+1)}.
\end{align}
The only symmetric deformation is $\Xeffhatnasi_{v_{\mathrm{corner}}}\Xeff_{\vhat_{\mathrm{corner}}}$\footnote{This expression is valid when restricted to the ground space, and does not realize the symmetry on the full Hilbert space. A precise representation of this operator is given by \begin{align}
    \Xeffhatnasi_{v_{\mathrm{corner}}}\Xeff_{\vhat_{\mathrm{corner}}} \frac{1}{2}\left\{1+\prod_{\vhat'\in\adj{v_{\mathrm{corner}}}\cap\dualvertex_{\gamma}}\left(\Xhat_{\vhat'}\prod_{v'\in\adj{\vhat'}}Z_{v'}\right)\right\} \frac{1}{2}\left\{1+\prod_{\vhat\in\adj{\vhat_{\mathrm{corner}}}\cap\vertex_{\beta}}\left(X_{v'}\prod_{\vhat'\in\adj{v'}}\Zhat_{\vhat'}\right)\right\},
\end{align} which is \repdelike symmetric.}, which lifts the degeneracy from four to two.
Finally, we find
\begin{align}
    V_{\Dee}^{\mathrm{eff}} U_{v_{\mathrm{corner}}}^{\mathrm{eff}} = - U_{v_{\mathrm{corner}}}^{\mathrm{eff}} V_{\Dee}^{\mathrm{eff}},\;
    V_{\Dee}^{\mathrm{eff}} \Uhat_{\vhat_{\mathrm{corner}}}^{\mathrm{eff}} = - \Uhat_{\vhat_{\mathrm{corner}}}^{\mathrm{eff}} V_{\Dee}^{\mathrm{eff}},
\end{align}
each of which protects the doubly denerate corner mode.

\subsubsection{Duality argument}
As a consistency check, we verify that the models belong to different phases after the Kennedy-Tasaki transformation.
\paragraph{$\alpha$-model}
Applying the KT transformation to the $\alpha$-model, we obtain
\begin{align}
    H_{\alpha}^{\ztwothree} = - \sum_{v\in\vertex} \prod_{\vhat\in\adj{v}} \Zhat_{\vhat} - \sum_{\vhat\in\dualvertex} \prod_{v\in\adj{\vhat}} Z_v.
\end{align}
The ground states satisfy
\begin{align}
    \begin{split}
        \prod_{\vhat\in\adj{v}} \Zhat_{\vhat} &= +1,\\
        \prod_{v\in\adj{\vhat}} Z_v &= +1.
    \end{split}
\end{align}
\paragraph{$\beta$-model}
Applying the KT transformation to the $\beta$-model, we obtain
\begin{align}
    \begin{split}
        H_{\beta}^{\ztwothree} 
        = + \sum_{v\in\vertex} \prod_{\vhat\in\adj{v}} \Zhat_{\vhat} - \sum_{\vhat\in\dualvertex} \left(\prod_{v\in\adj{\vhat}} Y_v\right) \left\{1 + \prod_{v\in\adj{\vhat}}\left(\prod_{\vhat'\in\adj{v}}\Zhat_{\vhat'}\right)\right\}.
    \end{split}
\end{align}
The ground states satisfy
\begin{align}
    \label{eqtst}
    \begin{split}
        \prod_{\vhat\in\adj{v}} \Zhat_{\vhat} &= -1,\\
        \prod_{v\in\adj{\vhat}} Y_v &= +1.
    \end{split}
\end{align}
\paragraph{$\gamma$-model}
Applying the KT transformation to the $\gamma$-model, we obtain
\begin{align}
    H_{\gamma}^{\ztwothree} = - \sum_{v\in\vertex} \left(\prod_{\vhat\in\adj{v}} \Yhat_{\vhat}\right) \left\{1 + \prod_{\vhat\in\adj{v}}\left(\prod_{v'\in\adj{\vhat}} Z_{v'}\right)\right\} + \sum_{\vhat\in\dualvertex} \prod_{v\in\adj{\vhat}} Z_{v}.
\end{align}
The ground states satisfy
\begin{align}
    \label{eqtsf}
    \begin{split}
        \prod_{\vhat\in\adj{v}} \Yhat_{\vhat} &= +1,\\
        \prod_{v\in\adj{\vhat}} Z_{v} &= -1.
    \end{split}
\end{align}
\paragraph{}
Although all of them break the $\ztwoV\times\ztwodualV$ subsystem symmetry, the additional $\ztwoCZ$ symmetry distinguishes the phases they belong to:
\begin{description}
    \item[$\alpha$-model] $\ztwoCZ$ symmetry is preserved.
    \item[$\beta$-model] $\ztwoCZ$ symmetry is broken since $Y_v$ is effectively charged under $\VCZ$ due to the first equation in (\ref{eqtst}). However, as a small portion of $\ztwoV$, we can take a usual $\mathbb{Z}_2^{\vertex, \mathrm{all}}$ 0-form symmetry generated by $\prod_{v\in\vertex} X_v$. Then, the diagonal part of $\ztwoCZ \times \mathbb{Z}_2^{\vertex, \mathrm{all}}$ is preserved.
    \item[$\gamma$-model] $\ztwoCZ$ symmetry is broken since $\Yhat_{\vhat}$ is effectively charged under $\VCZ$ due to the second equation in (\ref{eqtsf}). However, as a small portion of $\ztwodualV$, we can take a usual $\mathbb{Z}_2^{\dualvertex, \mathrm{all}}$ 0-form symmetry generated by $\prod_{\vhat\in\dualvertex} \Xhat_{\vhat}$. Then, the diagonal part of $\ztwoCZ \times \mathbb{Z}_2^{\dualvertex, \mathrm{all}}$ is preserved.
\end{description}
Therefore, we conclude again that the original models are in distinct three SPT phases.

\subsection{Subsystem non-invertible SPTs for $\cub{3,0,2}$}
\label{Section_of_302}
Next, we discuss $\cub{3,0,2}$ models with subsystem non-invertible symmetry.
In the models, a qubit associated with $\vertex$ is placed at each site $n = (n_1,n_2,n_3)\in \mathbb{Z}^3$, and a qubit associated with $\dualvertex$ is placed on each plaquette.
We work on a three-dimensional torus, defined by the periodic boundary condition $(x_1,x_2,x_3) \sim (x_1+m_1L_1,x_2+m_2L_2,x_3+m_3L_3)$ for $m_1,m_2,m_3\in\mathbb{Z}$, where $L_i\in\mathbb{Z}$ denotes the system size in the $i$-th direction.
To summarize the above,
\begin{align}
    \begin{split}
        \vertex =& \left\{(n_1,n_2,n_3)\middle|n_i \in \{0,1,\cdots,L_i - 1\},i=1,2,3\right\}, \\
        \dualvertex =& \dualvertex_1 \cup \dualvertex_2 \cup \dualvertex_3,
    \end{split}
\end{align}
where
\begin{align}
    \begin{split}
        \dualvertex_{1} &= \left\{\left(n_1,n_2+\frac{1}{2},n_3+\frac{1}{2}\right)\middle|n_i \in \{0,1,\cdots,L_i - 1\},i=1,2,3\right\}, \\
        \dualvertex_{2} &= \left\{\left(n_1+\frac{1}{2},n_2,n_3+\frac{1}{2}\right)\middle|n_i \in \{0,1,\cdots,L_i - 1\},i=1,2,3\right\}, \\
        \dualvertex_{3} &= \left\{\left(n_1+\frac{1}{2},n_2+\frac{1}{2},n_3\right)\middle|n_i \in \{0,1,\cdots,L_i - 1\},i=1,2,3\right\}.
    \end{split}
\end{align}

For the $\cub{3,0,2}$ lattice, a planar subsystem $\ztwoV$ symmetry is generated by the following operators:
\begin{align}
    U_1(n) = \prod_{n_2,n_3} X_{(n,n_2,n_3)},\; U_2(n) = \prod_{n_1,n_3} X_{(n_1,n,n_3)},\;U_3(n) = \prod_{n_1,n_2} X_{(n_1,n_2,n)}.
\end{align}
The $\ztwodualV$ symmetry is generated by operators of the form
\begin{align}
    \label{302_ztwodualV}
    \Uhat_{i}(n;C_i) = \prod_{p \in C_i} \Xhat_{p},
\end{align}
where $C_i \,(i=1,2,3)$ is a closed loop supported on the mesh-like intersection between the rigid plane $x_i=n+\frac{1}{2}$ and the plaquettes of the cubic lattice.
Each such loop $C_i$ can be regarded as a subset of $\dualvertex\setminus\dualvertex_{i}$.
The \repdelike subsystem non-invertible symmetry in this case is discussed in \cite{Ebisu:2024lie}.

\subsubsection{Interface mode}
Here, we demonstrate that the interface between the $\alpha$-model and the $\beta$-model hosts a nontrivial 1+1d hinge mode protected by the symmetry.
\paragraph{$\alpha\beta$-interface}
We analyze the interface between the $\alpha$-model and the $\beta$-model, following the setup in Section \ref{section_of_interface}.
Let $L_1,L_2,L_3$ be even integers, and choose even integers $\ell_i$ such that $2 \leq \ell_i \leq L_i-2$ for $i=1,2$.
As in the previous cases, we define the pillar-like bulk of the $\beta$-model $\mathcal{D}$ as
\begin{align}
    \mathcal{D} = \left\{(x_1,x_2,x_3)\middle|\frac{1}{2} < x_k < \ell_k + \frac{1}{2},\; k=1,2\right\}.
\end{align}
We define the interface by
\begin{align}
    \begin{split}
        \Vhatinterface = \dualvertex \cap \partial\mathcal{D},
        \dualvertex_{\beta} = \dualvertex \cap \mathcal{D},
        \vertex_{\beta} = \vertex \cap \mathcal{D},\\
        \vertex_{\alpha} = \vertex \setminus \vertex_{\beta},
        \dualvertex_{\alpha} = \dualvertex \setminus (\dualvertex_{\beta}\cup \Vhatinterface).
    \end{split}
\end{align}
In this case, some of the qubits in $\Vhatinterface$ can be gapped out. We finally obtain
\begin{align}
    \Vhateff = \Vhateff^{(0,0)} \cup \Vhateff^{(0,1)}\cup \Vhateff^{(1,0)}\cup \Vhateff^{(1,1)},
\end{align}
where each $\Vhateff^{(m_1,m_2)}$ corresponds to a set of the qubits localized at the hinge labeled by $(m_1,m_2) \in \{0,1\} \times\{0,1\}$,
\begin{align}
    \Vhateff^{(m_1,m_2)} = \left\{\left(m_1 \ell_1 + \frac{1}{2}, m_2 \ell_2 + \frac{1}{2}, n_3 \right)\middle| n_3 \in \{0,1,\cdots L_3-1\}\right\}. 
\end{align}
We argue that each hinge $\Vhateff^{(m_1,m_2)}$ has a degenerate mode protected by the symmetry.
According to (\ref{uhatphataction_on_alphabeta_interface}), $\ztwodualV$ symmetry operators hitting the hinge act nontrivially on $\Vhateff$.
We can choose $\phat$ satisfying (\ref{phatcond}) so that $\Uhat_{\phat} \sim \Xeff_{\vhat_{1}} \Xeff_{\vhat_{2}}$ for any $\vhat_{1}, \vhat_{2} \in \Vhateff$.
We also have
\begin{align}
    \Dee \prod_{\vhat\in\Vhateff} \Zeff_{\vhat} \sim \prod_{\vhat\in\Vhateff} \Zeff_{\vhat} \Dee.
\end{align}
From the discussion in Appendix \ref{appendix_of_derivation_of_noninv}, the non-invertible symmetry acts on the interface as
\begin{align}
    \label{eq_of_noninv_302}
    \Dee \sim \left(\prod_{\vhat\in \Vhateff} \Zeff_{\vhat}\right)\left(\prod_{\vhat \in \Vhateff} \frac{1+\Xeff_{\vhat}}{2} + \prod_{\vhat \in \Vhateff} \frac{1-\Xeff_{\vhat}}{2}\right),
\end{align}
up to a nonzero constant factor.

Now, let us focus on the qubits at the hinge $\Vhateff^{(0,0)}$.
After the procedure shown in Appendix \ref{appendix_of_decomposing_symmetry}, we have two kinds of symmetries. One is generated by $\Uhat_{\vhat}^{\mathrm{eff}}=\Xeff_{\vhat}$ for each $\vhat \in \Vhateff^{(0,0)}$, coming from $\ztwodualV$ symmetry.
Due to the magnetic Gauss law, $\Uhat_{\vhat}^{\mathrm{eff}}$ is independent of the choice of $\vhat \in \Vhateff^{(0,0)}$.
The other $\mathbb{Z}_2$ symmetry is
\begin{align}
    \Uhat_{\Dee}^{\mathrm{eff}} = \prod_{\vhat \in \Vhateff^{(0,0)}}\Zeff_{\vhat},
\end{align}
coming from the non-invertible symmetry.
Finally, we have the projective algebra
\begin{align}
    \Uhat_{\Dee}^{\mathrm{eff}} \Uhat_{\vhat}^{\mathrm{eff}} = - \Uhat_{\vhat}^{\mathrm{eff}} \Uhat_{\Dee}^{\mathrm{eff}},
\end{align}
which protects the degenerate hinge mode.

\subsubsection{Duality argument}
We again apply the KT transformation to the models and compare the gapped phases they belong to.

\paragraph{$\alpha$-model}
The Hamiltonian after applying the KT transformation to the $\alpha$-model reads
\begin{align}
    H_{\alpha}^{\ztwothree} = - \sum_{s \in\vertex} \prod_{\partial p \ni s} \Zhat_{p} - \sum_{p\in\dualvertex} \prod_{s\in\partial p} Z_s - g\sum_{c} \sum_{i=1}^{3} \prod_{p\in \partial c \setminus \dualvertex_{i}}\Xhat_{p}.
\end{align}
The last term imposes the magnetic Gauss law associated with the $\ztwodualV$ symmetry (\ref{302_ztwodualV}).
The ground states satisfy
\begin{align}
    \begin{split}
        \prod_{\partial p \ni s} \Zhat_{p} &= +1,\; \prod_{p\in \partial c \setminus \dualvertex_{i}}\Xhat_{p} = +1,\\
        \prod_{s \in \partial p} Z_s &= +1.
    \end{split}
\end{align}
The $\dualvertex$ sector realizes the ground states for the X-cube model, which exhibits a fracton order and spontaneous breaking of the $\ztwodualV$ subsystem symmetry.
The $\vertex$ sector is in the fracton phase of a 3+1d analogue of the plaquette Ising model, which spontaneously breaks the $\ztwoV$ subsystem symmetry.
On the other hand, the ground states of the system preserve the $\ztwoCZ$ symmetry.

\paragraph{$\beta$-model}
The Hamiltonian after applying the KT transformation to the $\beta$-model is given by
\begin{align}
    \begin{split}
        H_{\beta}^{\ztwothree} 
        &= + \sum_{s\in\vertex} \prod_{\partial p \ni s} \Zhat_{p} - \sum_{p \in \dualvertex} \left(\prod_{s \in\partial p} Y_s \right) \left\{1 + \prod_{s \in\partial p}\left(\prod_{\partial p'\ni s}\Zhat_{p'}\right)\right\} - g\sum_{c} \sum_{i=1}^{3} \prod_{p\in \partial c \setminus \dualvertex_{i}}\Xhat_{p}.
    \end{split}
\end{align}
The ground states are specified by
\begin{align}
    \begin{split}
        \prod_{\partial p \ni s} \Zhat_{p} &= -1,\; \prod_{p\in \partial c \setminus \dualvertex_{i}}\Xhat_{p} = +1,\\
        \prod_{s \in\partial p} Y_s &= +1.
    \end{split}
\end{align}
Again, the $\dualvertex$ sector realizes the ground states for the X-cube model.
The $\vertex$ sector is in the fracton phase of a 3+1d analogue of the plaquette Ising model.
The $\ztwoCZ$ symmetry is also spontaneously broken.
However, the system preserves $\mathbb{Z}_2$ 0-form symmetry generated by $\VCZ\prod_{s\in\vertex} X_s$, which is a part of $\ztwoV\times\ztwoCZ$ symmetry.

\paragraph{}
To summarize the above, after the KT transformation, we have
\begin{description}
    \item[$\alpha$-model] The preserved internal symmetry is the $\ztwoCZ$ symmetry generated by $\VCZ$.
    \item[$\beta$-model] $\ztwoCZ$ is spontaneously broken. However, as a small portion of $\ztwoV$, we can take a usual $\mathbb{Z}_2^{\vertex, \mathrm{all}}$ 0-form symmetry generated by $\prod_{v\in\vertex} X_v$. Then, the diagonal part of $\ztwoCZ \times \mathbb{Z}_2^{\vertex, \mathrm{all}}$ is preserved.
\end{description}
Therefore, we conclude that the original $\alpha$- and $\beta$-models belong to distinct \repdelike SPT phases.

\subsection{Subsystem non-invertible SPTs for general $\cub{d,0,q}$}
\label{Section_of_d0q}
We have constructed lattice models with (subsystem) non-invertible SPTs on several lattices.
Let us summarize the results so far:
\begin{description}
    \item[$\cub{d,0,1}$] In Section \ref{Section_of_d01}, we have two models which belong to distinct non-invertible SPT phases. The interface mode appears on a codimension-$1$ boundary.
    \item[$\cub{d,0,d}$] In Section \ref{Section_of_d0d}, we have \emph{three} models which belong to distinct subsystem non-invertible SPT phases. The interface modes appear on a codimension-$d$ corner.
    \item[$\cub{3,0,2}$] In Section \ref{Section_of_302}, we have two models which belong to distinct subsystem non-invertible SPT phases. The interface mode appears on a codimension-$2$ corner ($1$-dimensional hinge).
\end{description}

Based on these results, it is natural to expect that for general $\cub{d,0,q}$,
\begin{description}
    \item[$\cub{d,0,q}$] We have two (three if $q=d$) models which belong to distinct subsystem non-invertible SPT phases. The interface mode appears on a codimension-$q$ corner.
\end{description}
The goal of this section is to justify this statement.

\paragraph{Duality argument}
We first consider the duality argument for the $\alpha$-model (\ref{firstdefofalpha}) and the $\beta$-model (\ref{firstdefofbeta}).
After applying the Kennedy-Tasaki transformation to the $\alpha$- and $\beta$-models with the \repdelike symmetry, we have the ground states for each model specified by
\begin{align}
    \label{d0q_duality_condition}
    \begin{split}
        \alpha \;:\; \prod_{v\in\adj{\vhat}} Z_v &= +1,\; \prod_{\vhat\in\adj{v}} \Zhat_{\vhat} = +1,\\
        \beta \;:\; \prod_{v\in\adj{\vhat}} Y_v &= +1,\; \prod_{\vhat\in\adj{v}} \Zhat_{\vhat} = -1.
    \end{split}
\end{align}
Here, we omit the magnetic Gauss law for the $\dualvertex$ sector. However, we emphasize that \emph{no} magnetic Gauss law is imposed in the $\vertex$ sector.\footnote{In the case of $\cub{d,p,q}$ with $p>0$, we have a nontrivial magnetic Gauss law in the $\vertex$ sector, and the argument presented here does not apply.}
Thus, the ground states of each model can be labeled by configurations of the eigenvalues of $Z_v$ in the $\alpha$-model and $Y_v$ in the $\beta$-model that satisfy the above conditions.

Both models break $\ztwoV\times\ztwodualV$ spontaneously.
For the $\alpha$-model, $\VCZ$ acts trivially on the ground states, so the system preserves the $\ztwoCZ$ symmetry.
For the $\beta$-model, the operator $Y_v$ is effectively charged under $\VCZ$ due to the second condition for the $\beta$-model in (\ref{d0q_duality_condition}), and since no magnetic Gauss law is imposed in the $\vertex$ sector, the $\ztwoCZ$ symmetry is spontaneously broken.
Therefore, the original models with the \repdelike symmetry belong to distinct SPT phases.

\paragraph{Interface mode}
Let us study the interface mode for general $\cub{d,0,q}$.
We have already discussed the $\alpha\beta$-interface in Section \ref{section_of_interface}.
To see an interface mode protected by the symmetry, the interface should be defined so that
\begin{enumerate}
    \item $\Vhateff$ (defined in (\ref{definition_of_vhateff})) is not empty.
    \item There is a choice of $\phat$, which (locally) has a nontrivial sign factor in (\ref{equation_of_projective_algebra}).
    \item $\Dee$ is non-zero on the effective Hilbert space, in which the bulk is completely gapped.
\end{enumerate}
The first condition requires a codimension-$q$ corner and is satisfied by such a choice, since $|\adj{\vhat}| = 2^q$.\footnote{For $\cub{d,p,q}$, $|\adj{\vhat}|=2^{q-p}\cdot \frac{q!}{p!(q-p)!}$.}.
The second condition is satisfied since the $\ztwodualV$ symmetry is generated by line operators (with some constraints), and there always exists a line operator which intersects a given codimension-$q$ corner exactly once.
The third condition can be satisfied by choosing an appropriate system size.
There exist interfaces satisfying these conditions.
Once such an interface is realized, it is straightforward to show that the interface hosts a symmetry-protected degenerate mode by repeating the arguments of the previous sections.

\section{Weak non-invertible SPTs in $\cub{2,0,1}$}
\label{Section_of_weak_SPT}
In this section, we explore $\cub{2,0,1}$ models focusing on the \repdelike symmetry and the lattice translational symmetry with along one direction.
Specifically, two lattice models, which we will call the $\alpha'$-model and the $\gamma$-model belong to distinct SPT phases protected by the symmetry.
First, we write down the \repdelike symmetric models to study in this section.
\paragraph{$\alpha$-model}
As usual, the $\alpha$-model is given by the Hamiltonian
\begin{align}
    H_{\alpha}^{\repde} = -\sum_{s} X_s \prod_{\partial\ell \ni s} \Zhat_{\ell} - \sum_{\ell} \Xhat_{\ell} \prod_{s\in\partial \ell} Z_s.
\end{align}
The unique gapped ground state is specified by
\begin{align}
    X_s \prod_{\partial\ell \ni s} \Zhat_{\ell} = +1,\;
    \Xhat_{\ell} \prod_{s\in\partial \ell} Z_s = +1.
\end{align}
\paragraph{$\alpha'$-model}
We define the $\alpha'$-model by the Hamiltonian
\begin{align}
    \label{weakdefofalphaprime}
    H_{\alpha'}^{\repde} = -\sum_{s} X_s \prod_{\partial\ell \ni s} \Zhat_{\ell} + \sum_{\ell} \Xhat_{\ell} \prod_{s\in\partial \ell} Z_s.
\end{align}
The sign of the second term is opposite to the one of the $\alpha$-model.
The unique gapped ground state is specified by
\begin{align}
    X_s \prod_{\partial\ell \ni s} \Zhat_{\ell} = +1,\;
    \Xhat_{\ell} \prod_{s\in\partial \ell} Z_s = -1.
\end{align}
\paragraph{$\gamma$-model}
As in Section \ref{Section_of_d0d}, we introduce the $\gamma$-model,
\begin{align}
    \label{weakdefofgamma}
    H_{\gamma}^{\repde} = 
    - \sum_{s\in \vertex} X_{s} \left(\prod_{\partial\ell\ni s} \Yhat_{\ell}\right) \left\{1 + \prod_{\partial\ell\ni s} \left(\Xhat_{\ell}\prod_{s'\in \partial\ell} Z_{s'}\right)\right\} + \sum_{\ell\in \dualvertex} \Xhat_{\ell} \prod_{s \in \partial\ell} Z_{s}.
\end{align}
This model can be regarded as the $\beta$-model for $\cub{2,1,2}$ after a translation by $(\tfrac{1}{2}, \tfrac{1}{2})$.
The unique gapped ground state is specified by
\begin{align}
    \begin{split}
        X_{s} \prod_{\partial\ell \ni s} \Yhat_{\ell} = +1,\;
        \Xhat_{\ell} \prod_{s\in \partial\ell} Z_{s} = -1.
    \end{split}
\end{align}
\paragraph{}
We consider the lattice translational symmetry along the $x_2$-direction $(x_1,x_2) \mapsto (x_1,x_2 + 1)$ denoted by $\Ztrans$.
As shown in Section \ref{Section_of_d01}, the $\alpha$-model and the $\beta$-model are in different \repdelike SPT phases. In this case, the translational symmetry $\Ztrans$ is unnecessary to distinguish them.
We concentrate on the $\alpha$-, $\alpha'$- and $\gamma$-models.
Although these models belong to the same \repdelike SPT phase, the classification becomes subtle if the translational symmetry is imposed.

Let us clarify our claims in this section:
\begin{itemize}
    \item The $\alpha$-model and the $\gamma$-model belong to distinct SPT phases protected by $\ztwodualV\times\Ztrans$. No additional symmetry is required.
    \item The $\alpha$-model and the $\alpha'$-model belong to distinct SPT phases protected by $\ztwodualV\times\Ztrans$. No additional symmetry is required.
    \item The $\alpha'$-model and the $\gamma$-model belong to distinct SPT phases protected by a combination of the \repdelike symmetry and the translational symmetry $\Ztrans$. The whole symmetry is required to distinguish them. This is the main claim of this section.
\end{itemize}
We demonstrate them in the following sections.

\subsection{Interfaces}
\label{section_of_weak_interface}
Here, we study the interfaces between the $\alpha$-, $\alpha'$- and $\gamma$-models.

\paragraph{$\alpha'\gamma$-interface}
We argue that the $\alpha'\gamma$-interface has an interface mode protected by a combination of the \repdelike non-invertible symmetry and the lattice translation symmetry.
In this section, we examine only the action of the symmetry on the interface and defer a detailed analysis of the anomaly to Section \ref{section_of_LSM_anomaly}.
Since the $\gamma$-model for $\cub{d,p,q}$ is equivalent to the $\beta$-model for $\cub{d,d-q,d-p}$, and the $\alpha$-model and the $\alpha'$-model are almost the same, the analysis follows almost the same steps as for the $\alpha\beta$-interfaces in Section \ref{Section_of_Non-invertible_SPTs}.

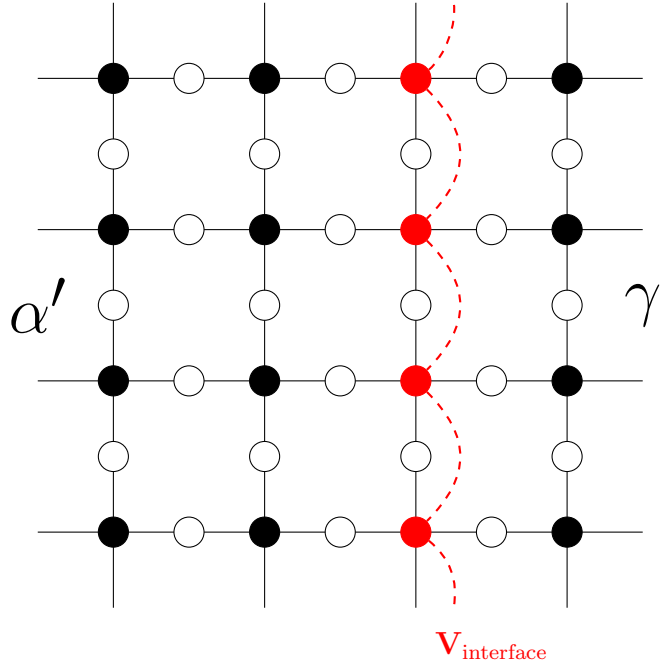
\begin{figure}[htbp]
    \begin{center}
    \begin{tikzpicture}[scale=2]
	\begin{pgfonlayer}{nodelayer}
		\node [style=none] (0) at (0, 0) {};
		\node [style=none] (2) at (0, 1) {};
		\node [style=none] (3) at (0, 2) {};
		\node [style=none] (4) at (1, 2) {};
		\node [style=none] (6) at (2, 1) {};
		\node [style=none] (7) at (2, 2) {};
		\node [style=none] (8) at (-2, 2) {};
		\node [style=none] (9) at (-1, 2) {};
		\node [style=none] (10) at (-2, 1) {};
		\node [style=none] (11) at (-2, 0) {};
		\node [style=none] (12) at (-1, 0) {};
		\node [style=none] (13) at (-2, -1) {};
		\node [style=none] (14) at (-2, -2) {};
		\node [style=none] (15) at (-1, -2) {};
		\node [style=none] (16) at (0, -2) {};
		\node [style=none] (17) at (0, -1) {};
		\node [style=none] (18) at (1, -2) {};
		\node [style=none] (19) at (2, -2) {};
		\node [style=none] (20) at (2, -1) {};
		\node [style=none] (21) at (1, 0) {};
		\node [style=none] (22) at (-3, 0) {};
		\node [style=none] (23) at (-3, 2) {};
		\node [style=none] (24) at (-2, 3) {};
		\node [style=none] (25) at (0, 3) {};
		\node [style=none] (26) at (2, 3) {};
		\node [style=none] (27) at (3, 2) {};
		\node [style=none] (28) at (3, -2) {};
		\node [style=none] (29) at (2, -3) {};
		\node [style=none] (30) at (-2, -3) {};
		\node [style=none] (31) at (-3, -2) {};
		\node [style=none] (32) at (3, 0) {};
		\node [style=none] (34) at (0, -3) {};
		\node [style=none] (35) at (2, 0) {};
		\node [style=none] (36) at (-2, -4) {};
		\node [style=none] (37) at (0, -4) {};
		\node [style=none] (38) at (2, -4) {};
		\node [style=none] (39) at (4, -4) {};
		\node [style=none] (40) at (4, -2) {};
		\node [style=none] (41) at (5, -2) {};
		\node [style=none] (42) at (4, 0) {};
		\node [style=none] (43) at (5, 0) {};
		\node [style=none] (44) at (4, 2) {};
		\node [style=none] (45) at (5, 2) {};
		\node [style=none] (46) at (-2, -5) {};
		\node [style=none] (47) at (0, -5) {};
		\node [style=none] (48) at (2, -5) {};
		\node [style=none] (49) at (4, -3) {};
		\node [style=none] (50) at (3, -4) {};
		\node [style=none] (51) at (5, -4) {};
		\node [style=none] (52) at (4, 3) {};
		\node [style=none] (53) at (4, -5) {};
		\node [style=none] (54) at (-1, -4) {};
		\node [style=none] (55) at (1, -4) {};
		\node [style=none] (56) at (-3, -4) {};
		\node [style=none] (57) at (4, 1) {};
		\node [style=none] (58) at (4, -1) {};
		\node [style=dualvertex] (59) at (-2, 1) {};
		\node [style=dualvertex] (60) at (-1, 2) {};
		\node [style=dualvertex] (61) at (0, 1) {};
		\node [style=dualvertex] (62) at (-1, 0) {};
		\node [style=dualvertex] (63) at (-2, -1) {};
		\node [style=dualvertex] (64) at (-1, -2) {};
		\node [style=dualvertex] (65) at (0, -1) {};
		\node [style=dualvertex] (66) at (-1, -4) {};
		\node [style=dualvertex] (67) at (-2, -3) {};
		\node [style=dualvertex] (68) at (0, -3) {};
		\node [style=dualvertex] (69) at (2, -3) {};
		\node [style=dualvertex] (70) at (1, -4) {};
		\node [style=dualvertex] (71) at (1, -2) {};
		\node [style=dualvertex] (72) at (1, 0) {};
		\node [style=dualvertex] (73) at (2, -1) {};
		\node [style=dualvertex] (74) at (1, 2) {};
		\node [style=dualvertex] (75) at (2, 1) {};
		\node [style=dualvertex] (76) at (3, 2) {};
		\node [style=dualvertex] (77) at (4, 1) {};
		\node [style=dualvertex] (78) at (3, 0) {};
		\node [style=dualvertex] (79) at (4, -1) {};
		\node [style=dualvertex] (80) at (3, -2) {};
		\node [style=dualvertex] (81) at (4, -3) {};
		\node [style=dualvertex] (82) at (3, -4) {};
		\node [style=vertex] (83) at (-2, 2) {};
		\node [style=vertex] (84) at (-2, 0) {};
		\node [style=vertex] (85) at (-2, -2) {};
		\node [style=vertex] (86) at (-2, -4) {};
		\node [style=vertex] (87) at (0, -4) {};
		\node [style=vertex] (88) at (0, -2) {};
		\node [style=vertex] (89) at (0, 0) {};
		\node [style=vertex] (90) at (0, 2) {};
		\node [style=boundaryvertex] (91) at (2, 2) {};
		\node [style=boundaryvertex] (92) at (2, 0) {};
		\node [style=boundaryvertex] (93) at (2, -2) {};
		\node [style=boundaryvertex] (94) at (2, -4) {};
		\node [style=vertex] (95) at (4, -4) {};
		\node [style=vertex] (96) at (4, -2) {};
		\node [style=vertex] (97) at (4, 0) {};
		\node [style=vertex] (98) at (4, 2) {};
		\node [style=none] (99) at (-3, -1) {\Huge $\alpha'$};
		\node [style=none] (100) at (5, -1) {\Huge $\gamma$};
		\node [style=none] (101) at (2.5, -5) {};
		\node [style=none] (102) at (2.5, 3) {};
		\node [style=none] (103) at (3, -5.5) {\color{red}$\mathbf{V}_{\mathrm{interface}}$};
	\end{pgfonlayer}
	\begin{pgfonlayer}{edgelayer}
		\draw (21.center) to (0.center);
		\draw (0.center) to (12.center);
		\draw (12.center) to (11.center);
		\draw (11.center) to (22.center);
		\draw (6.center) to (7.center);
		\draw (7.center) to (26.center);
		\draw (7.center) to (27.center);
		\draw (7.center) to (4.center);
		\draw (4.center) to (3.center);
		\draw (3.center) to (25.center);
		\draw (3.center) to (2.center);
		\draw (2.center) to (0.center);
		\draw (3.center) to (9.center);
		\draw (9.center) to (8.center);
		\draw (8.center) to (24.center);
		\draw (8.center) to (10.center);
		\draw (8.center) to (23.center);
		\draw (10.center) to (11.center);
		\draw (11.center) to (13.center);
		\draw (13.center) to (14.center);
		\draw (14.center) to (31.center);
		\draw (14.center) to (30.center);
		\draw (0.center) to (17.center);
		\draw (17.center) to (16.center);
		\draw (14.center) to (15.center);
		\draw (15.center) to (16.center);
		\draw (16.center) to (18.center);
		\draw (18.center) to (19.center);
		\draw (19.center) to (20.center);
		\draw (19.center) to (28.center);
		\draw (19.center) to (29.center);
		\draw (16.center) to (34.center);
		\draw (35.center) to (21.center);
		\draw (35.center) to (32.center);
		\draw (35.center) to (6.center);
		\draw (35.center) to (20.center);
		\draw (30.center) to (36.center);
		\draw (56.center) to (36.center);
		\draw (36.center) to (46.center);
		\draw (36.center) to (54.center);
		\draw (54.center) to (37.center);
		\draw (34.center) to (37.center);
		\draw (37.center) to (47.center);
		\draw (29.center) to (38.center);
		\draw (37.center) to (55.center);
		\draw (55.center) to (38.center);
		\draw (38.center) to (48.center);
		\draw (38.center) to (50.center);
		\draw (50.center) to (39.center);
		\draw (39.center) to (53.center);
		\draw (39.center) to (51.center);
		\draw (27.center) to (44.center);
		\draw (44.center) to (45.center);
		\draw (32.center) to (42.center);
		\draw (42.center) to (43.center);
		\draw (28.center) to (40.center);
		\draw (40.center) to (41.center);
		\draw (52.center) to (44.center);
		\draw (44.center) to (57.center);
		\draw (57.center) to (42.center);
		\draw (42.center) to (58.center);
		\draw (58.center) to (40.center);
		\draw (40.center) to (49.center);
		\draw (49.center) to (39.center);
		\draw [style=boundary, bend left=45, looseness=1.25] (91) to (92);
		\draw [style=boundary, bend left=45, looseness=1.25] (92) to (93);
		\draw [style=boundary, bend left=45, looseness=1.25] (93) to (94);
		\draw [style=boundary, bend left] (94) to (101.center);
		\draw [style=boundary, bend right, looseness=0.75] (91) to (102.center);
	\end{pgfonlayer}
\end{tikzpicture}
    \caption{The $\alpha'\gamma$-interface $x_1 = 0$ for $\cub{2,0,1}$.}
    \label{201alphaprimegammainterface}
    \end{center}
\end{figure}

We put the system on a torus with the periodic boundary condition $(x_1,x_2) \sim (x_1+L_1,x_2) \sim (x_1,x_2+L_2)$.
We set $L_1$ and $L_2$ to be even.
To characterize the interface, it is convenient to define the bulk of the $\gamma$-model $\mathcal{D}$ as
\begin{align}
    \mathcal{D} = \left\{(x_1,x_2)\middle|0 < x_1 < \ell_1 \right\},
\end{align}
where $\ell_1$ is an even\footnote{This assumption is not essential, but it affects the sign factor in Eq. (\ref{zzhenosayou}), which is irrelevant to our conclusions.} integer satisfying $1<\ell_1<L_1$.
We denote the surface of $\mathcal{D}$ as $\partial\mathcal{D}$.
Then, as illustrated in Figure \ref{201alphaprimegammainterface}, consider the $\alpha'\gamma$-interface specified by
\begin{align}
    \begin{split}
        \Vinterface = \vertex \cap \partial\mathcal{D},
        \vertex_{\gamma} = \vertex \cap \mathcal{D},
        \dualvertex_{\gamma} = \dualvertex \cap \mathcal{D},\\
        \vertex_{\alpha'} = \vertex \setminus (\vertex_{\gamma}\cup \Vinterface),
        \dualvertex_{\alpha'} = \dualvertex \setminus \dualvertex_{\gamma}.
    \end{split}
\end{align}
They satisfy
\begin{itemize}
    \item $\vertex = \vertex_{\alpha'} \sqcup \Vinterface \sqcup \vertex_{\gamma}$
    \item $\dualvertex = \dualvertex_{\alpha'} \sqcup\dualvertex_{\gamma}$
    \item $\vertex_{\alpha'}$ and $\dualvertex_{\gamma}$ are not directly connected by any edge.
    \item $\vertex_{\gamma}$ and $\dualvertex_{\alpha'}$ are not directly connected by any edge.
\end{itemize}
$\vertex_{\alpha'}$ and $\dualvertex_{\alpha'}$ belong to the bulk of the $\alpha'$-model.
Similarly, $\vertex_{\gamma}$ and $\dualvertex_{\gamma}$ are in the bulk of the $\gamma$-model.
$\Vinterface$ are the set of the qubits located at the interface.
We consider the bulk part of the Hamiltonian
\begin{align}
    \begin{split}
        H_{\alpha'|\gamma,0}^{\repde} =& -\sum_{s\in \vertex_{\alpha'}} X_s \prod_{\partial\ell\ni s} \Zhat_{\ell} + \sum_{\ell\in\dualvertex_{\alpha'}} \Xhat_{\ell} \prod_{s\in \partial\ell} Z_s \\
        &- \sum_{s\in \vertex_{\gamma}} X_{s} \left(\prod_{\partial\ell\ni s} \Yhat_{\ell}\right) \left\{1 + \prod_{\partial\ell\ni s} \left(\Xhat_{\ell}\prod_{s'\in \partial\ell} Z_{s'}\right)\right\} + \sum_{\ell\in \dualvertex_{\gamma}} \Xhat_{\ell} \prod_{s \in \partial\ell} Z_{s} .
    \end{split}
\end{align}
The ground states of $H_{\alpha'|\gamma,0}^{\repde}$ are specified by
\begin{align}
    \begin{split}
        X_s \prod_{\partial\ell' \ni s} \Zhat_{\ell'} &= +1,\\
        \Xhat_{\ell} \prod_{s'\in \adj{\ell}} Z_{s'} &= -1,
    \end{split}
\end{align}
for each $s \in \vertex_{\alpha'}$ and $\ell \in \dualvertex_{\alpha'}$, and
\begin{align}
    \begin{split}
        X_{s} \prod_{\partial\ell'\ni s} \Yhat_{\ell'} &= +1,\\
        \Xhat_{\ell} \prod_{s'\in\partial\ell} Z_{s'} &= -1,
    \end{split}
\end{align}
for each $s \in \vertex_{\gamma}$ and $\ell \in \dualvertex_{\gamma}$.

We study the action of the symmetry operators on the low-energy subspace which satisfies the conditions above.
The remaining degrees of freedom at this point are the effective qubits on the interface. Their Pauli operators are given by
\begin{align}
    \begin{split}
        \Xeffhatnasi_{s} &= X_{s} \left(\prod_{\ell_{\alpha}\in \dualvertex_{\alpha}\cap\adj{s}} \Zhat_{\ell_{\alpha}}\right) \left(\prod_{\ell_{\beta}\in \dualvertex_{\beta}\cap\adj{s}} Y_{\ell_{\beta}}\right),\\
        \Yeffhatnasi_{s} &= Y_{s} \left(\prod_{\ell_{\alpha}\in \dualvertex_{\alpha}\cap\adj{s}} \Zhat_{\ell_{\alpha}}\right) \left(\prod_{\ell_{\beta}\in \dualvertex_{\beta}\cap\adj{s}} Y_{\ell_{\beta}}\right),\\
        \Zeffhatnasi_{s} &= Z_{s},
    \end{split}
\end{align}
for each $s \in \Vinterface$.
The $\ztwoV$ symmetry acts as
\begin{align}
    U = \prod_{s\in\vertex} X_s \sim \prod_{s\in\Vinterface} \Xeffhatnasi_{s}.
\end{align}
On the other hand, the $\ztwodualV$ symmetry acts as
\begin{align}
    \Uhat(\hat{C}) = \prod_{\ell\in\hat{C}} \Xhat_{\ell} \sim 1,
\end{align}
since $L_1$ and $L_2$ are even and the length of any closed loop $\hat{C}$ running on the lattice is also even.
The non-invertible symmetry acts effectively on the interface qubits as
\begin{align}
    \begin{split}
    \Dee \Xeffhatnasi_{s} \sim -\Xeffhatnasi_{s} \Dee,
    \end{split}
\end{align}
for $s\in \Vinterface$.
We also have
\begin{align}
    \Dee \Zeffhatnasi_{(m\ell_1, n)} \Zeffhatnasi_{(m\ell_1, n+1)} &\sim - \Zeffhatnasi_{(m\ell_1, n)} \Zeffhatnasi_{(m\ell_1, n+1)} \Dee,\\
    \label{zzhenosayou}
    \Dee \Zeffhatnasi_{(0, n)} \Zeffhatnasi_{(\ell_1, n)} &\sim \Zeffhatnasi_{(0, n)} \Zeffhatnasi_{(\ell_1, n)} \Dee,
\end{align}
for $m=0,1,\,n = 1,2,\cdots,L_2$.
Equivalently,
\begin{align}
    \begin{split}
        \Dee \Zeffhatnasi_{v_{\mathrm{e},1}} \Zeffhatnasi_{v_{\mathrm{e},2}} &\sim \Zeffhatnasi_{v_{\mathrm{e},1}} \Zeffhatnasi_{v_{\mathrm{e},2}} \Dee,\;
        \Dee \Zeffhatnasi_{v_{\mathrm{o},1}} \Zeffhatnasi_{v_{\mathrm{o},2}} \sim \Zeffhatnasi_{v_{\mathrm{o},1}} \Zeffhatnasi_{v_{\mathrm{o},2}} \Dee,\\
        \Dee \Zeffhatnasi_{v_{\mathrm{e}}} \Zeffhatnasi_{v_{\mathrm{o}}} &\sim -\Zeffhatnasi_{v_{\mathrm{e}}} \Zeffhatnasi_{v_{\mathrm{o}}} \Dee,
    \end{split}
\end{align}
for $v_{\mathrm{e},1},v_{\mathrm{e},2},v_{\mathrm{e}}\in \Ve,\;v_{\mathrm{o},1},v_{\mathrm{o},2},v_{\mathrm{o}}\in \Vo$, where
\begin{align}
    \begin{split}
        \Ve &= \left\{(n_1,n_2)\in \Vinterface\middle| n_1 + n_2 \in 2\mathbb{Z}\right\},\\
        \Vo &= \left\{(n_1,n_2)\in \Vinterface\middle| n_1 + n_2 + 1 \in 2\mathbb{Z}\right\}.
    \end{split}
\end{align}
Furthermore, we have
\begin{align}
    \Dee^{2} \sim 1 + \prod_{s\in\Vinterface} \Xeffhatnasi_{s},
\end{align}
up to a nonzero constant factor.
Through the discussion in Appendix \ref{appendix_of_derivation_of_noninv}, the non-invertible symmetry on the interface is identified as
\begin{align}
    \label{eq_of_noninv_weak}
    \begin{split}
    \Dee &\sim \prod_{s\in\Vinterface} \Zeffhatnasi_{s} \left(\prod_{n=1,3,\cdots,L_2 - 1} \Xeffhatnasi_{(0,n)}\Xeffhatnasi_{(\ell_1,n)} + \prod_{n=2,4,\cdots,L_2} \Xeffhatnasi_{(0,n)}\Xeffhatnasi_{(\ell_1,n)}\right)\\
    &\propto \prod_{m=1,2,\cdots,\frac{L_2}{2}} \Yeffhatnasi_{(0,2m-1)}\Zeffhatnasi_{(0,2m)}\Yeffhatnasi_{(\ell_1,2m-1)}\Zeffhatnasi_{(\ell_1,2m)} + \prod_{m=1,2,\cdots,\frac{L_2}{2}} \Zeffhatnasi_{(0,2m-1)}\Yeffhatnasi_{(0,2m)}\Zeffhatnasi_{(\ell_1,2m-1)}\Yeffhatnasi_{(\ell_1,2m)},
    \end{split}
\end{align}
up to a nonzero constant factor.
We focus on the interface $x_1 = 0$ and study several symmetries on it. One is generated by $U^{\mathrm{eff}}=\prod_{n=1}^{L_2} \Xeffhatnasi_{(0,n)}$ coming from the $\ztwoV$ symmetry.
After the procedure shown in Appendix \ref{appendix_of_decomposing_symmetry}, we also have
\begin{align}
    \label{todome}
    \Uhat_{\Dee}^{\mathrm{eff}} = \prod_{m=1,\cdots,\frac{L_2}{2}} \Yeffhatnasi_{(0,2m-1)}\Zeffhatnasi_{(0,2m)},
\end{align}
coming from the non-invertible symmetry. Finally, we have the lattice translational symmetry.
In the next section, we argue that these symmetries enjoy a mixed anomaly, which implies the ingappability of the interface mode.

\paragraph{$\alpha\alpha'$-interface}
Here, we study the $\alpha\alpha'$-interface and it has an edge mode protected by the $\ztwodualV$ 1-form symmetry and the lattice translational symmetry.
We impose periodic boundary conditions with system sizes $L_1$ and $L_2$, and assume that $L_2$ is even.
As usual, we define the bulk of the $\alpha'$-model as
\begin{align}
    \mathcal{D} = \left\{(x_1,x_2)\middle|\frac{1}{2} < x_1 < \ell_1 + \frac{1}{2}\right\},
\end{align}
where $\ell_1$ is an integer with $1<\ell_1<L_1$.
Then, consider the $\alpha\alpha'$-interface specified by
\begin{align}
    \begin{split}
        \Vhatinterface = \dualvertex \cap \partial\mathcal{D},
        \dualvertex_{\alpha'} = \dualvertex \cap \mathcal{D},
        \vertex_{\alpha'} = \vertex \cap \mathcal{D},\\
        \vertex_{\alpha} = \vertex \setminus \vertex_{\alpha'},
        \dualvertex_{\alpha} = \dualvertex \setminus (\dualvertex_{\alpha'}\cup \Vhatinterface).
    \end{split}
\end{align}
The bulk Hamiltonian is given by
\begin{align}
    \begin{split}
        H_{\alpha|\alpha',0}^{\repde} =& -\sum_{s\in \vertex_\alpha} X_s \prod_{\partial\ell\ni s} \Zhat_{\ell} - \sum_{\ell\in\dualvertex_{\alpha}} \Xhat_{\ell} \prod_{s\in \partial\ell} Z_s \\
        & -\sum_{s\in \vertex_{\alpha'}} X_s \prod_{\partial\ell\ni s} \Zhat_{\ell} + \sum_{\ell\in\dualvertex_{\alpha'}} \Xhat_{\ell} \prod_{s\in \partial\ell} Z_s.
    \end{split}
\end{align}
The effective qubits localized on the interface are described by the Pauli operators
\begin{align}
    \begin{split}
        \Xeff_{\ell} &= \Xhat_{\ell} \prod_{s\in\partial\ell} Z_{s},\\
        \Yeff_{\ell} &= \Yhat_{\ell} \prod_{s\in\partial\ell} Z_{s},\\
        \Zeff_{\ell} &= \Zhat_{\ell},
    \end{split}
\end{align}
for $\ell \in \Vhatinterface$.
The magnetic Gauss law coming from the $\ztwodualV$ 1-form symmetry imposes the constraint
\begin{align}
    \label{eq_of_two_universes}
    \Xeff_{(\frac{1}{2},n)} \Xeff_{(\frac{1}{2},n+1)} = -1,
\end{align}
on the interface around $x_1 = \frac{1}{2}$.
Combining the locality of the theory, it also follows that $\Xeff_{(\frac{1}{2},n)}$ is a conserved quantity, which divides the Hilbert space of the interface theory into two universes labeled by its eigenvalue.
However, due to (\ref{eq_of_two_universes}), the one-site lattice translation swaps the two universes. Therefore, the ground states are doubly-degenerate.

\paragraph{$\alpha\gamma$-interface} The situation is the same as the $\alpha\alpha'$-interface.
The $\ztwodualV$ 1-form symmetry on the interface results in two universes and they are related by the lattice translation.
Therefore, all the spectrum including the ground states are doubly-degenerate. Note that only the translational symmetry and the $\ztwodualV$ 1-form symmetry are required.
This discussion works without any other symmetry.

\subsection{Exotic LSM anomaly on the $\alpha'\gamma$-interface}
\label{section_of_LSM_anomaly}
As shown in the previous section, the effective Hamiltonian for the $\alpha'\gamma$-interface is described by a 1+1d $S=1/2$ spin chain with $\mathbb{Z}_2^{YZ} \times \mathbb{Z}_2^{ZY}$ symmetry respectively generated by
\begin{align}
    U_{YZ} = \prod_{n} Y_{2n}Z_{2n+1},\,
    U_{ZY} = \prod_{n} Z_{2n}Y_{2n+1}
\end{align}
and the one-site translational symmetry denoted by $\Ztrans$ whose generator is $T$.
The translation operator acts as
\begin{align}
    T X_j T^{-1} = X_{j+1},\; T Y_j T^{-1} = Y_{j+1}, T Z_j T^{-1} = Z_{j+1},
\end{align}
and in particular,
\begin{align}
    T U_{YZ} T^{-1} = U_{ZY}, \; T U_{ZY} T^{-1} = U_{YZ}.
\end{align}
The total symmetry is described by the group
\begin{align}
    \label{Gtot}
    \mathrm{G}_{\mathrm{tot}} = \left(\mathbb{Z}_2^{YZ} \times \mathbb{Z}_2^{ZY}\right) \underset{}{\rtimes} \Ztrans.
\end{align}

The system enjoys a mixed anomaly between the internal symmetry and the translational symmetry, namely, Lieb-Schultz-Mattis (LSM) anomaly.
The LSM anomaly of $\mathrm{G}_{\mathrm{tot}}$ is an instance of anomalies of dipole-type symmetries pointed out in \cite{Seifnashri:2023dpa} by computing the lattice F-symbol.
Instead, we provide another simple argument which follows the one in \cite{Aksoy:2023hve}.
First, we gauge the diagonal subgroup $\mathbb{Z}_2^{\mathrm{diag}}$ of $\mathbb{Z}_2^{YZ} \times \mathbb{Z}_2^{ZY}$ generated by $\prod_j X_j$.
The dual symmetry $\mathbb{Z}_2^{\mathrm{dual}}$ then arises from the gauging.
If $\mathrm{G}_{\mathrm{tot}}$ is anomaly-free, the resulting symmetry after gauging is written as a direct product of the dual symmetry and the remaining symmetry\cite{Tachikawa:2017gyf}.
However, the symmetry after the gauging, $\hat{\mathrm{G}}_{\mathrm{tot}}$, has the same operator implementation.
Consequently, the extension
\begin{align}
    0 \to \mathbb{Z}_2^{\mathrm{dual}} \to \hat{\mathrm{G}}_{\mathrm{tot}} \to \mathbb{Z}_2 \times \Ztrans \to 0
\end{align}
is nontrivial, which implies that $\mathrm{G}_{\mathrm{tot}}$ must be anomalous.\footnote{This argument is consistent with the fact that $\mathbb{Z}_2^{YZ} \times \mathbb{Z}_2^{ZY}$ without $\Ztrans$ is anomaly-free, since the extension becomes trivial. Due to the same reason, $\mathbb{Z}_2^{YZ} \times \mathbb{Z}_2^{ZY}$ with \emph{two}-sites translation is also anomaly-free. In fact, the Hamiltonian $H=\sum_n (Y_{2n}Z_{2n+1}+Z_{2n}Y_{2n+1})$ has a trivially gapped ground state. Therefore, the \emph{one}-site translational symmetry is essential for the anomaly.}

\subsection{Duality argument}
\label{section_of_weak_duality_argument}
\paragraph{$\Deight$-type models}
First, we compare the $\alpha$-, $\alpha'$- and $\gamma$-models in the form of $\Dee_8$(-like) symmetric systems introduced in Section \ref{subsection_of_Deight}.
In other words, we gauge the $\ztwodualV$ 1-form symmetry of the models with the \repdelike symmetry and discuss the phases the resulting models belong to.
After gauging $\ztwodualV$, we obtain
\begin{align}
    \begin{split}
    \alpha &\;:\; H_\alpha^{\Deight} = -\sum_{s\in \vertex} X_{s} \Xtilde_{s} - \sum_{\ell \in \dualvertex} \prod_{s\in \partial\ell} Z_s \Ztilde_{s},\\
    \alpha' &\;:\; H_{\alpha'}^{\Deight} = -\sum_{s\in \vertex} X_{s} \Xtilde_{s} + \sum_{\ell \in \dualvertex} \prod_{s\in \partial\ell} Z_s \Ztilde_{s},\\
    \gamma &\;:\; H_{\gamma}^{\Deight} = - \sum_{s\in\vertex} X_s \Xtilde_{s} \left(\prod_{\partial\ell \in s}\prod_{s'\in\partial\ell}Z_{s'}+\prod_{\partial\ell\ni s}\prod_{s'\in\partial\ell}\Ztilde_{s'}\right) + \sum_{\ell\in \dualvertex} \prod_{s\in\partial\ell} Z_s \Ztilde_{s} .
    \end{split}
\end{align}
The ground states for each model are specified by
\begin{align}
    \begin{split}
        \alpha &\;:\; X_s \Xtilde_{s} = +1,\; \begin{tikzpicture}
	\begin{pgfonlayer}{nodelayer}
		\node [style=none] (0) at (-1, 0) {$Z \tilde{Z}$};
		\node [style=none] (1) at (1, 0) {$Z \tilde{Z}$};
		\node [style=none] (2) at (0, 0) {};
	\end{pgfonlayer}
	\begin{pgfonlayer}{edgelayer}
		\draw (0.center) to (2.center);
		\draw (2.center) to (1.center);
	\end{pgfonlayer}
\end{tikzpicture}
 = +1, \\
        \alpha' &\;:\; X_s \Xtilde_{s} = +1,\; \begin{tikzpicture}
	\begin{pgfonlayer}{nodelayer}
		\node [style=none] (0) at (-1, 0) {$Z \tilde{Z}$};
		\node [style=none] (1) at (1, 0) {$Z \tilde{Z}$};
		\node [style=none] (2) at (0, 0) {};
	\end{pgfonlayer}
	\begin{pgfonlayer}{edgelayer}
		\draw (0.center) to (2.center);
		\draw (2.center) to (1.center);
	\end{pgfonlayer}
\end{tikzpicture}
 = -1, \\
        \gamma &\;:\; \begin{tikzpicture}
	\begin{pgfonlayer}{nodelayer}
		\node [style=none] (0) at (0, 0) {$X \tilde{X}$};
		\node [style=none] (2) at (0, 1) {};
		\node [style=none] (3) at (0, 2) {$Z$};
		\node [style=none] (4) at (1, 2) {};
		\node [style=none] (6) at (2, 1) {};
		\node [style=none] (7) at (2, 2) {};
		\node [style=none] (8) at (-2, 2) {};
		\node [style=none] (9) at (-1, 2) {};
		\node [style=none] (10) at (-2, 1) {};
		\node [style=none] (11) at (-2, 0) {$Z$};
		\node [style=none] (12) at (-1, 0) {};
		\node [style=none] (13) at (-2, -1) {};
		\node [style=none] (14) at (-2, -2) {};
		\node [style=none] (15) at (-1, -2) {};
		\node [style=none] (16) at (0, -2) {$Z$};
		\node [style=none] (17) at (0, -1) {};
		\node [style=none] (18) at (1, -2) {};
		\node [style=none] (19) at (2, -2) {};
		\node [style=none] (20) at (2, -1) {};
		\node [style=none] (21) at (1, 0) {};
		\node [style=none] (22) at (-3, 0) {};
		\node [style=none] (23) at (-3, 2) {};
		\node [style=none] (24) at (-2, 3) {};
		\node [style=none] (25) at (0, 3) {};
		\node [style=none] (26) at (2, 3) {};
		\node [style=none] (27) at (3, 2) {};
		\node [style=none] (28) at (3, -2) {};
		\node [style=none] (29) at (2, -3) {};
		\node [style=none] (30) at (-2, -3) {};
		\node [style=none] (31) at (-3, -2) {};
		\node [style=none] (32) at (3, 0) {};
		\node [style=none] (34) at (0, -3) {};
		\node [style=none] (35) at (2, 0) {$Z$};
	\end{pgfonlayer}
	\begin{pgfonlayer}{edgelayer}
		\draw (21.center) to (0.center);
		\draw (0.center) to (12.center);
		\draw (12.center) to (11.center);
		\draw (11.center) to (22.center);
		\draw (6.center) to (7.center);
		\draw (7.center) to (26.center);
		\draw (7.center) to (27.center);
		\draw (7.center) to (4.center);
		\draw (4.center) to (3.center);
		\draw (3.center) to (25.center);
		\draw (3.center) to (2.center);
		\draw (2.center) to (0.center);
		\draw (3.center) to (9.center);
		\draw (9.center) to (8.center);
		\draw (8.center) to (24.center);
		\draw (8.center) to (10.center);
		\draw (8.center) to (23.center);
		\draw (10.center) to (11.center);
		\draw (11.center) to (13.center);
		\draw (13.center) to (14.center);
		\draw (14.center) to (31.center);
		\draw (14.center) to (30.center);
		\draw (0.center) to (17.center);
		\draw (17.center) to (16.center);
		\draw (14.center) to (15.center);
		\draw (15.center) to (16.center);
		\draw (16.center) to (18.center);
		\draw (18.center) to (19.center);
		\draw (19.center) to (20.center);
		\draw (19.center) to (28.center);
		\draw (19.center) to (29.center);
		\draw (16.center) to (34.center);
		\draw (35.center) to (21.center);
		\draw (35.center) to (32.center);
		\draw (35.center) to (6.center);
		\draw (35.center) to (20.center);
	\end{pgfonlayer}
\end{tikzpicture} =  +1, \; \begin{tikzpicture}
	\begin{pgfonlayer}{nodelayer}
		\node [style=none] (0) at (-1, 0) {$Z \tilde{Z}$};
		\node [style=none] (1) at (1, 0) {$Z \tilde{Z}$};
		\node [style=none] (2) at (0, 0) {};
	\end{pgfonlayer}
	\begin{pgfonlayer}{edgelayer}
		\draw (0.center) to (2.center);
		\draw (2.center) to (1.center);
	\end{pgfonlayer}
\end{tikzpicture}
 = -1.
    \end{split}
\end{align}
Due to the second conditions on $Z_{s}\Ztilde_{s}$ operators, the $\alpha$-model preserves the translational symmetry, while the other models do not.
Therefore, coming back to the original models, the $\alpha$-model is distinguished from $\alpha'$ and $\gamma$ only with the $\ztwodualV$ 1-form symmetry and the lattice translational symmetry.\footnote{Note that we gauged only the $\ztwodualV$ 1-form symmetry in this argument.}

\paragraph{$\ztwothree$-type models}
Next, we compare the $\alpha'$-model and the $\gamma$-model after applying the KT transformation:
\begin{align}
    \begin{split}
    \alpha' &\;:\; H_{\alpha'}^{\ztwothree} = - \sum_{s\in\vertex} \prod_{\partial\ell\ni s} \Zhat_{\ell} + \sum_{\ell\in\dualvertex} \prod_{s\in\partial\ell} Z_s - \sum_{p} \prod_{\ell\in \partial p} \Xhat_{\ell},\\
    \gamma &\;:\; H_{\gamma}^{\ztwothree} = - \sum_{s\in\vertex} \left(\prod_{\partial\ell\ni s} \Yhat_{\ell}\right) \left\{1 + \prod_{\partial\ell\ni s}\left(\prod_{s'\in\partial\ell} Z_{s'}\right)\right\} + \sum_{\ell\in\dualvertex} \prod_{s\in\partial\ell} Z_{s} - \sum_{p} \prod_{\ell\in \partial p} \Xhat_{\ell}.
    \end{split}
\end{align}
The ground states for each resulting model are given by
\begin{align}
    \label{eq_of_stabilizer_of_ztwothree201}
    \begin{split}
        \alpha' \;:\; \begin{tikzpicture}
	\begin{pgfonlayer}{nodelayer}
		\node [style=none] (0) at (0, 0) {};
		\node [style=none] (1) at (-2, 0) {};
		\node [style=none] (2) at (0, -2) {};
		\node [style=none] (3) at (0, -1) {$\hat{Z}$};
		\node [style=none] (4) at (-1, 0) {$\hat{Z}$};
		\node [style=none] (5) at (1, 0) {$\hat{Z}$};
		\node [style=none] (6) at (2, 0) {};
		\node [style=none] (7) at (0, 1) {$\hat{Z}$};
		\node [style=none] (8) at (0, 2) {};
	\end{pgfonlayer}
	\begin{pgfonlayer}{edgelayer}
		\draw (0.center) to (4.center);
		\draw (4.center) to (1.center);
		\draw (0.center) to (5.center);
		\draw (5.center) to (6.center);
		\draw (0.center) to (3.center);
		\draw (3.center) to (2.center);
		\draw (0.center) to (7.center);
		\draw (7.center) to (8.center);
	\end{pgfonlayer}
\end{tikzpicture} = +1,\; \begin{tikzpicture}
	\begin{pgfonlayer}{nodelayer}
		\node [style=none] (0) at (-1, 1) {};
		\node [style=none] (1) at (-1, -1) {};
		\node [style=none] (2) at (1, -1) {};
		\node [style=none] (3) at (1, 1) {};
		\node [style=none] (4) at (-1, 0) {$\Xhat$};
		\node [style=none] (5) at (0, -1) {$\Xhat$};
		\node [style=none] (6) at (1, 0) {$\Xhat$};
		\node [style=none] (7) at (0, 1) {$\Xhat$};
	\end{pgfonlayer}
	\begin{pgfonlayer}{edgelayer}
		\draw (0.center) to (4.center);
		\draw (4.center) to (1.center);
		\draw (1.center) to (5.center);
		\draw (5.center) to (2.center);
		\draw (2.center) to (6.center);
		\draw (6.center) to (3.center);
		\draw (3.center) to (7.center);
		\draw (7.center) to (0.center);
	\end{pgfonlayer}
\end{tikzpicture} = +1,\; \begin{tikzpicture}
	\begin{pgfonlayer}{nodelayer}
		\node [style=none] (0) at (-1, 0) {$Z$};
		\node [style=none] (1) at (1, 0) {$Z$};
		\node [style=none] (2) at (0, 0) {};
	\end{pgfonlayer}
	\begin{pgfonlayer}{edgelayer}
		\draw (0.center) to (2.center);
		\draw (2.center) to (1.center);
	\end{pgfonlayer}
\end{tikzpicture}
 = -1,\\
        \gamma \;:\; \begin{tikzpicture}
	\begin{pgfonlayer}{nodelayer}
		\node [style=none] (0) at (0, 0) {};
		\node [style=none] (1) at (-2, 0) {};
		\node [style=none] (2) at (0, -2) {};
		\node [style=none] (3) at (0, -1) {$\hat{Y}$};
		\node [style=none] (4) at (-1, 0) {$\hat{Y}$};
		\node [style=none] (5) at (1, 0) {$\hat{Y}$};
		\node [style=none] (6) at (2, 0) {};
		\node [style=none] (7) at (0, 1) {$\hat{Y}$};
		\node [style=none] (8) at (0, 2) {};
	\end{pgfonlayer}
	\begin{pgfonlayer}{edgelayer}
		\draw (0.center) to (4.center);
		\draw (4.center) to (1.center);
		\draw (0.center) to (5.center);
		\draw (5.center) to (6.center);
		\draw (0.center) to (3.center);
		\draw (3.center) to (2.center);
		\draw (0.center) to (7.center);
		\draw (7.center) to (8.center);
	\end{pgfonlayer}
\end{tikzpicture} = +1,\; \begin{tikzpicture}
	\begin{pgfonlayer}{nodelayer}
		\node [style=none] (0) at (-1, 1) {};
		\node [style=none] (1) at (-1, -1) {};
		\node [style=none] (2) at (1, -1) {};
		\node [style=none] (3) at (1, 1) {};
		\node [style=none] (4) at (-1, 0) {$\Xhat$};
		\node [style=none] (5) at (0, -1) {$\Xhat$};
		\node [style=none] (6) at (1, 0) {$\Xhat$};
		\node [style=none] (7) at (0, 1) {$\Xhat$};
	\end{pgfonlayer}
	\begin{pgfonlayer}{edgelayer}
		\draw (0.center) to (4.center);
		\draw (4.center) to (1.center);
		\draw (1.center) to (5.center);
		\draw (5.center) to (2.center);
		\draw (2.center) to (6.center);
		\draw (6.center) to (3.center);
		\draw (3.center) to (7.center);
		\draw (7.center) to (0.center);
	\end{pgfonlayer}
\end{tikzpicture} = +1,\; \begin{tikzpicture}
	\begin{pgfonlayer}{nodelayer}
		\node [style=none] (0) at (-1, 0) {$Z$};
		\node [style=none] (1) at (1, 0) {$Z$};
		\node [style=none] (2) at (0, 0) {};
	\end{pgfonlayer}
	\begin{pgfonlayer}{edgelayer}
		\draw (0.center) to (2.center);
		\draw (2.center) to (1.center);
	\end{pgfonlayer}
\end{tikzpicture}
 = -1.
    \end{split}
\end{align}
For each model, the $\vertex$ sector is in the antiferromagnetic phase, while the $\dualvertex$ sector realizes the toric code ground states.
Therefore, both models spontaneously break the $\ztwoV$ 0-form symmetry, the $\ztwodualV$ 1-form symmetry and the lattice translational symmetry.
However, we argue that these models belong to distinct gapped phases enriched by the lattice translational symmetry and the $\ztwoCZ$ 0-form symmetry, following the discussions in \cite{Essin:2013rca,Barkeshli:2014cna,Cheng:2015kce}.
For this goal, we consider another 1-form symmetry for each model, which comes from SSB of $\ztwodualV$.
The $\alpha'$-model has $\mathbb{Z}_2^{\alpha'}$ 1-form symmetry generated by
\begin{align}
    V_{\alpha'}(\hat{C}) = \prod_{\ell \in \hat{C}} \Zhat_{\ell},
\end{align}
where $\hat{C}$ is a closed loop running on the \emph{dual} lattice, and $\prod_{\ell \in \hat{C}}$ is the product over the links intersecting with $\hat{C}$.
Similarly, the $\gamma$-model has $\mathbb{Z}_2^{\gamma}$ 1-form symmetry generated by
\begin{align}
    V_{\gamma}(\hat{C}) = \prod_{\ell \in \hat{C}} \Yhat_{\ell}.
\end{align}
Importantly, although these symmetries are not strictly enforced, they remain robust under local perturbations by fattening the symmetry operators.
Forgetting the $\vertex$ sector, we consider the two toric code states (\ref{eq_of_stabilizer_of_ztwothree201}) realized on the $\dualvertex$ sector of the $\cub{2,0,1}$ lattice with the periodic boundary condition $(x_1,x_2)\sim(x_1+L_1,x_2)\sim(x_1,x_2+L_2)$.
We now take $\hat{C}$ as a loop wrapping the $x_2$-direction.
Then we have
\begin{align}
    \VCZ V_{\alpha'}(\hat{C}) &\sim V_{\alpha'}(\hat{C})\VCZ,\\
    \label{equation_of_Vgamma_phase}
    \VCZ V_{\gamma}(\hat{C})  &\sim (-1)^{L_2} V_{\gamma}(\hat{C})\VCZ.
\end{align}
Note that $\Yhat_{\vhat}$ is effectively charged under $\ztwoCZ$, similarly to (\ref{eq_charged_under_VCZ}).
If $L_2$ is odd, which can be interpreted as an insertion of a translational symmetry defect\cite{Seifnashri:2023dpa,Seiberg:2024gek}, (\ref{equation_of_Vgamma_phase}) yields a nontrivial sign factor.
The two line operators, $V_{\alpha'}(\hat{C})$ and $V_{\gamma}(\hat{C})$, respond to a translational symmetry defect in different ways.
In other words, these models contain sets of anyons with different symmetry fractionalization patterns.
To be precise, if we set $L_2$ to be odd, we have to insert a $\ztwoV$ defect so that the third conditions of (\ref{eq_of_stabilizer_of_ztwothree201}) are consistent.
However, it only flips the $\ztwoCZ$ charge of $\Uhat(C)$ for a closed loop $C$ wrapping the $x_2$-direction.
In particular, the fermionic lines, which are obtained from the fusion of $\Uhat(C)$ and $V_{\alpha'/\gamma}(\hat{C})$, have different $\ztwoCZ$ charges if $L_2$ is odd.
Therefore, after applying the KT transformation, the $\alpha'$-model and the $\gamma$-model belong to distinct gapped phases enriched by the lattice translational symmetry and the $\VCZ$ 0-form symmetry.
We conclude that the original models belong to different SPT phases protected by both the \repdelike non-invertible symmetry and the lattice translational symmetry.

\section{Conclusion and outlook}
\label{Section_of_conclusion}
In this work, we have constructed and analyzed lattice models that realize subsystem/weak non-invertible SPT phases.

In Section \ref{Section_of_Non-invertible_SPTs}, we introduced a family of models exhibiting subsystem non-invertible symmetry. Focusing on the graph structure $\cub{d,0,q}$, we studied the $\alpha$-model (\ref{firstdefofalpha}) and the $\beta$-model (\ref{firstdefofbeta}), and showed that they realize distinct SPT phases protected by subsystem non-invertible symmetry.
For the special case $q = d$, discussed in Section \ref{Section_of_d0d}, we also constructed the $\gamma$-model (\ref{firstdefofgamma}), which realizes a third, inequivalent SPT phase.
We found that the interface between two such phases hosts anomalous modes localized on higher codimension subspaces.

In Section \ref{Section_of_weak_SPT}, we turned to weak non-invertible SPT phases and constructed explicit lattice models on $\cub{2,0,1}$, namely the $\alpha'$-model (\ref{weakdefofalphaprime}) and the $\gamma$-model (\ref{weakdefofgamma}).
Their phases are distinguished by a combination of non-invertible symmetry and lattice translational symmetry.

There are several future directions to study.
\begin{itemize}
    \item It is worth pursuing the complete classification and construction of our subsystem non-invertible SPT phases.
    Gapped phases with non-invertible symmetry have recently been classified using the methods of symmetry topological field theory (SymTFT)\cite{Bhardwaj:2023fca,Bhardwaj:2023idu,Bhardwaj:2024qiv,Bhardwaj:2025piv}.
    SymTFT for subsystem symmetry proposed in \cite{Cao:2023rrb} would help.
    In our case, the strategy of \cite{Cao:2025qhg} would be a powerful method, since our system can be mapped to \Deightlike symmetric system as discussed in Section \ref{subsection_of_Deight}.
    \item Similarly, general classification and construction of weak non-invertible SPTs are interesting. SymTFT-like approach incorporating translational symmetry is discussed in \cite{Pace:2024acq,Pace:2025hpb}.
    \item Another direction is to analyze the LSM anomaly discussed in Section \ref{section_of_LSM_anomaly} with concrete examples.
    For example, we can consider the Hamiltonian
    \begin{align}
        H = \sum_{j} \left(Y_j Z_{j+1} + Z_{j}Y_{j+1} + \lambda X_{j}X_{j+1}\right).
    \end{align}
    This model is equivalent to the XXZ spin chain, which is effectively described by the compact boson conformal field theory (CFT).
    However, if we map the Hamiltonian to the conventional representation of the XXZ model, the realization of the symmetry (\ref{Gtot}) becomes unconventional.
    The symmetry realized in the effective theory is expected to be the one denoted by $\mathbf{Vec}_{D_8}^{\gamma}$ in \cite{Diatlyk:2023fwf}, since the latter appears generally in compact boson CFTs and is self-dual under gauging $\mathbb{Z}_2^{\mathrm{diag}}$.\footnote{The author thanks Takamasa Ando for pointing this out.}
\end{itemize}

\paragraph{Note added}
After the submission of the previous version of this draft to arXiv, a related work with some overlaps appeared \cite{ParayilMana:2025nxw}.
The author would like to clarify that the two works have been carried out independently.

\section*{Acknowledgements}
The author thanks Takamasa Ando and Takuya Okuda for helpful discussions and valuable comments on the draft.
This work was supported by JST SPRING, Grant Number JPMJSP2108.

\appendix

\section{Detailed analysis of interface modes}
\label{appendix_interface}

\subsection{Derivation of low-energy effective expressions of non-invertible symmetries}
\label{appendix_of_derivation_of_noninv}
First, we derive the equations (\ref{eq_of_noninv_d01}), (\ref{eq_of_noninv_d0d}) and (\ref{eq_of_noninv_302}).
The equation (\ref{eq_of_noninv_d0d2}) is also shown by a discussion similar to the one in this appendix.
Although the setups of (\ref{eq_of_noninv_d01}), (\ref{eq_of_noninv_d0d}) and (\ref{eq_of_noninv_302}) look different,
these situations are actually reduced to the following problem.
Assume that we have the effective qubits on the interface labeled by $v \in \setto$, where $\setto$ is the set of labels of the qubits we have.
Each qubit labeled by $v\in\setto$ has the Pauli operators denoted by $X_v,Z_v$.
We also assume that the non-invertible operator $\Dee$, regarded as an operator acting on the low-energy effective Hilbert space, is nonzero\footnote{This condition is assured by choosing the system size so that the low-energy effective Hilbert space is compatible with $U_p=\Uhat_{\phat}=+1$, since $\Dee^2$ is nonzero and proportional to a summation over all possible $U_p$ and $\Uhat_{\phat}$\cite{Gorantla:2024ocs}.} and satisfies the following algebra:
\begin{align}
    \label{cond1}
    X_v \Dee &= -\Dee X_v\;\;(v\in \setto), \\
    \label{cond2}
    \Dee X_{v_1} X_{v_2} &= \Dee = X_{v_1} X_{v_2} \Dee \;\; (v_1,v_2\in \setto), \\
    \label{cond3}
    \prod_{v\in \setto} Z_{v} \Dee &= \Dee \prod_{v\in \setto} Z_{v}.
\end{align}
The first condition (\ref{cond1}) comes from Eq. (\ref{eq_of_D_flips_X}).
The second condition (\ref{cond2}) is associated with the fact that the non-invertible symmetry operator $\Dee$ absorbs $\ztwoV\times\ztwodualV$ symmetry generators as the equation (\ref{D_absorbs_U}).
The third condition (\ref{cond3}) implies a product of $Z$ (or $\Zhat$) operators for all qubits is not effectively charged under $\Dee$.
Under these assumptions, what we want to show is
\begin{align}
    \label{result_of_appendix1}
    \Dee = K \prod_{v\in\setto} Z_v \left(\prod_{v\in\setto}\frac{1+X_v}{2} + \prod_{v\in\setto}\frac{1-X_v}{2}\right),
\end{align}
where $K$ is a nonzero c-number.

From the first condition (\ref{cond1}), the non-invertible operator should have the following form:
\begin{align}
    \Dee = \left(\prod_{v\in\setto} Z_v \right) f(X),
\end{align}
where $f(X)$ is a polynomial of $X$s.
The second condition (\ref{cond2}) implies
\begin{align}
    f(X)X_{v_1}X_{v_2} = f(X),
\end{align}
for $v_1,v_2\in\setto$.
This implies that $f(X)=0$ on the subspace satisfying $X_{v_1} = - X_{v_2}$ for some $v_1,v_2\in\setto$.
Therefore, for fixed $u\in\setto$, we obtain
\begin{align}
    f(X) = (a X_{u} + b) \left(\prod_{v\in\setto}\frac{1+X_v}{2} + \prod_{v\in\setto}\frac{1-X_v}{2}\right),
\end{align}
where $a$ and $b$ are c-numbers.
Then, using the third condition (\ref{cond3}) yields $a=0$.
Since $\Dee$ is nonzero, we obtain (\ref{result_of_appendix1}).

We next derive (\ref{eq_of_noninv_weak}).
See Section \ref{section_of_weak_interface} for the setup.
In this case, we have a qubit for each element of $\Veff = \Ve \sqcup \Vo$.
Note that $|\Ve|=|\Vo|=L_2$, which is even.
$\Dee$ has the following properties:
\begin{align}
    \label{2cond1}
    \Xeffhatnasi_{v} \Dee &= - \Dee \Xeffhatnasi_{v} \;\; v\in\Veff,\\
    \label{2cond2}
    \Zeffhatnasi_{v_1}\Zeffhatnasi_{v_2} \Dee &= \Dee \Zeffhatnasi_{v_1}\Zeffhatnasi_{v_2}, \;\; v_1,v_2 \in \Ve,\\
    \label{2cond3}
    \Zeffhatnasi_{v_1}\Zeffhatnasi_{v_2} \Dee &= \Dee \Zeffhatnasi_{v_1}\Zeffhatnasi_{v_2}, \;\; v_1,v_2 \in \Vo,\\
    \label{2cond4}
    \Zeffhatnasi_{v_1}\Zeffhatnasi_{v_2} \Dee &= -\Dee \Zeffhatnasi_{v_1}\Zeffhatnasi_{v_2}, \;\; v_1 \in \Ve, v_2 \in \Vo,\\
    \label{2cond5}
    \Dee &= \Dee^{\dagger},\\
    \label{2cond6}
    \Dee^2 &= K' \left(1+\prod_{v\in\Veff}\Xeffhatnasi_{v}\right),
\end{align}
where $K'$ is a nonzero constant.
From the first condition (\ref{2cond1}), $\Dee$ can be written as
\begin{align}
    \Dee = \left(\prod_{v\in\Veff}\Zeffhatnasi_{v}\right) g(\Xeffhatnasi),
\end{align}
where $g(\Xeffhatnasi)$ is a polynomial of $\Xeffhatnasi$s.
Using the conditions (\ref{2cond2}), (\ref{2cond3}) and (\ref{2cond4}), $g(\Xeffhatnasi)$ must have the following form:
\begin{align}
    g(\Xeffhatnasi) = C_{\mathrm{e}} \prod_{v\in\Ve} \Xeffhatnasi_{v} + C_{\mathrm{o}} \prod_{v\in\Vo} \Xeffhatnasi_{v},
\end{align}
where $C_{\mathrm{e}}$ and $C_{\mathrm{o}}$ are c-numbers.
Equation (\ref{2cond5}) implies that $C_\mathrm{e}$ and $C_{\mathrm{o}}$ are real.
From Eq. (\ref{2cond6}), we find $C_\mathrm{e} = C_{\mathrm{o}} \neq 0$, and finally obtain (\ref{eq_of_noninv_weak}).

\subsection{Decomposition of a non-invertible symmetry into symmetries realized on spatially separated interfaces}
\label{appendix_of_decomposing_symmetry}
We discuss the way to extract the symmetry realized at the interface from the action of non-invertible symmetry on the effective qubits, such as (\ref{eq_of_noninv_general}).
We suppose effective qubits are spatially separated in two regions, $\mathrm{L}$ and $\mathrm{R}$\footnote{In more general case in which we have more regions $\mathrm{R}_i$ for $i=1,2,\cdots,N$ and we are interested in the region $\mathrm{R}_1$, we can set $\mathrm{L} = \mathrm{R}_1$ and $\mathrm{R} = \bigcup_{i=2}^{N}\mathrm{R}_i$.}.
The original Hamiltonian is written as a sum of symmetric local terms.
In the low-energy limit, the degrees of freedom in the bulk are frozen, and
the low-energy effective Hilbert space is given by $\mathcal{H} = \mathcal{H}_{\mathrm{L}}\otimes \mathcal{H}_{\mathrm{R}}$, where $\mathcal{H}_{\mathrm{L/R}}$ is the Hilbert space for effective qubits localized on the region $\mathrm{L/R}$\footnote{We consider only bosonic systems.}.
The effective Hamiltonian on the interfaces can be written as
\begin{align}
    H = H_{\mathrm{L}} \otimes \mathrm{id}_{\mathrm{R}} + \mathrm{id}_{\mathrm{L}} \otimes H_{\mathrm{R}}.
\end{align}
Importantly, $H_{\mathrm{L}/\mathrm{R}}$ is determined only by the information around the region $\mathrm{L}/\mathrm{R}$, and they do not affect each other.
Therefore, $H_{\mathrm{L}} \otimes \mathrm{id}_{\mathrm{R}}$ and $\mathrm{id}_{\mathrm{L}} \otimes H_{\mathrm{R}}$ must be symmetric, independently of each other.
Now, as done in \cite{Seifnashri:2024dsd}, we consider a non-invertible symmetry operator acting on the effective qubits,
\begin{align}
    \Dee = \sum_{I} \Dee_{\mathrm{L}}^{(I)} \otimes \Dee_{\mathrm{R}}^{(I)},
\end{align}
where $\Dee_{\mathrm{L/R}}^{(I)}$ acts on $\mathcal{H}_{\mathrm{L/R}}$. We take $\{\Dee_{\mathrm{R}}^{(I)}\}_{I}$ such that they are linearly independent.

Let us focus on $\mathrm{L}$.
Since $[\Dee,H_{\mathrm{L}} \otimes \mathrm{id}_{\mathrm{R}}]=0$, we have
\begin{align}
    \left[\sum_{I} \left(\Dee_{\mathrm{R}}^{(I)}\right)_{ij} \Dee_{\mathrm{L}}^{(I)}, H_{\mathrm{L}}\right] = 0,
\end{align}
where $\left(\Dee_{\mathrm{R}}^{(I)}\right)_{ij}$ is a matrix element of $\Dee_{\mathrm{R}}^{(I)}$ with some basis.
Since $\{\Dee_{\mathrm{R}}^{(I)}\}_{I}$ is linearly independent,
$\Dee_{\mathrm{L}}^{(I)}$ can be regarded as a symmetry operator on $\mathcal{H}_\mathrm{L}$. Note that they are not independent in general.
In the case of (\ref{eq_of_noninv_weak}), we have
\begin{align}
    \begin{split}
    \Dee_{\mathrm{L}}^{(1)} &= \prod_{m=1,2,\cdots,\frac{L_2}{2}} \Yeffhatnasi_{(0,2m-1)}\Zeffhatnasi_{(0,2m)},\; \Dee_{\mathrm{R}}^{(1)} = \prod_{m=1,2,\cdots,\frac{L_2}{2}} \Yeffhatnasi_{(\ell_1,2m-1)}\Zeffhatnasi_{(\ell_1,2m)},\\
    \Dee_{\mathrm{L}}^{(2)} &= \prod_{m=1,2,\cdots,\frac{L_2}{2}} \Zeffhatnasi_{(0,2m-1)}\Yeffhatnasi_{(0,2m)},\; \Dee_{\mathrm{R}}^{(2)} = \prod_{m=1,2,\cdots,\frac{L_2}{2}} \Zeffhatnasi_{(\ell_1,2m-1)}\Yeffhatnasi_{(\ell_1,2m)}.
    \end{split}
\end{align}
$\Dee_{\mathrm{L}}^{(1)}$ and $\Dee_{\mathrm{L}}^{(2)}$ are symmetries realizing on the interface $\mathrm{L}$ around $x_1=0$. Then, we obtain the symmetry (\ref{todome}).
\newpage

\bibliography{bibnoninvSPT}
\bibliographystyle{ytphys}

\end{document}